\documentclass[preprintnumbers,nofootinbib,
superscriptaddress,amsmath,floatfix,prd]{revtex4}

\usepackage{amssymb}  % \gtrsim, \geqslant, etc: see amsguide.ps
\usepackage{graphicx}
\usepackage{exscale}
\usepackage{textcomp}
\usepackage{enumerate}
\usepackage{color}

\usepackage[normalem]{ulem}  % \sout{old text} for strikeout

\newcommand{\non}{\nonumber\\}

\newcommand{\be}{\begin{equation}}
\newcommand{\ee}{\end{equation}}
\newcommand{\bea}{\begin{eqnarray}}
\newcommand{\eea}{\end{eqnarray}}
\newcommand{\ba}[1]{\begin{array}{#1}}
\newcommand{\ea}{\end{array}}

\newcommand{\Tr}{{\rm Tr}}

\newcommand{\uk}{\hat{\mathbf{k}}}

\begin{document}

\title{From a complex scalar field to the two-fluid picture of superfluidity}

\author{Mark G.\ Alford}
\email{alford@wuphys.wustl.edu}
\affiliation{Department of Physics, Washington University St Louis, MO, 63130, USA}

\author{S. Kumar Mallavarapu}
\email{kumar.s@go.wustl.edu}
\affiliation{Department of Physics, Washington University St Louis, MO, 63130, USA}

\author{Andreas Schmitt}
\email{aschmitt@hep.itp.tuwien.ac.at}
\affiliation{Institut f\"{u}r Theoretische Physik, Technische Universit\"{a}t Wien, 1040 Vienna, Austria}

\author{Stephan Stetina}
\email{stetina@hep.itp.tuwien.ac.at}
\affiliation{Institut f\"{u}r Theoretische Physik, Technische Universit\"{a}t Wien, 1040 Vienna, Austria}

\date{25 July 2013}

\begin{abstract}

The hydrodynamic description of a superfluid is usually based on
a two-fluid picture. We compute the basic properties of the 
relativistic two-fluid
system from the underlying microscopic physics of a
relativistic $\varphi^4$ complex scalar field theory.
We work at nonzero but small temperature and weak coupling, and we neglect dissipation.
We clarify the relationship between different 
formulations of the two-fluid model, and how they
are parameterized
in terms of partly redundant current and momentum 4-vectors. As an application, we compute the velocities of first and second sound
at small temperatures and in the presence of a superflow. 
While our results are of a very general nature,
we also comment on their interpretation as a step towards the hydrodynamics of the color-flavor locked state of quark matter, which, in particular 
in the presence of kaon condensation, appears to be a complicated multi-component fluid.

\end{abstract}

%\pacs{12.38.Mh,24.85.+p}

\maketitle

\section{Introduction}
\label{intro}

A superfluid at finite temperature can be described ``macroscopically'' in terms
of a two-fluid model involving a superfluid and a normal fluid. It can also be
described ``microscopically'' in terms of a condensate associated
with the spontaneous breaking of a $U(1)$ global symmetry in an underlying
field theory.
In this paper, we show how to relate the macroscopic hydrodynamics of a 
non-dissipative relativistic superfluid to the microscopic field-theoretic description.

\subsection{Macroscopic two-fluid description}
The two-fluid picture was suggested by Tisza \cite{tisza38} shortly after 
the superfluid nature of $^4$He was established experimentally \cite{1938Natur.141...74K,1938Natur.141...75A}. At zero temperature, there is only a superfluid 
component, while at nonzero temperatures below the critical temperature, both fluid components coexist.
Landau \cite{landau41} predicted a ``second sound'' mode where
the {\it relative} densities of the two fluids oscillate, resulting in a temperature (or entropy) wave. This is in 
contrast to an ordinary sound wave, where the {\it total} density (or pressure) oscillates.

The generalization to relativistic superfluids was introduced by Lebedev and Khalatnikov \cite{1982PhLA...91...70K,1982ZhETF..83.1601L}
and Carter 
\cite{carter89}. The approaches in these works---termed ``potential'' and ``convective'' variational approaches, respectively---differ in their formulation but are equivalent and can be translated into each other \cite{1992PhRvD..45.4536C,Andersson:2006nr}. The convective approach uses two four-currents,
corresponding to total conserved charge and 
entropy (which is also conserved in the dissipationless case).
The potential approach uses two conjugate four-momenta, which correspond to the 
superfluid four-velocity (multiplied by the chemical potential) and a thermal four-vector whose temporal component is the temperature.
The conventional presentation of the two-fluid model,
although it looks like two currents (super current and normal current)
is really based on one current (the entropy current)
and one momentum (proportional to the superfluid velocity). 
We explain how the various formalisms are related via Legendre transforms
of two Lorentz scalar quantities, a generalized pressure and a generalized energy density.
We calculate all these quantities in terms of the microscopic theory.

\subsection{Microscopic field-theoretic description}

The main physical ingredient of a superfluid is a Bose-Einstein condensate, which is responsible for the frictionless flow. This was realized 
by London, who pointed out the connection between the phenomenology of $^4$He and the quantum statistics of bosons \cite{1938Natur.141..643L}. 
In a more
modern, field-theoretic language, the superfluid state spontaneously breaks a $U(1)$ symmetry of the underlying Lagrangian. As a consequence,
there is a Goldstone mode whose dispersion is linear for small momenta. Using the terminology introduced in the context of $^4$He, this mode
is called a phonon. Roughly speaking, thermal excitations of this mode constitute  the normal component in the two-fluid
picture, while the superfluid component corresponds to the quantum mechanical
ground state. We will clarify this 
distinction below (see Sec.~\ref{sec:T0}).

Our microscopic model will be a relativistic $\varphi^4$ field theory with a
$U(1)$ symmetry that is spontaneously broken by a Bose-Einstein condensate of
the fundamental scalars. We derive the two-fluid picture from this
model. We express the basic currents and momenta
of the two-fluid formalism in terms of field-theoretic
quantities, and calculate the effective action at nonzero temperature 
in the presence of a homogeneous, static superflow. 
% As a consequence, the dispersion of the Goldstone mode becomes anisotropic
% since it will depend on the angle between the superflow and the momentum of
% the excitation.

\subsection{Previous work and future applications}

Our approach is related to several existing studies in the literature.
For instance, in Ref.~\cite{Carter:1995if} the two-fluid formalism is connected with a simple statistical approach to the phonon contribution, 
Ref.~\cite{Comer:2002dm} connects it with a Walecka model describing nuclear matter in a neutron star, and Ref.\ \cite{Nicolis:2011cs} 
makes the connection to a very general 
effective field theory. In Ref.~\cite{Son:2002zn}, a hydrodynamic
interpretation of a field-theoretic effective action was discussed for zero temperature. For the simplified case of a dissipationless, 
homogeneous fluid, our study is a generalization of this work to nonzero temperatures. In that reference, also an effective Lagrangian for
the phonons is formulated which has been employed to study transport properties of quark matter 
\cite{Manuel:2004iv,Alford:2007rw,Manuel:2007pz,Mannarelli:2008jq,Alford:2008pb,Mannarelli:2009ia}. 

%This effective theory, however, is difficult to translate into the 
%two-fluid picture commonly used in the discussion of neutron star physics. 

Our study is very general since we do not have to specify the system for which our ``microscopic'' theory, the $\varphi^4$ model, is an 
effective description. As a specific application, however, we are ultimately interested in a complete hydrodynamic description of the color-flavor 
locked (CFL) phase of dense quark matter \cite{Alford:1998mk}.
The hydrodynamics of dense matter, be it dense nuclear or quark matter, is very important for the understanding of the physics of compact 
stars.  One prominent example that relates the hydrodynamic properties to an observable -- in this case the rotation frequency of the star -- 
is the $r$-mode instability \cite{Andersson:1997xt}, for recent developments relating the instability to microscopic physics see
Refs.~\cite{Jaikumar:2008kh,Andersson:2010sh,Alford:2010fd,Alford:2011pi}. Other examples include pulsar glitches, magnetohydrodynamics, 
and asteroseismology; see for instance Refs.~\cite{1988ApJ...333..880E,1995AA...303..515L,1998MNRAS.297.1189L,Haskell:2012vp,Glampedakis:2011yw} for applications of 
multi-fluid hydrodynamics in the astrophysical context. Since the densities in the cores of compact stars can be as 
large as several times nuclear ground state density, it is conceivable that the cores are made of deconfined quark matter. The CFL phase is
the ground state of three-flavor, asymptotically dense quark matter and therefore also of fundamental theoretical interest. Whether it persists 
down to densities that exist inside a compact star is unknown, and many other candidate phases have been discussed \cite{Alford:2007xm}.
Taking into account the effect of a non-negligible strange quark mass, the next phase down in density appears to be the kaon-condensed CFL phase 
(CFL-$K^0$) \cite{BedaqueSchaefer}, whose superfluid properties are expected to be nontrivial.
Firstly, the quark Cooper pair condensate of CFL spontaneously breaks the $U(1)$ associated with baryon number conservation, making CFL a 
baryon superfluid. Secondly, the 
kaon condensate in CFL-$K^0$ spontaneously breaks the $U(1)$ associated with strangeness conservation. This $U(1)$, however, is explicitly 
broken since weak interactions have to be taken into account in the astrophysical context. As a consequence, the corresponding Goldstone mode 
has a mass of the order of 
$50\,{\rm keV}$ \cite{Son:2001xd} which is small compared to the critical temperature for kaon condensation, estimated to be of the order of tens of 
MeV \cite{Alford:2007qa}.  Therefore, the question arises how many (super)fluid components CFL-$K^0$ has, and how the two-fluid
picture is modified (or whether superfluidity survives in any sense) when there is a small {\it explicit} symmetry breaking of the $U(1)$.

\subsection{Overview}
In Sec.~\ref{sec:zeroT} we start with the zero-temperature case, where the relation between field theory and 
superfluid hydrodynamics is well established. We elaborate on this relation, which sets the notation and prepares us for the nonzero-temperature
case. The remaining paper is divided into four parts: the microscopic nonzero-temperature calculation, Sec.~\ref{sec:nonzeroT}, introducing 
the two-fluid formalism, Sec.~\ref{sec:2fluid}, and connecting both, Sec.~\ref{sec:translate}; then we use our results to compute the sound velocities
in the presence of a superflow in Sec.~\ref{sec:sound}.

Our convention for the metric tensor is $g^{\mu\nu}={\rm diag}(1,-1,-1,-1)$. Our units are $\hbar=c=k_B=1$. We work in the imaginary-time 
formalism, i.e., $T/V \sum_k \equiv T \sum_n \int d^3{\bf k}/(2\pi)^3$, where $n$ labels the Matsubara 
frequencies $\omega_n \equiv i k_0$. For bosons $\omega_n=2n \pi T$, for fermions $\omega_n=(2n+1) \pi T$.

\section{Zero temperature: single fluid}
\label{sec:zeroT}
\subsection{Lagrangian}

Our starting point is the Lagrangian
\be \label{lagr}
{\cal L} = \partial_\mu\varphi\partial^\mu\varphi^* -m^2|\varphi|^2-\lambda|\varphi|^4 \, ,
\ee
with the complex scalar field $\varphi$, the mass parameter $m>0$, and the coupling constant $\lambda>0$. The Lagrangian is obviously invariant 
under global $U(1)$ rotations $\varphi\to e^{i\alpha}\varphi$ which leads to a conserved charge\footnote{
In the context of CFL-$K^0$, this Lagrangian can be read as an effective
theory for (neutral) kaons (neglecting the small explicit symmetry breaking
due to the weak interactions).  In this case, $m$ and $\lambda$ can be
expressed in terms of the effective kaon chemical potential and the effective
kaon mass which, in turn, can be related to high-density QCD via the
high-density effective theory \cite{BedaqueSchaefer}.}.
Later, in our explicit low-temperature results, we shall set $m=0$ for simplicity. 
Note that, since $m^2>0$, the quadratic term of the classical potential 
is positive, and spontaneous
symmetry breaking can only occur if we introduce a chemical potential
associated to the conserved charge. If this chemical potential is larger
than $m$, a Bose-Einstein condensate of the particles is formed.
In our formalism, we shall introduce such a chemical potential
through the time dependence of the phase of the order parameter.

We allow for Bose-Einstein condensation in the usual way by separating
the condensate, 
\be
\varphi(x) = \frac{e^{i\psi(x)}}{\sqrt{2}}\left[\rho(x)+\varphi'_1(x)+i
\varphi'_2(x)\right] \, .
\ee
Here, $\rho(x)$ is the modulus and $\psi(x)$ the phase of the
condensate. For convenience, we have introduced the transformed
fluctuation field $\varphi'(x)$, which 
we have written in terms of its real and imaginary parts. Inserting
this into the Lagrangian yields
\be
{\cal L} = -U + {\cal L}^{(1)}+ {\cal L}^{(2)}+ {\cal L}^{(3)}+ {\cal L}^{(4)} \, , 
\ee
with the tree-level potential
\be \label{Utree}
U= -\frac{1}{2}\partial_\mu\rho\,\partial^\mu\rho-\frac{\rho^2}{2}(\partial_\mu\psi \partial^\mu\psi-m^2)+\frac{\lambda}{4}\rho^4 \, , 
\ee
and the fluctuation terms, listed by their order in the fluctuation from linear to quartic,
\begin{subequations}
\bea
{\cal L}^{(1)}&=& \partial_\mu\psi(\rho\partial^\mu\varphi_2'-\varphi_2'\partial^\mu\rho)+\rho(\partial_\mu\psi\partial^\mu\psi-m^2-\lambda\rho^2)\varphi_1'
+\partial_\mu\varphi_1'\partial^\mu\rho \, , \\[2ex]
{\cal L}^{(2)}&=&\frac{1}{2}\Big[\partial_\mu\varphi_1'\partial^\mu\varphi_1'+\partial_\mu\varphi_2'\partial^\mu\varphi_2'
+(\varphi_1'^2+\varphi_2'^2)(\partial_\mu\psi\partial^\mu\psi-m^2) \non [2ex]
&& +\,2\partial_\mu\psi(\varphi_1'\partial^\mu\varphi_2'
-\varphi_2'\partial^\mu\varphi_1')-\lambda\rho^2(3\varphi_1'^2+\varphi_2'^2)\Big] \, , \label{L2}\\[2ex]
{\cal L}^{(3)}&=& -\lambda\rho\varphi_1'(\varphi_1'^2+\varphi_2'^2) \, , \\[2ex]
{\cal L}^{(4)}&=& -\frac{\lambda}{4}(\varphi_1'^2+\varphi_2'^2)^2 \, .
\eea
\end{subequations}
In this paper, we shall only need the tree-level potential and the terms quadratic in the fluctuations. We neglect the cubic and quartic terms 
because we restrict ourselves to the one-loop effective action, see discussion at the beginning of Sec.~\ref{sec:nonzeroT}. The linear terms ${\cal L}^{(1)}$
can be rewritten such that they are -- up to a total derivative term -- proportional to the equations of motion (\ref{EOMs}), and thus they do not contribute to the 
on-shell action.

\subsection{Zero-temperature hydrodynamics}
\label{sec:T0}

Let us first discuss the tree-level potential and ignore all fluctuations. This allows us to present the translation to the 
hydrodynamic equations and quantities in the simplest case, based on Ref.~\cite{Son:2002zn}. Later we shall include fluctuations to 
generalize this translation to nonzero temperatures.

The classical equations of motion are
\begin{subequations} \label{EOMs}
\bea
\Box\rho &=& \rho(\sigma^2-m^2-\lambda\rho^2) \, , \\[2ex]
\partial_\mu(\rho^2\partial^\mu\psi) &=& 0 \label{eompsi} \, , 
\eea
\end{subequations}
where we have abbreviated 
\be \label{sigma}
\sigma^2 \equiv \partial_\mu\psi\partial^\mu\psi \, .
\ee
The field-theoretic expressions for the conserved charge current and 
the stress-energy tensor are
\begin{subequations}\label{jTmicro}
\bea 
j^\mu &=& \frac{\partial {\cal L}}{\partial(\partial_\mu\psi)} = \rho^2 \partial^\mu\psi  \, , \label{jmicro}\\[2ex]
T^{\mu\nu} &=& \frac{2}{\sqrt{-g}}\frac{\delta(\sqrt{-g}\,{\cal L})}{\delta g_{\mu\nu}} = 
2\frac{\partial {\cal L}}{\partial g_{\mu\nu}} - g^{\mu\nu}{\cal L} = \partial^\mu\rho\,\partial^\nu\rho +\rho^2 \partial^\mu\psi\partial^\nu\psi 
- g^{\mu\nu}{\cal L} \, , \label{Tmicro}
\eea
\end{subequations}
For the stress-energy tensor, we have used the gravitational definition by formally introducing a general metric $g$ which, after taking
the derivatives, we set to be the metric of flat Minkowski space. 
This definition guarantees that $T^{\mu\nu}$ is symmetric and 
conserved. Since $j^\mu$ is also conserved, we have 
\be \label{conserve}
\partial_\mu j^\mu = 0 \, ,\qquad \partial_\mu T^{\mu\nu} = 0 \, , 
\ee
which are the hydrodynamic equations. 
As we expect from Noether's theorem,
current conservation is identical to the equation of motion (\ref{eompsi})
for $\psi$.

%The reason is that $\psi$ only appears through its derivatives in the 
%Lagrangian, as it should be for the radial mode in the presence of an 
%exact $U(1)$ symmetry.

As an ansatz for the solution of the equations of motion we may choose
\begin{subequations} \label{ansatzpsirho}
\bea
\psi &=& p_\mu x^\mu +
{\rm Re}\sum_{\bf k} \delta\psi_{\bf k}\,e^{i(\omega t-{\bf k}\cdot{\bf x})} \, , \\[2ex]
\rho &=& \sqrt{\frac{p^2-m^2}{\lambda}}
 + {\rm Re}\sum_{\bf k} \delta \rho_{\bf k}\,e^{i(\omega t-{\bf k}\cdot{\bf x})} \, .
\eea
\end{subequations}
In this ansatz, we assume $\partial^\mu\psi$ and $\rho$ are each
composed of a static part plus small oscillations around it.
The static part of the solution is the superfluid mode,
corresponding to an infinite and uniformly flowing superfluid.
The density and flow are specified by the
values of the components of $p_\mu$, which are
pure numbers, not functions of $x$, and are
{\em not} constrained by the equation of motion.
The value of $p_\mu$ is determined by the boundary conditions, which
specify the topology of the field configuration, namely the number of times 
the phase winds around as we traverse 
the space-time region in which the superfluid resides.
The oscillations around the static solution yield two modes whose
dispersion relations $\omega({\bf k})$ are 
determined by the equations of motion. One of the modes is the 
massless Goldstone mode, the other is a massive ``radial'' mode.
The oscillatory modes can be thermally populated to yield the normal fluid. 
The boundary conditions place no constraint on the
amplitudes of the oscillatory modes because they are topologically trivial.

In this section we will assume uniform density and flow of the superfluid, but
in general one could obtain a space-time-dependent superfluid flow
by giving nonzero classical background
values to the $\delta\psi_{\bf k}$ and $\delta\rho_{\bf k}$. The normal
fluid would then consist of a thermal population on top of that background.

In Sec.~\ref{sec:effact} we will use the poles of the tree-level propagator
to determine the dispersion relation of the oscillatory modes. This
yields the same result as solving the equations of motion (\ref{EOMs}) with the ansatz 
(\ref{ansatzpsirho}). Oscillations in $\psi$ will also be relevant to
the calculation of the sound velocities in Sec.~\ref{sec:sound}. In fact, the coefficient of the linear term in 
the massless mode $\omega({\bf k})$ is nothing but the velocity of first sound. 
For now, we set $\delta\psi_{\bf k}=\delta\rho_{\bf k}=0$. 
In this case, $p^\mu=\partial^\mu\psi$, and we can write 
\be \label{rho}
\rho = \sqrt{\frac{\sigma^2-m^2}{\lambda}} \, . 
\ee
Inserting this solution back into the Lagrangian gives
\be \label{LU}
{\cal L} = -U = \frac{(\sigma^2-m^2)^2}{4\lambda} \, .
\ee
It is clear from the usual definition of the pressure in terms of the partition function that, in our current tree-level treatment, 
this Lagrangian is identical to the pressure.

We now explore the connection between field-theoretic quantities and
hydrodynamics.
The hydrodynamic approach involves writing the current and stress-energy
tensor in terms of fluid-mechanical quantities
\begin{subequations} \label{jTmacro}
\bea
j^\mu &=& n_s v^\mu \,, \\[2ex]
T^{\mu\nu} &=& (\epsilon_s+P_s)v^\mu v^\nu -g^{\mu\nu} P_s \, ,
\eea
\end{subequations}
where $v^\mu$ is the {\it superfluid velocity}; by definition $v_\mu v^\mu=1$.
To identify $n_s$, $\epsilon_s$, and $P_s$ we go to the
superfluid rest frame $v^\mu = (1,0,0,0)$, where there 
is no flow of charge, energy, or momentum: $j^i = T^{0i} = 0$. Then we have $j^0 = n_s$, $T^{00} = \epsilon_s$, $T^{ij}=\delta^{ij} P_s$,  
so $n_s$, $\epsilon_s$, and $P_s$ are the charge density, energy density, and 
pressure in the superfluid rest frame; the pressure is isotropic. 
(We will discuss the superfluid density $\rho_s$ below). 
They can be expressed covariantly via appropriate contractions,
\be \label{proj}
n_s=\sqrt{j^\mu j_\mu} = v^\mu j_\mu\, , \qquad
\epsilon_s = v_\mu v_\nu T^{\mu\nu} \, , \qquad P_s = -\frac{1}{3}(g_{\mu\nu}-v_\mu v_\nu)T^{\mu\nu} \, .
\ee

To relate these hydrodynamical quantities to the microscopic physics we can use
Eqs.\ (\ref{jTmicro}). For $n_s$ and $v^\mu$ we obtain immediately from
(\ref{jmicro}),
\be  \label{vmu}
n_s = \sigma\frac{\sigma^2-m^2}{\lambda} \, , \qquad
v^\mu = \frac{\partial^\mu\psi}{\sigma}  =\gamma\,(1,{\bf v}_s) \ ,
\ee
where ${\bf v}_s$ is the superfluid 3-velocity with 
Lorentz factor $\gamma=1/\sqrt{1-{\bf v}_s^2}$. The velocity is proportional to the gradient
of a scalar field,
${\bf v}_s = -\nabla\psi / \partial_0\psi$, which is what one expects
since a superfluid without vortices is irrotational. [The minus sign appears because the 3-velocity ${\bf v}_s$ corresponds to the spatial components of the 
contravariant 4-vector $v^\mu$, while the operator $\nabla$ corresponds to the spatial components of the covariant 4-vector $\partial_\mu$, i.e., 
$\partial^\mu = (\partial_t,-\nabla)$.]

For $\epsilon_s$ and $P_s$ we first use (\ref{proj}) and (\ref{Tmicro})
to obtain partly microscopic expressions
\bea \label{epssPs}
\epsilon_s &=& v_\mu\partial^\mu\psi \, n_s -{\cal L} \, , \qquad P_s = {\cal L}+(v_\mu\partial^\mu\psi-\sigma)n_s \, . \label{epsrest}
\eea
From contracting the relation for $v^\mu$ in (\ref{vmu}) with $\partial_\mu\psi$ we know that $v^\mu\partial_\mu\psi=\sigma$; as a consequence, we obtain 
the expected relation $P_s={\cal L}$. We can now identify the physical meaning of $\sigma$: using the
zero-temperature thermodynamic relation in the superfluid rest frame 
$\epsilon_s + P_s = \mu_s n_s$, we find
\be
\mu_s = \sigma = v^\mu\partial_\mu\psi \, ,
\ee  
so $\sigma$ is identified with $\mu_s$,
the chemical potential in the superfluid rest frame. 
Going back to the solution for the modulus of the condensate
(\ref{rho}) we see that, as expected, Bose condensation only occurs for $\mu_s^2>m^2$. 
Using (\ref{LU}) and (\ref{epssPs}) 
we finally obtain fully microscopic expressions for the energy density and the pressure,
\be
P_s = \frac{(\sigma^2-m^2)^2}{4\lambda} \, , \qquad \epsilon_s = \frac{(3\sigma^2+m^2)(\sigma^2-m^2)}{4\lambda} \, .
\ee
Note that $m$ is the only mass scale in our Lagrangian. The trace of the stress-energy tensor $T^\mu{}_{\mu} = \epsilon_s-3P_s = m^2\rho^2$
vanishes for $m=0$. 

Finally, the {\it superfluid density} $\rho_s$ is defined 
via the expansion in small three-velocities of the momentum
and energy densities
\be
T^{0i} = \rho_s {\bf v}_{si} + {\cal O}(|{\bf v}_s|^3) \, , \qquad T^{00} = \frac{\epsilon_s + P_s{\bf v}_s^2}{1-{\bf v}_s^2} = \epsilon_s + 
\rho_s {\bf v}_s^2 + {\cal O}(|{\bf v}_s|^4) \, . 
\ee
From these expansions we obtain its microscopic form
\be \label{rhosT0}
\rho_s = \epsilon_s + P_s = \sigma^2\frac{\sigma^2-m^2}{\lambda} \, .
\ee

\subsection{Single-fluid formalism}
\label{sec:1fluid}

Let us now, following 
Refs.~\cite{1982PhLA...91...70K,1982ZhETF..83.1601L,carter89,1992PhRvD..45.4536C,Andersson:2006nr}, introduce the formalism that 
we shall use below to describe the two fluid components at nonzero temperature. 
Although this might seem unnecessary for the zero-temperature case where there is only a single fluid, it is a useful preparation for the 
nonzero temperature case. 

In the hydrodynamic formalism, the basic variables that describe one fluid are
$j^\mu$, which is the {\em conserved current density}, and $\partial^\mu\psi$, which we will identify as the {\em conjugate momentum}.
Motivated by Eqs.~(\ref{jTmicro}), we write the stress-energy tensor as
a function of these variables
\be \label{Tmunusingle}
T^{\mu\nu} =  - g^{\mu\nu}\Psi + j^\mu\partial^\nu\psi \, , 
\ee
where we call $\Psi$ the {\it generalized pressure}, and we define the {\it generalized energy density}
\be \label{Lambdasingle}
\Lambda \equiv  T^\mu_{\;\;\mu} + 3\Psi = -\Psi + j^\mu\partial_\mu\psi \, .
\ee
By comparison with the expressions from the previous subsection it is easy to see that in the single-fluid case $\Psi$ and $\Lambda$ are simply 
pressure and energy density in the fluid rest frame, $\Psi=P_s={\cal L}$, $\Lambda=\epsilon_s$. 
It is then natural to treat $\Lambda$ and $\Psi$ as Legendre 
transforms of each other:  $\Psi$ depends only on $\partial^\mu\psi$, and
we assume that $\Lambda$ depends only on $j^\mu$ (the fact that $\partial\Lambda/\partial(\partial_\mu\psi)=0$ follows from (\ref{Lambdasingle}) and (\ref{jmicro})).
So
\be
d\Lambda = \partial_\mu\psi\, dj^\mu \, ,\qquad d\Psi = j^\mu d(\partial_\mu\psi) \, . 
\ee
Since $\Lambda$ and $\Psi$ are
Lorentz scalars, they depend on the Lorentz scalars $j^2 = j_\mu j^\mu = n_s^2$ and $\sigma^2 = \partial_\mu\psi\partial^\mu\psi$, 
respectively. This allows us to write
\be
\partial_\mu\psi =  \frac{\partial\Lambda}{\partial j^\mu} = {\cal B}\, j_\mu \,, \qquad {\cal B}\equiv 2\frac{\partial\Lambda}{\partial j^2} \, , 
\ee
and
\be
j^\mu = \frac{\partial\Psi}{\partial(\partial_\mu\psi)} = \overline{\cal B}\, \partial^\mu\psi \, , \qquad \overline{\cal B}\equiv 
2\frac{\partial\Psi}{\partial \sigma^2} \, .
\ee
These relations say that the current is proportional to the conjugate momentum with a coefficient given by the underlying 
microscopic physics. In this one-fluid system, we simply read off 
\be
\overline{\cal B} = {\cal B}^{-1} = \frac{\sigma^2-m^2}{\lambda} \, .
\ee
This formalism will 
allows us to introduce the so-called {\it entrainment coefficients} as the coefficients in a two-fluid system which relate the 
current of the first fluid with the conjugate momentum of the second fluid and vice versa. But before we do so, we go back to our 
microscopic calculation to ``prepare'' it for the translation into the two-fluid picture.

\section{Nonzero temperature: microscopic calculation}
\label{sec:nonzeroT}

\subsection{Effective action and anisotropic dispersions}
\label{sec:effact}

In our context, the nonzero-temperature calculation is complicated for two reasons: $(i)$ we are interested in the situation of a nonzero 
(homogeneous) superflow; this makes the system and in particular the dispersions of the Goldstone mode anisotropic and $(ii)$ in order to 
compute the properties of the system for all temperatures up to the critical temperature, a self-consistent formalism such 
as the Cornwall-Jackiw-Tomboulis (CJT) formalism \cite{Luttinger:1960ua,Baym:1962sx,Cornwall:1974vz} is needed. (For a detailed discussion
of this formalism in the same context, but without superflow, see Ref.~\cite{Alford:2007qa}.)  

Here we are mostly interested in analytical results to discuss the translation to hydrodynamics thoroughly. Therefore, we shall restrict
ourselves to small temperatures and ignore the complications from point $(ii)$. This means that we shall not determine the condensate self-consistently,
and rather work with its zero-temperature value. Starting from the full self-consistent formalism, one can show that the temperature-dependence
of the condensate includes an additional power of the coupling constant compared to the terms we are keeping. We are thus
working in the low-temperature, weak-coupling approximation.  Note, however, that 
the melting of the condensate, as well as other contributions from the full action may induce terms proportional to $\lambda\mu^2T^2$ in the 
pressure, while we shall only keep terms proportional to $T^4$ and higher order in $T$. This approximation is only consistent 
if $T^2\gg \lambda\mu^2$, i.e., 
strictly speaking, our approximation leaves a ``gap'' between $T=0$ and the small temperatures we are discussing, although 
this gap can be made arbitrarily small by choosing the coupling constant $\lambda$ sufficiently small. 
We defer a careful discussion of the self-consistent CJT calculation with superflow,
including a numerical evaluation up to the critical temperature to a forthcoming publication \cite{future}.

Within this approximation, we can work with the simple one-loop effective action 
\be \label{effact}
\Gamma= -\frac{V}{T}\,U(\rho,\sigma) - \frac{1}{2} \sum_k\Tr\ln \frac{S^{-1}(k)}{T^2} \, , 
\ee
with the tree-level potential $U$ from Eq.~(\ref{Utree}) and the inverse tree-level propagator in momentum space $S^{-1}(k)$ 
which can be read off from the 
terms quadratic in the fluctuations in Eq.~(\ref{L2}). The trace is taken over the $2\times 2$ space of the two real degrees of freedom of the
complex scalar field, and the sum is taken over four-momenta $k_\mu = (k_0,{\bf k})$, $k_0=-i\omega_n$ with the bosonic Matsubara frequencies 
$\omega_n$. The sum over three-momenta, here written for a finite volume $V$ and thus over discrete momenta, becomes an integral in the 
thermodynamic limit.  At the stationary point, i.e., with Eq.~(\ref{rho}), we have
\be
S^{-1}(k) = \left(\begin{array}{cc} -k^2+2(\sigma^2-m^2) &  2ik\cdot\partial\psi 
\\[2ex] -2ik\cdot\partial\psi & -k^2 \end{array}\right) \, .
\ee
Here, $k^2\equiv k_\mu k^\mu$, $k\cdot\partial\psi =k_\mu\partial^\mu\psi$
and $\sigma$ given in Eq.~(\ref{sigma}). In all 
our microscopic calculations we evaluate thermal fluctuations in the
background of a {\em uniform} superflow, so we take
$\partial_\mu\psi$ (and hence $\sigma$) to be space-time independent.
Since we neglect the melting of the condensate, we
have inserted the zero-temperature solution into the propagator.

In order to compute the effective action, one may employ partial integration with respect to the $|{\bf k}|$ integral. The effective action density then becomes
\bea \label{partial}
\frac{T}{V}\Gamma
%&=& \frac{(\sigma^2-m^2)^2}{4\lambda} +\frac{1}{6}\frac{T}{V}\sum_k|{\bf k}|
%\Tr\left[S(k)\frac{\partial S^{-1}(k)}{\partial |{\bf k}|}\right]
%\non[2ex] 
&=&\frac{(\sigma^2-m^2)^2}{4\lambda} -\frac{2}{3}\frac{T}{V}\sum_k \frac{{\bf k}^2(k^2-\sigma^2+m^2)+2k\cdot\partial\psi\,{\bf k}
\cdot\nabla\psi}{{\rm det}\,S^{-1}(k)} \, , 
\eea
where we have used the explicit form of the tree-level propagator and have performed the $2\times 2$ trace. 
We have also inserted the solution (\ref{rho}) into the potential $U$. The evaluation of the Matsubara sum and the simplification of the result
is explained in detail
in appendix \ref{App0}. Neglecting the thermal contribution of the massive mode, we can write the result as 
\bea \label{Gammaeps}
\frac{T}{V}\Gamma &\simeq& \frac{(\sigma^2-m^2)^2}{4\lambda}-\int\frac{d^3{\bf k}}{(2\pi)^3} \frac{F(\epsilon_{1,{\bf k}},{\bf k})}
{(\epsilon_{1,{\bf k}}+\epsilon_{1,-{\bf k}})(\epsilon_{1,{\bf k}}+\epsilon_{2,-{\bf k}})(\epsilon_{1,{\bf k}}-\epsilon_{2,{\bf k}})}\,
\coth\frac{\epsilon_{1,{\bf k}}}{2T} \, , 
\eea
where $F(k_0,{\bf k})$ denotes the numerator in the momentum sum of Eq.\ (\ref{partial}), 
\be \label{Fk0}
F(k_0,{\bf k}) \equiv  -\frac{2}{3}\left[{\bf k}^2(k^2-\sigma^2+m^2)+2k\cdot\partial\psi\,{\bf k} \cdot\nabla\psi\right] \, ,
\ee
and where $\epsilon_{1/2,{\bf k}}$ are complicated excitation energies whose small-momentum approximations are
\begin{subequations} \label{smallk}
\bea
\epsilon_{1,{\bf k}} &=& \sqrt{\frac{\sigma^2-m^2}{3\sigma^2-m^2}}\;\zeta(\uk)\,|{\bf k}| + {\cal O}(|{\bf k}|^3) \, , 
\label{eps12}\\[2ex]
\epsilon_{2,{\bf k}} &=& \sqrt{2}\sqrt{3\sigma^2-m^2+2(\nabla\psi)^2} + {\cal O}(|{\bf k}|) \, . 
\eea
\end{subequations}
Here we have abbreviated 
\be \label{zeta}
\zeta(\uk)\equiv \left[\sqrt{1+2\frac{(\nabla\psi)^2-(\nabla\psi\cdot\uk)^2}{3\sigma^2-m^2}}-\frac{2\partial_0\psi\nabla\psi\cdot\uk}{\sqrt{\sigma^2
-m^2}\sqrt{3\sigma^2-m^2}}\right]\left[1+\frac{2(\nabla\psi)^2}{3\sigma^2-m^2}\right]^{-1} \, .
\ee
%with $\theta$ being the angle between ${\bf k}$ and $\nabla\psi$. 
The physically relevant low-energy excitation is $\epsilon_{1,{\bf k}}$.
This is the Goldstone mode which, as Eq.~(\ref{eps12}) confirms, is massless and linear in the momentum for small momenta, as it should be. 
The coefficient 
in front of the linear term is the speed of (first) sound. For the case without superflow we have $\zeta(\uk)=1$, and thus, in the limit 
$m=0$, we recover the well-known $1/\sqrt{3}$. The superflow introduces an angular dependence $\zeta(\uk)$ into the Goldstone dispersion and 
thus also into the sound velocity. This angular-dependent function also shows that, in contrast to the 
zero-temperature case, we cannot write the result in a covariant way since the temporal and spatial components of $\partial_\mu\psi$
appear separately. Another observation is the complicated factor in the integrand of Eq.~(\ref{Gammaeps}). This factor indicates
a mixing between the original modes of the complex field (between particles and antiparticles, essentially) due to condensation. Such a factor 
appears also, although considerably simpler, for the case without superflow. It is analogous to a Bogoliubov coefficient in the case of 
Cooper pairing which, in that case,  accounts for the mixing between fermions and fermion-holes. 

We shall further evaluate the effective action for small temperatures in Sec.~\ref{sec:smallT}.

\subsection{Stress-energy tensor and current}
\label{sec:Tmunu}

The stress-energy tensor and the current are given by 
\be \label{Tmunudef}
T^{\mu\nu} = \left\langle\frac{2}{\sqrt{-g}}\frac{\delta(\sqrt{-g}\,{\cal L})}{\delta g_{\mu\nu}}\right\rangle \, ,
\qquad j^\mu = \left\langle\frac{\partial{\cal L}}{\partial(\partial_\mu\psi)}\right\rangle \, , 
\ee
where the angular brackets denote the expectation value in the finite-temperature ensemble of the microscopic theory,
\be
\langle A\rangle \equiv \frac{1}{Z}\int{\cal D}\varphi_1'{\cal D}\varphi_2' \,A\, 
\exp\left(\int d^4x\,{\sqrt{-g}\,{\cal L}}\right) \, ,
\ee
with the partition function $Z$ defined so that $\langle 1 \rangle=1$. 
After performing the functional integral we can write the results as 
\begin{subequations}
\bea
T^{\mu\nu} &=& -\left(2\frac{\partial U}{\partial g_{\mu\nu}}-g^{\mu\nu}U\right) 
- \frac{T}{V}\sum_k\Tr\left[S(k)\frac{\partial S^{-1}(k)}{\partial g_{\mu\nu}}-
\frac{g^{\mu\nu}}{2} \right] \,  , \label{dlnZ} \\[2ex]
j^\mu &=& 
\partial^\mu\psi\,\frac{\sigma^2-m^2}{\lambda}- \frac{1}{2} \frac{T}{V}\sum_k\Tr\left[S(k)\frac{\partial S^{-1}(k)}{\partial (\partial_\mu\psi)}
\right] \, .
\eea
\end{subequations}
In the case of the stress-energy tensor, one can check explicitly that the sum over the Matsubara frequencies leads to an infinite result. 
A renormalization is thus required. As a renormalization condition, we require for the case without superflow the obvious 
interpretation of the diagonal components $T^{\mu\nu}$ in terms of the energy density $\epsilon$ and the pressure $P$,
\be \label{condition}
T^{00} = \epsilon  \, ,\qquad T^{ij} = \delta^{ij}P \, .
\ee
Then, switching on a nonzero superflow, does not yield any additional divergences. 
These conditions can be implemented on a very general level, without explicit evaluation of the stress-energy tensor. We do so in detail
in appendix \ref{AppA}. The calculation in this appendix also leads to a very useful formulation of the effective action, the 
stress-energy tensor, and the current, which will later facilitate the interpretation 
in terms of the hydrodynamic two-fluid picture. We define (for motivation
see Eq.~(\ref{PsiPsik}))
\be \label{PsiK}
\Psi_k \equiv \Psi_k(\sigma^2,k\cdot \partial\psi,k^2) \equiv  - \frac{1}{2}\Tr\ln \frac{S^{-1}(k)}{T^2} \, ,
\ee
and 
\begin{subequations} \label{dPsiK}
\bea
A_k&\equiv&\frac{\partial\Psi_k}{\partial(k\cdot\partial\psi)} = 4\frac{k\cdot\partial\psi}{{\rm det}\,S^{-1}} \, , \\[2ex]
B_k&\equiv&2\frac{\partial\Psi_k}{\partial\sigma^2} = \frac{2k^2}{{\rm det}\,S^{-1}} \, , \\[2ex] 
C_k&\equiv&2\frac{\partial\Psi_k}{\partial k^2} = -\frac{2(k^2-\sigma^2+m^2)}{{\rm det}\,S^{-1}} \, .
\eea
\end{subequations}
The reason to introduce these quantities and the choice of notation will become clear after we have discussed the two-fluid formalism.
In the zero-temperature discussion of the single-fluid dynamics we have already encountered the generalized pressure $\Psi$, of which 
the notation $\Psi_k$ is reminiscent. The notations $A_k$, $B_k$, $C_k$ are chosen in analogy to $\overline{\cal A}$, $\overline{\cal B}$, 
$\overline{\cal C}$, to be introduced in Sec.~\ref{sec:entrain} (we already know the coefficient $\overline{\cal B}$ from the single-fluid
treatment in Sec.~\ref{sec:1fluid}). With the help of these quantities we can write  the action as (see appendix \ref{AppA})
\bea  \label{gamCA}
\frac{T}{V}\Gamma = -U - \frac{1}{3}(g^{\mu\nu}-u^\mu u^\nu)\frac{T}{V}\sum_k\left(C_k k_\mu k_\nu+A_k k_\mu \partial_\nu\psi\right) \, ,
\eea
where we have abbreviated $u^\mu=(1,0,0,0)$, while the renormalized stress-energy tensor and the current become 
\begin{subequations} \label{Tmunujmu}
\bea  \label{Tmunu}
T^{\mu\nu} &=& -\left(2\frac{\partial U}{\partial g_{\mu\nu}}-g^{\mu\nu}U\right)+ 
\frac{T}{V}\sum_k\left[C_kk^\mu k^\nu+B_k\partial^\mu\psi\partial^\nu\psi+A_k(k^\mu\partial^\nu\psi+k^\nu\partial^\mu\psi)
+2u^\mu u^\nu\right] \, , \\[2ex]
\label{currABC}
j^\mu &=& \partial^\mu\psi\,\frac{\sigma^2-m^2}{\lambda} + \frac{T}{V}\sum_k\left(B_k\partial^\mu\psi+A_k k^\mu\right) \, .
\eea
\end{subequations}
%These expressions are useful because, firstly, they are written in a covariant way and, secondly, they anticipate the two-fluid
%formulation which relies on two basic four-vectors. One can be chosen to be $\partial^\mu\psi$; the other one is of course not $k^\mu$, which
%is a purely microscopic quantity, but we shall see that this formulation is close enough to the two-fluid formulation that it can easily 
%be cast in a more ``macroscopic'' form. 
It is instructive to compute the trace of the stress-energy tensor. With the help of Eq.~(\ref{one}) we immediately find
\be \label{trace}
T^\mu_{\;\;\mu} = m^2\left[\rho^2+2\frac{T}{V}\sum_k\frac{k^2}{{\rm det}\,S^{-1}(k)} \right] \, ,
\ee
i.e., the trace vanishes when we set $m=0$, as expected. 

\subsection{Evaluation for small temperatures}
\label{sec:smallT}

We want to compute the effective action, the components of the stress-energy tensor, and the current for small 
temperatures explicitly. To this end, we have to approximate momentum sums of the form 
\be
\frac{T}{V}\sum_k \frac{F(k)}{{\rm det}\,S^{-1}(k)} \, ,
\ee
where, for the case of the effective action, $F(k)$ is given by Eq.~(\ref{Fk0}), and for the stress-energy tensor and the current 
we need to replace $F(k)$ by 
\begin{subequations}\label{FF}
\bea
F^{\mu\nu}(k) 
&\equiv& 2\left[-(k^2-\sigma^2+m^2)k^\mu k^\nu +k^2\partial^\mu\psi\partial^\nu\psi+2(k\cdot\partial\psi)(k^\mu\partial^\nu\psi
+k^\nu\partial^\mu\psi)+u^\mu u^\nu{\rm det}\,S^{-1}(k)\right] \, , \\[2ex]
F^\mu(k) &\equiv& 2(k^2\partial^\mu\psi+2k\cdot\partial\psi\,k^\mu)  \, ,
\eea
\end{subequations}
which can be read off from Eqs.~(\ref{Tmunujmu}). 

For all three cases, we can write the result of the Matsubara sum in the form (\ref{Gammaeps}). Then, we write
$\coth[\epsilon_{1,{\bf k}}/(2T)] = 1+2f(\epsilon_{1,{\bf k}})$ with the Bose distribution function $f(x)=1/(e^{x/T}-1)$. The integral over the 
first term is, in our approximation, temperature independent. It is divergent and has to be renormalized by subtracting the vacuum 
contribution. After doing so, a finite term remains which however is suppressed by one power of $\lambda$ compared to the 
zero-temperature term $\propto\lambda^{-1}$ we have already computed, for instance in Eq.~(\ref{Gammaeps}). We shall thus neglect this 
contribution and only keep the thermal 
contribution, i.e., the integral over the second term that contains the Bose function. This contribution is finite and unaffected by 
the renormalization.\footnote{The subleading zero-temperature contribution
we are neglecting here gives rise to the difference between the {\it superfluid density} and the {\it condensate density}.
At zero temperature, the charge density is always identical to the superfluid density, $n(T=0)=n_s$, while the condensate density, defined as $\mu\rho^2$ with the 
modulus of the condensate $\rho$, might be smaller. In our leading-order approximation, condensate and superfluid densities are identical.}

We explain the small-temperature expansion in detail in appendix \ref{AppB}. Before we come to the main results, let us discuss the simpler 
example without superflow, $\nabla\psi=0$. In this case, $\epsilon_{1,{\bf k}}=\epsilon_k^+$, $\epsilon_{2,{\bf k}}=\epsilon_k^-$ with
\be \label{eps0}
\epsilon_k^\pm = \sqrt{k^2+3\mu^2-m^2\mp\sqrt{4\mu^2k^2+(3\mu^2-m^2)^2}} \, ,
\ee
where we have set $\partial_0\psi=\mu$. Here, $\epsilon_k^+$ is the Goldstone mode, while $\epsilon_k^-$ is the massive mode. 
This yields the small-temperature result for the pressure 
to order $T^6$ (see appendix \ref{AppB}), 
\bea \label{Piso}
P=\frac{T}{V}\Gamma(\nabla\psi=0) &\simeq & \frac{(\mu^2-m^2)^2}{4\lambda} + \frac{(3\mu^2-m^2)^{3/2}}{(\mu^2-m^2)^{3/2}}\frac{\pi^2 T^4}{90}-
\frac{\mu^6(3\mu^2-m^2)^{1/2}}{(\mu^2-m^2)^{7/2}}\frac{4\pi^4T^6}{63\mu^2} \, .
\eea
The $T^4$ term has two interesting properties related to the resulting charge density, which is obtained by taking the 
derivative with respect to $\mu$. Firstly, if we set $m=0$, the $\mu$-dependence drops out, such that there is no $T^4/\mu$ contribution to the 
charge density in this case. Secondly, in the presence of a finite $m$, one finds that the $T^4/\mu$ contribution to the charge density 
is negative, i.e., for small temperatures the density {\it decreases} with temperature. 
(The second derivative $\partial^2 P/\partial\mu^2$ is positive, 
i.e., the system is thermodynamically stable.) The $T^6$ term has neither of these properties, it
contributes to the charge density even for $m=0$ and gives rise to a positive $T^6/\mu^3$ term in the density. 
%%%%%%%%%%%%%%%%%%%%%%%%%%%%%%%%%%%%%%%%%%%%%%%%%%%%%%%%%%%%%%%%%%%%%%%
\begin{table*}[t]
\begin{tabular}{|c||c|c|c|} 
\hline
\rule[-1.5ex]{0em}{6ex} 
 & $\;\;\displaystyle{\frac{\mu^4}{4\lambda}}(1-{\bf v}_s^2)\;\;$   & $\;\;\displaystyle{\frac{\pi^2T^4}{10\sqrt{3}}\,\frac{1-{\bf v}_s^2}{(1-3{\bf v}_s^2)^3}}\;\;$ 
& $\;\;\displaystyle{\frac{4\pi^4T^6}{105\sqrt{3}\,\mu^2}\,\frac{1-{\bf v}_s^2}{(1-3{\bf v}_s^2)^6}}\;\;$ 
\\[2ex] \hline\hline
\rule[-1.5ex]{0em}{6ex} 

$\;\;\displaystyle{\frac{T}{V}\Gamma}\;\;$ & $\;\;\displaystyle{1-{\bf v}_s^2}\;\;$ & 
$\displaystyle{\;\;(1-{\bf v}_s^2)(1-3{\bf v}_s^2)\;\;}$ & 
$\displaystyle{\;\;-(1-{\bf v}_s^2)(1-3{\bf v}_s^2)(5+30{\bf v}_s^2+9{\bf v}_s^4)\;\;}$ \\[2ex] \hline\hline
\rule[-1.5ex]{0em}{6ex} 

$\;\;\displaystyle{T^{00}}\;\;$ & $\;\;\displaystyle{3+{\bf v}_s^2}\;\;$ & 
$\displaystyle{\;\;3-20{\bf v}_s^2+9{\bf v}_s^4\;\;}$ & 
$\displaystyle{\;\;- (15-160{\bf v}_s^2-774{\bf v}_s^4+432{\bf v}_s^6+135{\bf v}_s^8)\;\;}$ \\[2ex] \hline
\rule[-1.5ex]{0em}{6ex} 

$\;\;\displaystyle{T^{0i}}\;\;$ & $\;\;\displaystyle{4{\bf v}_{si}}\;\;$ & 
$\displaystyle{\;\;-8{\bf v}_{si}\;\;}$ & 
$\displaystyle{\;\;2{\bf v}_{si} (95+243{\bf v}_s^2-135{\bf v}_s^4-27{\bf v}_s^6)\;\;}$ \\[2ex] \hline
\rule[-1.5ex]{0em}{6ex} 

$\;\;\displaystyle{T_{\perp}}\;\;$ & $\;\;\displaystyle{1-{\bf v}_s^2}\;\;$ & 
$\displaystyle{\;\;(1-{\bf v}_s^2)(1-3{\bf v}_s^2)\;\;}$ & 
$\displaystyle{\;\;-(1-{\bf v}_s^2)(1-3{\bf v}_s^2)(5+30{\bf v}_s^2+9{\bf v}_s^4)\;\;}$ \\[2ex] \hline
\rule[-1.5ex]{0em}{6ex} 

$\;\;\displaystyle{T_{||}}\;\;$ & $\;\;\displaystyle{1+3{\bf v}_s^2}\;\;$ & 
$\displaystyle{\;\;1-12{\bf v}_s^2+3{\bf v}_s^4\;\;}$ & 
$\displaystyle{\;\;- (5-180{\bf v}_s^2-582{\bf v}_s^4+324{\bf v}_s^6+81{\bf v}_s^8)\;\;}$ \\[2ex] \hline\hline
\rule[-1.5ex]{0em}{6ex} 

$\;\;\displaystyle{\mu\, j^0}\;\;$ & $\;\;\displaystyle{4}\;\;$ & 
$\displaystyle{\;\;-8{\bf v}_s^2\;\;}$ & 
$\displaystyle{\;\;2 (5+105{\bf v}_s^2+147{\bf v}_s^4-81{\bf v}_s^6)\;\;}$ \\[2ex] \hline
\rule[-1.5ex]{0em}{6ex} 

$\;\;\displaystyle{\mu\, {\bf j}}\;\;$ & $\;\;\displaystyle{4{\bf v}_s}\;\;$ & 
$\displaystyle{\;\;-8{\bf v}_s\;\;}$ & 
$\displaystyle{\;\;2 {\bf v}_s (95+243{\bf v}_s^2-135{\bf v}_s^4-27{\bf v}_s^6)\;\;}$ \\[2ex] \hline

\end{tabular}
\caption{Microscopic results for the effective action $\Gamma$, the stress-energy tensor $T^{\mu\nu}$, and the conserved charge current $j^\mu$ up to order $T^6$
for $m=0$.
Each row is the result for the quantity given in the left column; this quantity is a sum of the $\mu^4$, $T^4$, and $T^6/\mu^2$ terms given in the top row, each 
multiplied by the specific entry of the table. For example, the first row is equivalent to Eq.\ (\ref{tvgam}). These results are obtained without making any assumptions
about the magnitude of the superfluid velocity ${\bf v}_s$. We see that there is a divergence in all nonzero temperature results 
for ${\bf v}_s^2\to 1/3$, indicating an instability of the superflow,
in accordance with Landau's critical velocity for superfluidity.
}
\label{table0}
\end{table*}
%%%%%%%%%%%%%%%%%%%%%%%%%%%%%%%%%%%%%%%%%%%%%%%%%%%%%%%%%%%%%%%%%%%%%%%
The case with superflow is of course more complicated. In particular, the momentum integration now involves a nontrivial angular integral over the angle 
between the momentum and $\nabla\psi$. Nevertheless, it turns out that this angular integral can be performed analytically for all cases we consider. 
For brevity, we set $m=0$ in the following. Moreover, we denote $\mu=\partial_0\psi$. At this point, this may be viewed as a mere 
notation; we shall see below that $\mu$ is the chemical potential in the normal-fluid rest frame, which is defined below Eqs.\ (\ref{sn}). Then, we obtain for the 
effective action density to order $T^6$
\be \label{tvgam}
\frac{T}{V}\Gamma \simeq \frac{\mu^4}{4\lambda}(1-{\bf v}_s^2)^2+\frac{\pi^2T^4}{10\sqrt{3}}\,\frac{(1-{\bf v}_s^2)^2}{(1-3{\bf v}_s^2)^2} 
- \frac{4\pi^4T^6}{105\sqrt{3}\,\mu^2}\,\frac{(1-{\bf v}_s^2)^2}{(1-3{\bf v}_s^2)^5}(5+30{\bf v}_s^2+9{\bf v}_s^4) \, .
\ee
%\be \label{tvgam}
%\frac{T}{V}\Gamma \simeq \frac{\mu^4}{4\lambda}(1-2{\bf v}_s^2)+\frac{\pi^2}{10\sqrt{3}}(1+4{\bf v}_s^2)\,T^4-\frac{4\pi^4}{21\sqrt{3}}
%(1+19{\bf v}_s^2)\,\frac{T^6}{\mu^2} \, .
%\ee
Obviously, for ${\bf v}_s=0$ we recover the $m=0$ limit of Eq.~(\ref{Piso}). Note that the result is valid to all orders in the superfluid velocity.
We have not applied any expansion in $|{\bf v}_s|$. Analogously, we compute the components of the stress-energy tensor and the current. 
For the spatial components of the stress-energy tensor we define
\be
T_\perp\equiv \frac{1}{2}\left[\delta_{ij}-\frac{\partial_i\psi\partial_j\psi}{(\nabla\psi)^2}\right]T_{ij} \, ,\qquad
T_{||} \equiv \frac{\partial_i\psi\partial_j\psi}{(\nabla\psi)^2} T_{ij} \, , 
\ee
which are the spatially transverse and longitudinal components with respect to ${\bf v}_s$. We collect all results in Table \ref{table0}.

%\begin{subequations} \label{Tcomps}
%\bea
%T^{00}&\simeq& \frac{\mu^4}{4\lambda}(3-2{\bf v}_s^2)+\frac{\pi^2}{10\sqrt{3}}(3+4{\bf v}_s^2)\,T^4
%-\frac{4\pi^4}{21\sqrt{3}}(3+19{\bf v}_s^2)\,\frac{T^6}{\mu^2} \, , \\[2ex]
%T^{0i}&\simeq& {\bf v}_{si}\left(\frac{\mu^4}{\lambda}-\frac{4\pi^2}{5\sqrt{3}}\,T^4
%+\frac{152\pi^4}{21\sqrt{3}}\,\frac{T^6}{\mu^2}\right) \, , \\[2ex]
%T_\perp&\simeq& \frac{\mu^4}{4\lambda}(1-2{\bf v}_s^2)+\frac{\pi^2}{10\sqrt{3}}(1+4{\bf v}_s^2)\,T^4
%-\frac{4\pi^4}{21\sqrt{3}}(1+19{\bf v}_s^2)\,\frac{T^6}{\mu^2} \, , \\[2ex]
%T_{||}&\simeq& \frac{\mu^4}{4\lambda}(1+2{\bf v}_s^2)+\frac{\pi^2}{10\sqrt{3}}(1-4{\bf v}_s^2)\,T^4
%-\frac{4\pi^4}{21\sqrt{3}}(1-19{\bf v}_s^2)\,\frac{T^6}{\mu^2} \, , 
%\eea
%\end{subequations}
%The current becomes 
%\begin{subequations} \label{jcomps}
%\bea
%j^0 &\simeq& \frac{\mu^3}{\lambda}(1-{\bf v}_s^2)-\frac{4\pi^2}{5\sqrt{3}}{\bf v}_s^2\,\frac{T^4}{\mu} +\frac{8\pi^4}{21\sqrt{3}}(1+38{\bf v}_s^2)
%\,\frac{T^6}{\mu^3} \, , \label{j0}\\[2ex]
%{\bf j} &\simeq& {\bf v}_{s}\left(\frac{\mu^3}{\lambda}-\frac{4\pi^2}{5\sqrt{3}}\,\frac{T^4}{\mu} +\frac{152\pi^4}{21\sqrt{3}}\,\frac{T^6}{\mu^3}\right) 
%\, . 
%\eea
%\end{subequations}
All expressions in the table are written in terms of frame-dependent quantities. At zero temperature 
it is very natural to write them in terms of the Lorentz scalar $\sigma=\mu\sqrt{1-{\bf v}_s^2}$, 
for instance $T_\perp = \sigma^4/(4\lambda)$, as we have done in Sec.\ \ref{sec:zeroT}. 
Such a formulation is less obvious for the nonzero temperature terms. As we shall discuss in 
Sec.\ \ref{sec:translate}, the microscopic calculation has a preferred rest frame which we shall identify as the normal-fluid rest frame, and 
all thermodynamic variables in the above expressions, i.e., $\mu$, $T$, and ${\bf v}_s$, are measured in this particular frame.
As a consequence, the velocity-dependence shown here is a mixture of trivial Lorentz boosts and complicated effects of the superflow on the 
collective excitations.

\section{Relativistic two-fluid formalism}
\label{sec:2fluid}

So far we have discussed the zero-temperature limit of the microscopic theory and its formulation in hydrodynamic terms, as well as the 
generalization to nonzero temperatures of our field theory. Before translating the latter into the language of hydrodynamics we need to 
introduce the two-fluid formalism. As indicated in the introduction, a superfluid becomes a two-fluid system for temperatures larger than zero and
smaller than the critical temperature. At zero temperature there is only the superfluid component, while at the 
critical temperature (and above) there is only the so-called normal component. The discussion in this entire section is general in the sense that we do not 
assume a uniform superflow; we shall get back to this assumption only in the next section where the microscopic results are needed.
The most straightforward way to add a second component to the single-fluid hydrodynamics seems to be
\begin{subequations} \label{sn}
\bea
j^\mu &=& n_nu^\mu + n_s \frac{\partial^\mu\psi}{\sigma} \,, \label{jmacroT} \\[2ex]
T^{\mu\nu} &=& (\epsilon_n + P_n)u^\mu u^\nu -g^{\mu\nu} P_n  + (\epsilon_s+P_s) \frac{\partial^\mu\psi\partial^\nu\psi}{\sigma^2} -g^{\mu\nu}P_s
 \, .\label{tmunuson}
\eea
\end{subequations}
Here we have simply taken the zero-temperature expression (\ref{jTmacro}) with the superfluid velocity (\ref{vmu}) and added analogous terms for the 
normal component. In particular, we have introduced the normal fluid velocity $u^\mu$, with $u^\mu u_\mu=1$. As a consequence, we can now 
also define the {\it normal-fluid rest frame} by $u^\mu=(1,0,0,0)$. Obviously, if we go to either the superfluid or the normal-fluid rest frame, there
will be a nonvanishing three-current. In other words, in contrast to the zero-temperature (i.e., single fluid) case, there is now in general 
no  frame where the pressure is isotropic. The decomposition in terms of superfluid and normal components in the form 
(\ref{sn}) can be found for instance in Refs.~\cite{Son:2000ht,Gusakov:2006ga,Gusakov:2007px,Herzog:2008he,Herzog:2009md}.\footnote{There
is a slight difference in Refs.~\cite{Gusakov:2006ga,Gusakov:2007px}, where the stress-energy tensor is written as (adapted to our
notation) 
\be
T^{\mu\nu} = (\mu n+Ts)u^\mu u^\nu -g^{\mu\nu}P+\sigma n_s\left[\frac{w^\mu w^\nu}{\sigma^2}+\frac{\mu}{\sigma}\left(u^\mu \frac{w^\nu}{\sigma}+
u^\nu \frac{w^\mu}{\sigma}\right)\right]\, , \nonumber
\ee
where $n=n_n+\frac{\mu}{\sigma}n_s$ is the charge density, measured in the normal-fluid rest frame, and $P\equiv P_s+P_n$. 
To obtain this stress energy tensor from 
Eq.~(\ref{tmunuson}), one defines $w^\mu\equiv \partial^\mu\psi-\mu u^\mu$  and uses $\epsilon_s+P_s=\sigma n_s$, 
$\epsilon_n+P_n=\mu n_n+Ts$.} In order to make a first connection to field theory, let us go to the normal-fluid rest frame (as we shall argue
later, this is the frame in which our microscopic calculation is performed). In this frame, take for instance
the spatial components of the current (\ref{jmacroT}), and contract the equation with $\nabla\psi$. This yields 
\be \label{ns1}
n_s = - \sigma\frac{\nabla\psi\cdot{\bf j}}{(\nabla\psi)^2} \, .
\ee
The right-hand side is now defined in terms of the field-theoretic quantities. We have thus found a microscopic
expression for the superfluid number density $n_s$. Note that $n_s$ is the superfluid number density measured in the superfluid rest frame; 
from the temporal component of Eq.~(\ref{jmacroT}) we see that the superfluid number density measured in the normal-fluid rest frame is $\mu n_s/\sigma$, 
with $\mu=\partial^0\psi$ being the chemical potential in the normal-fluid rest frame. As we know from Eq.\ (\ref{vmu}), the factor 
$\mu/\sigma = 1/\sqrt{1-{\bf v}_s^2}$ is a usual Lorentz factor. 

In the  two-fluid formulation (\ref{sn}) the charge current is decomposed into two non-conserved currents, so that each fluid component
is characterized by a non-conserved current. 
This contrasts with
the formulation of Refs.~\cite{1982PhLA...91...70K,1982ZhETF..83.1601L,carter89}, where the two components of the fluid are given 
by the conserved current $j^\mu$ and the entropy current $s^\mu$ (the latter being conserved in the dissipationless case). We shall
explain this formalism now. This is a generalization of the one-fluid discussion of Sec.~\ref{sec:1fluid}.

\subsection{Two currents, entrainment, and hydrodynamic equations}
\label{sec:entrain}

Generalizing the zero-temperature, single-fluid stress-energy tensor (\ref{Tmunusingle}), we write 
\be \label{tmunucomer}
T^{\mu\nu} = -g^{\mu\nu}\Psi +j^\mu\partial^\nu\psi + s^\mu\Theta^\nu \, , 
\ee
where, as before, $j^\mu$ is the total conserved current (now including thermal contributions), and $\partial^\mu\psi$
the corresponding conjugate momentum which, as we know, is related to the superfluid velocity. Analogously, we have introduced the 
entropy current $s^\mu$ and the corresponding conjugate momentum $\Theta^\mu$. The entropy current is related to the four-velocity 
of the normal fluid via 
\be
s^\mu=su^\mu \, ,
\ee
where $s\equiv (s^\mu s_\mu)^{1/2}$. In the normal-fluid rest frame, $s^\mu=(s^0,0,0,0)$, so $s=s^0$ is the entropy density in the normal-fluid rest frame.
(Of course, in the superfluid rest frame, $s^0\neq s$). Remember that the generalized pressure $\Psi$ was, in the single-fluid case, 
identical to the pressure in the superfluid rest frame. Now we may relate $\Psi$ to the pressures of the normal 
and superfluid components as follows. From Eq.~(\ref{tmunucomer}) we can derive the following expression for the generalized pressure  (without using any 
properties of the various four-vectors) \cite{Comer:2002dm}
\be \label{Psi}
\Psi = \frac{1}{2}\left[\frac{s\cdot \partial\psi(s_\mu\partial_\nu\psi + s_\nu\partial_\mu\psi)-s^2\partial_\mu\psi\partial_\nu\psi
-\sigma^2s_\mu s_\nu}{(s\cdot \partial\psi)^2-s^2\sigma^2}-g_{\mu\nu}\right]T^{\mu\nu}   \, .
\ee
Inserting the stress-energy tensor (\ref{tmunuson}) on the right-hand side of 
this equation, we find $\Psi=P_s+P_n$. 
Consequently, the generalized pressure is the sum of the pressures of the superfluid and normal components,
each measured in their respective rest frames. 
%(If, say, the pressure of the superfluid component, was measured in the normal-fluid rest frame, the 
%result would be anisotropic.)

According to its definition $\Lambda \equiv T^\mu_{\;\;\mu}+3\Psi$ (\ref{Lambdasingle}), the generalized energy density is now 
\be \label{Lamgen}
\Lambda  =-\Psi + j\cdot\partial\psi + s\cdot\Theta \, , 
\ee
which can be read as a generalized, covariant thermodynamic relation between pressure and energy density. It follows that in the normal fluid rest frame
$\Theta^0$ is the temperature, although because of entrainment (see
(\ref{psiTheta})) the spatial components $\Theta^i$ may not vanish.
Analogously to the pressure, 
one can show that $\Lambda=\epsilon_s+\epsilon_n$. As already discussed in the zero-temperature case, Eq.~(\ref{Lamgen})
is a Legendre transform between $\Lambda$ and $\Psi$ that changes the dependence from the (now two) currents to the conjugate momenta.
The energy density is a function of the currents $j^\mu$, $s^\mu$, while the pressure is a function of the momenta,  
\be \label{dLambda}
d\Lambda = \partial_\mu\psi\,dj^\mu + \Theta_\mu ds^\mu \, , \qquad d\Psi = j_\mu d(\partial^\mu\psi) + s_\mu d\Theta^\mu \, .
\ee
Since $\Lambda$ and $\Psi$ are Lorentz scalars, they must depend on the Lorentz scalars built from the currents,
$s^2$, $j^2$, $s\cdot j$, and the momenta, $\sigma^2$, $\Theta^2$, $\Theta\cdot\partial\psi$, respectively. Therefore, 
we can write 
\begin{subequations}\label{psiTheta}
\bea
\partial^\mu\psi &=& \frac{\partial\Lambda}{\partial j_\mu} = {\cal B} j^\mu + {\cal A} s^\mu  \, ,\label{psiTheta1}\\[2ex]
\Theta^\mu &=& \frac{\partial \Lambda}{\partial s_\mu} = {\cal A} j^\mu + {\cal C} s^\mu  \, ,\label{Theta}
\eea
\end{subequations}
where
\be
{\cal A}\equiv \frac{\partial \Lambda}{\partial (j\cdot s)} \, , \qquad {\cal B}\equiv 2\frac{\partial \Lambda}{\partial j^2}
\,, \qquad {\cal C}\equiv 2\frac{\partial \Lambda}{\partial s^2} \, ,
\ee
in the notation introduced in Ref.~\cite{Carter:1995if} (where ${\cal C}$, ${\cal B}$, ${\cal A}$ are called caloric, bulk, and anomaly
coefficients).
In general, the momenta are not four-parallel to their corresponding currents, but also receive a contribution from the
other current. This effect manifests itself in the coefficient ${\cal A}$ which therefore is also called {\it entrainment coefficient}, 
see for instance Ref.~\cite{Andersson:2011zza}.

For completeness, and because it will turn out to be useful for the discussion in the subsequent section, we also introduce a notation
for the inverse transformation,
\begin{subequations}\label{js}
\bea
j^\mu &=& \frac{\partial\Psi}{\partial (\partial_\mu\psi)} = \overline{\cal B}\, \partial^\mu\psi + \overline{\cal A}\, \Theta^\mu  \, ,
\label{js1} \\[2ex]
s^\mu &=& \frac{\partial \Psi}{\partial \Theta_\mu} = \overline{\cal A}\, \partial^\mu\psi + \overline{\cal C}\, \Theta^\mu  \, ,
\label{js2}
\eea
\end{subequations}
with
\be \label{entrainbar}
\overline{\cal A}\equiv \frac{\partial \Psi}{\partial(\Theta\cdot\partial\psi)} \, , \qquad 
\overline{\cal B}\equiv 2\frac{\partial \Psi}{\partial \sigma^2}
\,, \qquad \overline{\cal C}\equiv 2\frac{\partial \Psi}{\partial\Theta^2} \, .
\ee
Obviously, the coefficients are related by a simple matrix inversion,
\be \label{AAbar}
\overline{\cal C} = \frac{{\cal B}}{{\cal B}{\cal C}-{\cal A}^2} \, , \qquad \overline{\cal B} = \frac{{\cal C}}{{\cal B}{\cal C}-{\cal A}^2} \, , 
\qquad \overline{\cal A} = -\frac{{\cal A}}{{\cal B}{\cal C}-{\cal A}^2} \, .
\ee
%By contracting Eqs.~(\ref{psiTheta}) with $s_\mu$ and $j_\mu$, we can derive expressions for the coefficients ${\cal A}$, ${\cal B}$, 
%${\cal C}$ in terms of various contractions of $s^\mu$, $j^\mu$, $T^{\mu\nu}$ and the 
%Lorentz scalar $\Lambda$ \cite{Comer:2002dm},
%\bea \label{calB}
%{\cal A} &=& -\frac{j^\mu s^\nu T_{\mu\nu}-(j\cdot s)\Lambda}{(s\cdot j)^2-s^2j^2} \, , 
%\qquad
%{\cal B} =\frac{s^\mu s^\nu T_{\mu\nu}-s^2\Lambda}{(s\cdot j)^2-s^2j^2}  \, , 
%\qquad
%{\cal C} =\frac{j^\mu j^\nu T_{\mu\nu}-j^2\Lambda}{(s\cdot j)^2-s^2j^2} \, .
%\eea
%Analogously,
%\begin{subequations}
%\bea
%\overline{\cal A}
%&=& -\frac{\partial^\mu\psi\Theta^\nu T_{\mu\nu}-(\Theta\cdot\partial\psi)\Lambda}{(\partial\psi\cdot\Theta)^2-\sigma^2\Theta^2} \, , 
%\qquad
%\overline{\cal B}
%= \frac{\Theta^\mu\Theta^\nu T_{\mu\nu}-\Theta^2\Lambda}{(\partial\psi\cdot\Theta)^2-\sigma^2\Theta^2} \, , 
%\qquad
%\overline{\cal C}= \frac{\partial^\mu\psi\partial^\nu \psi T_{\mu\nu}-\sigma^2\Lambda}{(\partial\psi\cdot\Theta)^2-\sigma^2\Theta^2} \, .
%\eea
%\end{subequations}
In summary, we have written the stress-energy tensor (\ref{tmunucomer}) in terms of the Lorentz-scalar $\Psi$ and 4 four-vectors, two
of which are independent. Usually, one chooses either the two four-currents or the two four-momenta as independent variables. The formulation in terms of 
normal and superfluid components (\ref{sn}) is a ``mixed'' form, which uses one of the currents and one of the momenta as its 
basic variables, namely $s^\mu$ (which corresponds to $u^\mu$) and $\partial^\mu\psi$. To translate between these two formulations (see also 
appendix A in Ref.\ \cite{Herzog:2008he}), we rewrite  the current and stress-energy tensor with the help of Eq.\ (\ref{psiTheta}),
\begin{subequations}
\bea
j^\mu &=&\frac{1}{\cal B}\,\partial^\mu\psi - \frac{\cal A}{\cal B}\,s^\mu \, , \\[2ex]  
T^{\mu\nu}
&=& -g^{\mu\nu} \Psi + \frac{1}{{\cal B}}\,\partial^\mu \psi \partial^\nu\psi +\frac{{\cal B}{\cal C}-{\cal A}^2}{\cal B} \,s^\mu s^\nu  \, .
\eea
\end{subequations}
[As an aside, from this form we see that the stress-energy tensor is symmetric, which is not obvious from the form (\ref{tmunucomer}).] 
Comparing with Eqs.~(\ref{sn}), we can identify 
\be \label{identify}
n_s=\frac{\sigma}{\cal B} \, , \qquad n_n = -\frac{{\cal A}s}{\cal B} \, ,\qquad 
\epsilon_s+P_s = \frac{\sigma^2}{\cal B} \, , \qquad \epsilon_n + P_n = \frac{{\cal B}{\cal C}-{\cal A}^2}{\cal B}\,  s^2 \, .
\ee
With the stress-energy tensor (\ref{tmunucomer}) we can also rewrite the hydrodynamical conservation equations (\ref{conserve}). We find
\bea
0=\partial_\mu T^{\mu\nu} = \partial^\nu\psi\partial_\mu j^\mu+  j_\mu(\partial^\mu\partial^\nu\psi - \partial^\nu\partial^\mu\psi) + \Theta^\nu \partial_\mu s^\mu 
+ s_\mu (\partial^\mu\Theta^\nu-\partial^\nu\Theta^\mu) \, ,
\label{dThydro}
\eea
where we have used $\partial^\nu \Psi = j_\mu\partial^\nu\partial^\mu\psi
+s_\mu\partial^\nu\Theta^\mu$ [because of Eq.~(\ref{dLambda})]. The first term on the right-hand side vanishes due to the current conservation.
In our particular case of a superfluid, the second term also vanishes (up to this point, the specific form of the momentum $\partial_\mu\psi$ has not been used, i.e.,
the result holds for arbitrary two-fluid systems). As a consequence, at zero temperature where there is no entropy current, the energy-momentum 
conservation is automatically fulfilled due to the conserved, curl-free current. Contracting the 
remaining two terms with $s_\nu$ shows that, due to the antisymmetry of the so-called {\it vorticity} 
$\partial^\mu\Theta^\nu-\partial^\nu\Theta^\mu$, the entropy current is also conserved (provided that $s_\nu\Theta^\nu\neq 0$). 
Consequently, the conservation equations are equivalent 
to\footnote{The equation for the vorticity is sometimes (see for instance Ref.~\cite{Carter:1995if})
written with the help of the inverse temperature four-vector
\be
\beta^\mu \equiv -\frac{s^\mu}{s\cdot\Theta} \, .\nonumber
\label{jstheta}
\ee
One can easily check that $s_\mu (\partial^\nu\Theta^\mu-\partial^\mu\Theta^\nu) = 
s\cdot\Theta(\beta_\mu\partial^\mu\Theta^\nu+\Theta_\mu\partial^\nu\beta^\mu)$, and thus the 
the vorticity equation becomes 
\be
\beta_\mu\partial^\mu\Theta^\nu+\Theta_\mu\partial^\nu\beta^\mu = 0 \, .\nonumber
\ee
For a physical interpretation of the vorticity equation, see for instance Sec.\ 6 in Ref.\ \cite{Andersson:2006nr}.
}
\be
\partial_\mu j^\mu = 0 \, , \qquad \partial_\mu s^\mu = 0 \,, \qquad  s_\mu (\partial^\mu\Theta^\nu-\partial^\nu\Theta^\mu) = 0 \, .
\ee

It is instructive to write down the current and the stress-energy tensor in two preferred rest frames, the normal-fluid rest frame, defined by 
$s^\mu=(s^0,0,0,0)$, and the superfluid rest frame, defined by $\partial^\mu \psi = (\partial^0\psi,0,0,0)$. The various
components are listed in Table \ref{table1} and can be interpreted as follows. 

The components of the current are written in terms of $n_n$ and $n_s$. (Alternatively, due to the translation given in Eq.~(\ref{identify}), 
we could have written them in terms of the coefficients ${\cal A}$ and ${\cal B}$.) 
Since $n_n$ and $n_s$ are measured in their respective rest frames, the charge density of the other fluid component contains an explicit 
Lorentz factor, i.e., $n_s$ is multiplied by $\partial^0\psi/\sigma$ in the normal-fluid rest frame, and $n_n$ is multiplied by $s^0/s$ 
in the superfluid rest frame. The spatial components of the currents are given by the respective number densities
times the three-velocities of the other fluid.

The components of the stress-energy tensor are written in terms of the Lorentz scalars $\Psi$ and $\Lambda$. The last two lines of the 
table illustrate the meaning of these quantities. The transverse pressure, i.e., the pressure measured in the spatially orthogonal 
direction with respect to the fluid velocity of the other current, is identical to the generalized pressure $\Psi$. The energy density 
$T^{00}$ contains the kinetic energy from the other fluid component. This is exactly the term that distinguishes the transverse 
from the longitudinal pressure. Therefore, the combination of the frame-dependent quantities $T^{00}+T_\perp-T_{||}$ is identical to the 
generalized energy density $\Lambda$.

We may use this table for a comparison with our microscopic 
low-temperature results from Table \ref{table0}. One can for instance check from the explicit results, that the difference between 
longitudinal and transverse pressure is indeed $-{\bf j}\cdot\nabla\psi=\mu\,{\bf j}\cdot{\bf v}_s$, and that the momentum density $T^{i0}$ is indeed 
$\partial^0\psi=\mu$ times the current 
$j^i$. More importantly, we see that $T_\perp=\frac{T}{V}\Gamma$, which, since $T_\perp=\Psi$, suggests that the generalized
pressure should be identified with the effective action. We shall confirm this on a more general level in the next section.

%%%%%%%%%%%%%%%%%%%%%%%%%%%%%%%%%%%%%%%%%%%%%%%%%%%%%%%%%%%%%%%%%%%%%%%
\begin{table*}[t]
\begin{tabular}{|c||c|c|} 
\hline
\rule[-1.5ex]{0em}{5ex} 
 & $\;\;$ normal-fluid rest frame $\;\;$  & superfluid rest frame \\[1ex] \hline\hline
\rule[-1.5ex]{0em}{6ex} 

charge density $j^0$ & $\;\;\displaystyle{n_n+n_s\frac{\partial^0\psi}{\sigma}}\;\;$ & 
$\displaystyle{\;\;n_s+n_n\frac{s^0}{s}\;\;}$ \\[2ex] \hline
\rule[-1.5ex]{0em}{6ex} 

spatial current ${\bf j}$ & $\;\;\displaystyle{\frac{\partial^0\psi}{\sigma}n_s {\bf v}_s }\;\;$ & 
$\displaystyle{\;\;\frac{s^0}{s}n_n{\bf v}_n\;\;}$ \\[2ex] \hline
\rule[-1.5ex]{0em}{6ex} 

%thermal vector  ${\bf \Theta}$ & $\;\;\displaystyle{\frac{n_n}{s_0}\nabla\psi}\;\;$ & 
%$\displaystyle{\;\;\frac{n_n\partial_0\psi+s\Theta_0}{s_0}\frac{{\bf s}}{s}\;\;}$ \\[2ex] \hline
%\rule[-1.5ex]{0em}{5ex} 

energy density $T^{00}$ & $\;\;\displaystyle{\Lambda-{\bf j}\cdot\nabla\psi}\;\;$ & 
$\displaystyle{\;\;\Lambda+{\bf s}\cdot{\bf \Theta}\;\;}$ \\[1ex] \hline
\rule[-1.5ex]{0em}{5ex} 

$\;\;$momentum density $T^{0i}$$\;\;$ & $\;\;\displaystyle{j^i\partial^0\psi}\;\;$ & 
$\displaystyle{\;\;\Theta^i s^0\;\;}$ \\[1ex] \hline
\rule[-1.5ex]{0em}{5ex} 

long.\ pressure $T_{||}$ & $\;\;\displaystyle{\Psi-{\bf j}\cdot\nabla\psi}\;\;$ & 
$\displaystyle{\;\;\Psi+{\bf s}\cdot{\bf \Theta}\;\;}$ \\[1ex] \hline
\rule[-1.5ex]{0em}{5ex} 

transv.\ pressure $T_{\perp}$ & $\;\;\displaystyle{\Psi}\;\;$ & 
$\displaystyle{\;\;\Psi\;\;}$ \\[1ex] \hline
\rule[-1.5ex]{0em}{5ex} 

$T^{00}+T_\perp-T_{||}$  & $\;\;\displaystyle{\Lambda}\;\;$ & 
$\displaystyle{\Lambda}$ \\[1ex] \hline
%\rule[-1.5ex]{0em}{4ex} 

%superfluid density $\rho_s$  & $\;\;\displaystyle{\mu\frac{n_s}{\sqrt{1-v_s^2}}}\;\;$ & 
%$\displaystyle{\;\;\sigma n_s\;\;}$ \\[1ex] \hline
%\rule[-1.5ex]{0em}{4ex} 

%normal density $\rho_n$  & $\;\;\displaystyle{\mu n_n + sT}\;\;$ & 
%$\displaystyle{\;\;\sigma\frac{n_n}{\sqrt{1-v_n^2}}+\frac{s_\mu\partial^\mu\psi}{\sigma}\frac{\Theta_\mu\partial^\mu\psi}{\sigma}\;\;}$ 
%\\[2ex] \hline

\end{tabular}
\caption{Components of the current and the stress-energy tensor in the normal and superfluid rest frames. In each frame, $\partial^0\psi$
is the chemical potential, $s^0$ the entropy, and $\Theta^0$ the temperature, while $n_n$ and $n_s$ are the normal and superfluid number densities,
measured in their respective rest frames and ${\bf v}_n={\bf s}/s^0$ (${\bf v}_s=-\nabla\psi/\partial^0\psi$) the three-velocities of the 
normal (superfluid) component, measured in the superfluid (normal) rest frame. Longitudinal and transverse pressures are defined with respect to
the three-direction of the velocity of the other fluid component.}
\label{table1}
\end{table*}
%%%%%%%%%%%%%%%%%%%%%%%%%%%%%%%%%%%%%%%%%%%%%%%%%%%%%%%%%%%%%%%%%%%%%%%

Finally, let us use the relations summarized in the table to define superfluid and normal densities $\rho_s$ and 
$\rho_n$. These {\it energy} densities are defined in the low-velocity limit, in generalization
of the mass densities in the nonrelativistic framework (and in contrast to $n_s$, $n_n$ which are {\it number} densities).
To define the superfluid density $\rho_s$ we first have to go into the normal-fluid rest frame. With $T^{0i}=j^i\partial^0\psi$, ${\bf j}=-n_s\nabla\psi/\sigma$, 
and ${\bf v}_s = -\nabla\psi/\partial^0\psi$, we obtain 
\be
T^{0i} = (\partial^0\psi)^2\frac{n_s}{\sigma} {\bf v}_{si}= \rho_s {\bf v}_{si} + {\cal O}(|{\bf v}_s|^3) \, , 
\ee
with the Lorentz scalar
\be 
\rho_s = \sigma n_s \, .
\ee
(Alternatively, we can write $\rho_s = \sigma^2/{\cal B}$.) This expression for the superfluid density is obviously in agreement with the zero-temperature
result (\ref{rhosT0}). The superfluid density appears also in the energy density as part of the kinetic energy,
$T^{00} = \Lambda + \rho_s {\bf v}_s^2+ {\cal O}(|{\bf v}_s|^4)$. The normal fluid density is defined analogously: in the superfluid rest frame, we have 
$T^{i0}=j^i\partial^0\psi+s^i\Theta^0$. Inserting $j^i=n_ns^i/s$ and using ${\bf v}_n={\bf s}/s^0$, we obtain
\be
T^{0i} = \left(\partial^0\psi\frac{s^0n_n}{s}+s^0\Theta^0\right) {\bf v}_{ni} = \rho_n {\bf v}_{ni} + {\cal O}(|{\bf v}_n|^3) \, , 
\ee
with 
\be
\rho_n = \sigma n_n + s \Theta \, .
\ee
(Alternatively, $\rho_n = -\sigma s{\cal A}/{\cal B} + s\Theta$.) 

%We define the superfluid density measured in the normal-fluid rest frame via $T^{0i}=\rho_s {\bf v}_{s,i}$, 
%and the normal density measured in the superfluid rest frame via $T^{0i}=\rho_n {\bf v}_{n,i}$. 
%(Note that the superfluid density also appears in the other components, $T^{00} = \Lambda + \rho_s {\bf v}_s^2$, 
%$T_{||} = \Psi + \rho_s {\bf v}_s^2$, and analogously for the normal fluid density.)
%In the normal-fluid rest frame, with $T^{0i}=j^i\partial^0\psi$, $j^i=n_s\partial^i\psi/\sigma$, and ${\bf v}_s = 
%\nabla\psi/\partial^0\psi$, we read off
%\be
%\rho_s = \mu\frac{\mu n_s}{\sigma} \, ,
%\ee
%which is $\mu$ times the superfluid charge density measured in the normal-fluid rest frame. In the superfluid rest frame, we have $T^{0i}=s^0\Theta^i
%=T^{i0}=j^i\partial^0\psi+s^i\Theta^0$. Inserting $j^i=n_ns^i/s$ into the latter expression and using ${\bf v}_n={\bf s}/s$, we read off
%\be
%\rho_n = \partial_0\psi\frac{s_0 n_n}{s}+s_0\Theta_0 \, .
%\ee
%The first term is exactly analogous to $\rho_s$ since it is the chemical potential times the normal fluid number density (both measured in the 
%superfluid rest frame). The second term is entropy $s^0$ times temperature $\Theta^0$.  

\section{Two fluids in terms of microscopic variables}
\label{sec:translate}

How is the two-fluid formalism explained in the previous subsection related to the microscopic calculation of Sec.~\ref{sec:nonzeroT}?
In other words, what is the field-theoretic definition of the basic quantities of the two-fluid formalism? 
We have seen in Sec.~\ref{sec:1fluid} 
that for zero temperature the field-theoretic quantities could be translated straightforwardly into (single-fluid) hydrodynamics. 
For instance, the generalized pressure was simply given by the Lagrangian. In particular, 
the microscopic formulation was covariant, even in the presence of a chemical potential. 
The microscopic 
results from Sec.~\ref{sec:nonzeroT}, however, show that this covariance is lost when we go to nonzero temperatures, at least in the explicit 
results, see for instance the dispersion relations from Eq.~(\ref{smallk}). Nevertheless, we shall now
show that, at least in our static, homogeneous, dissipationless scenario, we may map the field-theoretic formulation onto 
the hydrodynamic one. Our covariant formulation from Sec.~\ref{sec:Tmunu} {\it before} performing the Matsubara sum will turn out to be 
very useful for this purpose. 

\subsection{Generalized pressure corresponds to effective action}

The first important step is to realize that the microscopic calculation is, in the terminology of the previous subsection, 
performed in the normal-fluid rest frame $u^\mu=(1,0,0,0)$, which is the rest frame of the heat bath. 
For discussions how a general frame can be introduced in relativistic 
thermodynamics and thermal field theory in a covariant way see for instance Refs.~\cite{1981PhyA..106..204I,Weldon:1982aq}. 
We are not interested in reintroducing a general $u^\mu$ in our calculation, which would be complicated conceptually 
because we are using the imaginary time formalism and also complicated technically since there is already one four-vector in our calculation, 
namely $\partial^\mu\psi$. 

Our first goal is to identify the generalized pressure $\Psi$. With the stress-energy tensor (\ref{tmunucomer}) and using $u^\mu = s^\mu/s$
we can write 
\be \label{Tortho1}
\Psi = \frac{1}{3}(g^{\mu\nu}-u^\mu u^\nu)(j_\mu \partial_\nu\psi-T_{\mu\nu}) \, .
\ee
Let us now compute the right-hand side with the microscopic expression for $T^{\mu\nu}$. 
With $T^{\mu\nu}$ and $j^\mu$ from Eqs.~(\ref{Tmunujmu}) we compute 
\bea \label{towardsPsi}
\frac{1}{3}(g^{\mu\nu}-u^\mu u^\nu)(j_\mu \partial_\nu\psi-T_{\mu\nu}) = 
-U - \frac{1}{3}(g^{\mu\nu}-u^\mu u^\nu)\frac{T}{V}\sum_k\left(C_k k_\mu k_\nu+A_k k_\mu \partial_\nu\psi\right) \, .
\eea
Here we have used $u^\mu=(1,0,0,0)$, while Eq.~(\ref{Tortho1}) is a general relation for arbitrary four-velocities $u^\mu$. 
It is important to keep in mind that we cannot simply promote $u^\mu$ to
an arbitrary four-velocity in the microscopic calculation. It would occur additionally in different places in the calculation, which we have not 
identified in our present treatment. But, of course, since $\Psi$ is a Lorentz scalar, 
the normal-fluid rest frame is as good as any other frame to compute $\Psi$. By comparing with Eq.~(\ref{gamCA}) we see that the right-hand side 
of Eq.~(\ref{towardsPsi}) is exactly 
($T/V$ times) the effective potential. Therefore, we have derived our first important result,
\be \label{PsiGam}
\Psi = \frac{T}{V}\Gamma \, .
\ee
This relation is somewhat expected since we already know that, at zero temperature (and at tree-level), $\Psi$ corresponds to the Lagrangian, 
which in this case gives the microscopic pressure. At nonzero temperature (without superflow), the effective action gives the (isotropic) pressure. 
Therefore, the relation (\ref{PsiGam}) is a natural generalization to the anisotropic case with a nonzero superflow. This motivates our choice of notation in
Eq.~(\ref{PsiK}) because now we have 
\be \label{PsiPsik}
\Psi = - U + \frac{T}{V}\sum_k\Psi_k \, .
\ee
Next, let us discuss the generalized thermodynamic relation $\Lambda = -\Psi + j\cdot\partial\psi + s\cdot\Theta$. In the normal-fluid rest frame, 
$s^\mu=(s^0,0,0,0)$ and thus $s\cdot\Theta = s^0\Theta^0$, which is the product of entropy and temperature, measured in this 
particular frame. To confirm this microscopically, we use the thermodynamical definition of the entropy density, 
\be \label{TPsiT}
s = 
\frac{\partial\Psi}{\partial T} = \frac{1}{V}\sum_k\left(\Psi_k+2+C_kk_0^2+A_k k_0\partial_0\psi\right) \, ,
\ee
with $A_k$ and $C_k$ defined in Eqs.~(\ref{dPsiK}).
Note that the first two terms ($\Psi_k$ and 2) come from the explicit $T$-dependence in the prefactor $T/V$ and in the $1/T^2$ within the logarithm.
On the other hand, we can compute $s\cdot\Theta$ via the generalized thermodynamic relation. We find 
\bea
s\cdot\Theta &=& \Lambda+\Psi - j\cdot \partial\psi \non[2ex]
 &=& \frac{T}{V}\sum_k\left[\Psi_k+C_kk_0^2+A_k k_0\partial_0\psi -(C_k k^2+B_k\sigma^2+2A_k k\cdot\partial\psi)+\frac{2m^2k^2}{{\rm det}\,S^{-1}}
\right] \, ,
\eea
where we have used $\Lambda= T^\mu_{\;\;\mu} + 3\Psi$ (\ref{Lambdasingle}), the trace of the stress-energy tensor (\ref{trace}), the 
effective action (which is $\Psi$) (\ref{gamCA}), and the current (\ref{currABC}). With the help of the identity (\ref{one}) we see that this is indeed the same as $T$ 
times the entropy from Eq.~(\ref{TPsiT}). We have thus identified all terms in the generalized thermodynamic relation microscopically.

\subsection{Entrainment and superfluid density from field theory}

The independent degrees of freedom of our microscopic calculation are the chemical potential $\mu=\partial^0\psi$, the temperature
$T=\Theta^0$, and the superfluid three-velocity ${\bf v}_s = -\nabla\psi/\mu$, all measured in the normal-fluid rest frame, where the entropy current 
vanishes by definition, $s^i=0$. We have thus given 8 independent components out of the 
16 components of the 4 four-vectors $j^\mu$, $s^\mu$, $\Theta^\mu$, $\partial^\mu\psi$ of the two-fluid formalism. The other 8 components
are $j^\mu$, $s^0$, and $\Theta^i$. For the Noether current $j^\mu$ and the entropy $s^0$ we have field-theoretic and thermodynamic definitions.
It remains to compute the spatial components of the thermal four-vector
$\Theta^\mu$. Additionally, we have to compute the coefficients $\overline{\cal A}$, $\overline{\cal B}$, $\overline{\cal C}$. 
They are defined as the derivatives of $\Psi$ with respect to the Lorentz scalars $\sigma^2$, $\Theta^2$, and $\Theta\cdot\partial\psi$. 
However, this is not the form in which our $\Psi$ is given. Therefore, we need to find a different way to compute these coefficients. 
This can be done with the help of Eqs.~(\ref{js}). First, we solve the spatial part of Eq.~(\ref{js2})
for $\Theta^i$ and insert the result into the spatial part of Eq.~(\ref{js1}). Together with the temporal components, this yields
three equations for the three variables $\overline{\cal A}$, $\overline{\cal B}$, $\overline{\cal C}$, whose solutions are listed in Table
\ref{table2}, where, for completeness, we also give the coefficients  ${\cal A}$, ${\cal B}$, ${\cal C}$, which are obtained from the 
inverse of Eqs.~(\ref{AAbar}).
%%%%%%%%%%%%%%%%%%%%%%%%%%%%%%%%%%%%%%%%%%%%%%%%%%%%%%%%%%%%%%%%%%%%%%%
\begin{table*}[t]
\begin{tabular}{|c|c||c|c|} 
\hline
\rule[-1.5ex]{0em}{6ex} 
$\;\;\overline{\cal A}\;\;$ & $\;\; \displaystyle{\frac{s^0}{\partial^0\psi}\frac{{\bf v}_s^2j^0\partial^0\psi+{\bf j}\cdot\nabla\psi}{{\bf v}_s^2
j^0\partial^0\psi+{\bf j}\cdot\nabla\psi +{\bf v}_s^2s^0\Theta^0}}\;\;$ & $\;\;{\cal A}\;\;$ & 
$\displaystyle{\frac{\partial^0\psi}{s^0{\bf j}\cdot\nabla\psi}({\bf v}_s^2j^0\partial^0\psi+{\bf j}\cdot\nabla\psi)}$ \\[2ex] \hline
\rule[-1.5ex]{0em}{6ex} 
$\overline{\cal B}$ & $\;\; \displaystyle{\frac{1}{(\partial^0\psi)^2}\frac{j^0\partial^0\psi({\bf v}_s^2j^0\partial^0\psi+{\bf j}\cdot\nabla\psi)
-{\bf j}\cdot\nabla\psi s^0\Theta^0}{{\bf v}_s^2j^0\partial^0\psi+{\bf j}\cdot\nabla\psi+{\bf v}_s^2s^0\Theta^0}}\;\;$& ${\cal B}$ & 
$\displaystyle{-\frac{(\nabla\psi)^2}{{\bf j}\cdot\nabla\psi}}$ \\[2ex] \hline
\rule[-1.5ex]{0em}{6ex} 
$\overline{\cal C}$ & $\;\; \displaystyle{\frac{{\bf v}_s^2(s^0)^2}{{\bf v}_s^2j^0\partial^0\psi+{\bf j}\cdot\nabla\psi
+{\bf v}_s^2s^0\Theta^0}}\;\;$& ${\cal C}$ & 
$\;\;\displaystyle{-\frac{j^0\partial^0\psi({\bf v}_s^2j^0\partial^0\psi+{\bf j}\cdot\nabla\psi)-{\bf j}\cdot\nabla\psi
s^0\Theta^0}{(s^0)^2\,{\bf j}\cdot\nabla\psi}}\;\;$ \\[2ex] \hline
\end{tabular}
\caption{Coefficients that relate the currents $j^\mu$, $s^\mu$ with the conjugate momenta $\partial^\mu\psi$, $\Theta^\mu$, given 
in terms of ``microscopic'' quantities: the Noether current $j^\mu$, the space-time derivative of the phase of the condensate 
$\partial^\mu\psi$, the temperature $\Theta^0$, and the entropy density $s^0$, all measured in the normal-fluid rest frame. 
%The total conserved 
%current $j^\mu$ can be decomposed into a term parallel to the corresponding momentum, $\overline{\cal B}\partial^\mu\psi$, plus 
%a term parallel to the momentum associated withe the entropy current, $\overline{\cal A}\Theta^\mu$, with the so-called
%entrainment coefficient $\overline{\cal A}$. Analogously, the entropy current is a sum of a term parallel to its momentum, 
%$\overline{\cal C}\Theta^\mu$ and an entrainment term $\overline{\cal A}\partial^\mu\psi$. The coefficients ${\cal A}$, ${\cal B}$, ${\cal C}$
%appear in the inverse decomposition of the momenta in terms of the currents. 
The low-temperature approximations for $\overline{\cal A}$, 
$\overline{\cal B}$, $\overline{\cal C}$ are given in Eqs.~(\ref{resultABC}). }
\label{table2}
\end{table*}
%%%%%%%%%%%%%%%%%%%%%%%%%%%%%%%%%%%%%%%%%%%%%%%%%%%%%%%%%%%%%%%%%%%%%%%
With these results we immediately find
\be
{\bf \Theta} = -\frac{\cal A}{\cal B}\nabla\psi = \frac{\nabla\psi}{s^0}\left[j^0+\partial^0\psi\frac{{\bf j}\cdot\nabla\psi}{(\nabla\psi)^2}
\right] \, .
\ee
Since all results are expressed in terms of quantities accessible from our microscopic calculation, we can for instance compute (for $m=0$ and in the 
low-temperature limit) 
\begin{subequations} \label{resultABC}
\bea
\overline{\cal A} &\simeq& \frac{4\pi^2T^3}{15\sqrt{3}\,\mu}\,\frac{1-{\bf v}_s^2}{(1-3{\bf v}_s^2)^2} 
-\frac{16\pi^4T^5}{315\sqrt{3}\,\mu^3}\,\frac{1-{\bf v}_s^2}{(1-3{\bf v}_s^2)^5}(25+78{\bf v}_s^2-27{\bf v}_s^4)\, ,\\[2ex]
\overline{\cal B} &\simeq& \frac{\mu^2}{\lambda}(1-{\bf v}_s^2)-\frac{4\pi^2T^4}{15\sqrt{3}\,\mu^2}\,\frac{1-{\bf v}_s^2}{(1-3{\bf v}_s^2)^3} 
+\frac{8\pi^4T^6}{315\sqrt{3}\,\mu^4}\frac{65+256{\bf v}_s^2-402{\bf v}_s^4+81{\bf v}_s^8}{(1-3{\bf v}_s^2)^6} \, ,\\[2ex]
\overline{\cal C} &\simeq& \frac{2\pi^2T^2}{15\sqrt{3}}\frac{1-{\bf v}_s^2}{1-3{\bf v}_s^2} +\frac{8\pi^4T^4}{315\sqrt{3}\,\mu^2}
\frac{5-59{\bf v}_s^2+27{\bf v}_s^4+27{\bf v}_s^6}{(1-3{\bf v}_s^2)^4} \, ,
\eea
\end{subequations}
and
\be
{\bf \Theta} \simeq -\frac{2{\bf v}_s T}{1-3{\bf v}_s^2}\left[1-\frac{8(1-{\bf v}_s^2)(5+9{\bf v}_s^2)}{7(1-3{\bf v}_s^2)^3}\left(\frac{\pi T}{\mu}\right)^2\right]\, .
\ee
%\begin{subequations} \label{resultABC}
%\bea
%\overline{\cal A} &\simeq& \frac{4\pi^2}{15\sqrt{3}}\left(1+\frac{89}{18}{\bf v}_s^2\right)\frac{T^3}{\mu} 
%-\frac{80\pi^4}{63\sqrt{3}}\left(1+\frac{1381}{81}{\bf v}_s^2\right)\frac{T^5}{\mu^3}\, ,\\[2ex]
%\overline{\cal B} &\simeq& \frac{\mu^2}{\lambda}(1-{\bf v}_s^2)-\frac{4\pi^2}{15\sqrt{3}}\left(1+\frac{143}{18}{\bf v}_s^2\right)\frac{T^4}{\mu^2}
%+\frac{104\pi^4}{63\sqrt{3}}\left(1+\frac{23044}{1053}{\bf v}_s^2\right)\frac{T^6}{\mu^4}\, ,\\[2ex]
%\overline{\cal C} &\simeq& \frac{2\pi^2}{15\sqrt{3}}\left(1+\frac{19}{9}{\bf v}_s^2\right)T^2 +\frac{8\pi^4}{63\sqrt{3}}
%\left(1-\frac{41}{81}{\bf v}_s^2\right)
%\frac{T^4}{\mu^2} \, ,
%\eea
%\end{subequations}
%and
%\be
%{\bf \Theta} \simeq -2{\bf v}_s T\left[1-\frac{40}{7}\left(\frac{\pi T}{\mu}\right)^2-\frac{4000}{441}\left(\frac{\pi T}{\mu}\right)^4\right] \, .
%\ee
We see that the coefficients $\overline{\cal A}$ and $\overline{\cal C}$ vanish at $T=0$. This is in accordance to our zero-temperature 
discussion, where only $\overline{\cal B}$ was nonzero.

The connection between the coefficients ${\cal A}$, ${\cal B}$ and the number densities $n_s$, $n_n$ is given in Eq.~(\ref{identify}).
We can insert ${\cal A}$ and ${\cal B}$ as functions of $n_s$, $n_n$ into the temporal component of Eq.~(\ref{Theta}) to get also ${\cal C}$ 
as a function of $n_s$, $n_n$. The result is the useful translation
\bea
{\cal A} &=& -\frac{\sigma n_n}{s n_s} \, , \qquad {\cal B} = \frac{\sigma}{n_s} \, , \qquad 
{\cal C} = \frac{\sigma n_n^2}{s^2 n_s}+\frac{\mu n_n+sT}{s^2} \, .
\eea
%(To avoid confusion: $s$, $T$, $\mu$, $n_n$ are quantities measured in the normal frame, while $n_s$ is the superfluid density measured in the 
%superfluid frame. The superfluid density measured in the normal frame is $n_s\mu/\sigma$.) 
As a check, we see that ${\cal B}$ given in 
Table \ref{table2} is indeed the same as ${\cal B} = \sigma/n_s$ with $n_s$ from Eq.~(\ref{ns1}). It is now of course straightforward to 
also express $\overline{\cal A}$, $\overline{\cal B}$, $\overline{\cal C}$ in terms of $n_s$ and $n_n$. 

In the small-temperature approximation, the superfluid and normal number densities, measured in the normal-fluid rest frame, 
become
\begin{subequations} \label{nsnn}
\bea
n_s\frac{\mu}{\sigma} &\simeq&  \frac{\mu^3}{\lambda}(1-{\bf v}_s^2)-\frac{4\pi^2T^4}{5\sqrt{3}\,\mu}\,\frac{1-{\bf v}_s^2}{(1-3{\bf v}_s^2)^3}
+\frac{8\pi^4T^6}{105\sqrt{3}\,\mu^3}\,\frac{1-{\bf v}_s^2}{(1-3{\bf v}_s^2)^6}(95+243{\bf v}_s^2-135{\bf v}_s^4-27{\bf v}_s^6) \, , \\[2ex]
n_n &\simeq& 
\frac{4\pi^2T^4}{5\sqrt{3}\,\mu}\,\frac{(1-{\bf v}_s^2)^2}{(1-3{\bf v}_s^2)^3}
-\frac{16\pi^4T^6}{35\sqrt{3}\,\mu^3}\,\frac{(1-{\bf v}_s^2)^2}{(1-3{\bf v}_s^2)^6}
(15+38{\bf v}_s^2-9{\bf v}_s^4) \, . \label{nnexpl}
\eea
\end{subequations}
%\begin{subequations} \label{nsnn}
%\bea
%n_s\frac{\mu}{\sigma} &=& \frac{\mu}{{\cal B}} 
%\simeq  \frac{\mu^3}{\lambda}(1-{\bf v}_s^2)-\frac{4\pi^2}{5\sqrt{3}}\left(1+\frac{47}{6}{\bf v}_s^2\right)\frac{T^4}{\mu}
%+\frac{152\pi^4}{21\sqrt{3}}\left(1+\frac{9904}{513}{\bf v}_s^2\right)\frac{T^6}{\mu^3} \, , \\[2ex]
%n_n &=& -\frac{{\cal A}s}{\cal B}  
%\simeq 
%\frac{4\pi^2}{5\sqrt{3}}
%\left(1+\frac{41}{6}{\bf v}_s^2\right)\frac{T^4}{\mu}-\frac{48\pi^4}{7\sqrt{3}}
%\left(1+\frac{4439}{243}{\bf v}_s^2\right)\frac{T^6}{\mu^3} \, . \label{nnexpl}
%\eea
%\end{subequations}
(Remember that $n_s$ is the superfluid density is the superfluid rest frame; the Lorentz
factor $\mu/\sigma=1/\sqrt{1-{\bf v}_s}$ transforms it to the normal-fluid rest frame.)
One can check that the sum of both densities gives the total charge density $j^0$ from Table \ref{table0}.
%Had we only been interested in the superfluid and normal densities, we could simply have computed $n_s$ from Eq.~(\ref{ns1}) and $n_n$
%from $n_n=j^0-\mu n_s/\sigma$, and the calculation of the coefficients $\overline{\cal A}$, $\overline{\cal B}$, $\overline{\cal C}$ would
%not have been necessary. 
%The results for $n_s$ and $n_n$ can be interpreted as follows. 
As expected, the normal density vanishes for $T=0$ and 
begins to increase with increasing temperature, while the superfluid density decreases. In a more complete treatment, we would expect the 
superfluid density to vanish at the critical temperature because the condensate melts. Remember that this melting is, in our one-loop effective 
action, a higher-order effect in the coupling constant, which we have neglected. The decrease of $n_s$ is therefore only due to the 
interaction between the two fluids. 
As an aside, note that for ${\bf v}_s=0$ the $T^4$ contributions in superfluid and normal densities cancel each other exactly. 
We have made this observation already below Eq.~(\ref{Piso}) where we have seen that in the $m=0$ limit there is no $T^4$ contribution
to the density.

Finally, we may express the generalized pressure $\Psi$ in terms of Lorentz scalars. This reformulation is instructive because it gives $\Psi$ in the form that 
is usually assumed in the two-fluid formalism. However, for field-theoretic calculations that start from an underlying Lagrangian or an 
effective action, this form of $\Psi$ can only be found a posteriori and is thus not of particular practical use. 
Our quantities in the normal-fluid rest frame $T$, $\mu$, ${\bf v}_s$ are translated into the relevant Lorentz scalars $\sigma^2$, $\Theta^2$, $\partial\psi\cdot\Theta$
via 
\begin{subequations}
\bea
\sigma^2 &=& \mu^2 - (\nabla\psi)^2 = \mu^2(1-{\bf v}_s^2) \\[2ex]
\Theta^2 &=& T^2 - \frac{{\cal A}^2}{{\cal B}^2}(\nabla\psi)^2 = \frac{(1-{\bf v}_s^2)(1-9{\bf v}_s^2)}{(1-3{\bf v}_s^2)^2}\,T^2 + {\cal O}(T^4) \, , \label{lor2}\\[2ex]
\partial\psi\cdot\Theta &=& \mu T - \frac{{\cal A}}{{\cal B}}(\nabla\psi)^2 = \frac{1-{\bf v}_s^2}{1-3{\bf v}_s^2}\,\mu T + {\cal O}(T^3) \, , \label{lor3}
\eea
\end{subequations}
We solve these equations for $T$, $\mu$, and ${\bf v}_s$ and insert the result into the effective action (\ref{tvgam}). Then, up to fourth order in the temperature 
we can write the generalized pressure as
\be \label{Psonic}
\Psi(\sigma^2,\Theta^2,\Theta\cdot\partial\psi) \simeq \frac{\sigma^4}{4\lambda}+\frac{\pi^2}{90\sqrt{3}}
\left[\Theta^2+2\frac{(\partial\psi\cdot\Theta)^2}{\sigma^2}\right]^2 \, . 
\ee
The term in the square brackets can be written as ${\cal G}^{\mu\nu}\Theta_\mu\Theta_\nu$ with the ``sonic metric'' ${\cal G}^{\mu\nu}\equiv g^{\mu\nu}+2v^\mu v^\nu$,
see Eq.\ (8.9) of Ref.\ \cite{Carter:1995if}. In other words, the Lorentz invariant $T^4$ term of the pressure in the presence of a superflow is obtained by 
replacing $T^2\to{\cal G}^{\mu\nu}\Theta_\mu\Theta_\nu$ in the $T^4$ term of the pressure in the absence of a superflow. In principle, we can use the higher order
terms in Eqs.\ (\ref{lor2}), (\ref{lor3}) to write the $T^6$ contribution in terms of Lorentz scalars. However, we have not found a compact way of writing this
contribution. But, we have checked that it is not simply given by the same replacement as for the $T^4$ term. This is no surprise since the sonic metric 
is constructed solely for systems for a Goldstone mode with linear dispersion. The $T^6$ term however, knows about the cubic term in the dispersion.

\section{Sound velocities}
\label{sec:sound}

As an application of our results we compute the two sound velocities of the superfluid. Related calculations can be found in the recent literature in the 
nonrelativistic context of superfluid atomic gases \cite{2009PhRvA..80e3601T,2009PhRvA..80d3613A,2010NJPh...12d3040H,2010PhRvL.105o0402B}, where the experimental 
observation of both sound modes is in principle possible, although challenging \cite{2009PhRvA..80d3605M,2011PhRvA..84e3612A}. 
Our results will give the sound velocities in the presence of an arbitrary superflow, i.e., an arbitrary relative velocity between the superfluid and the normal 
fluid (limited by a critical velocity, as our results will show). In particular, they will depend on the angle between the direction of the sound wave and the 
direction of the superflow. A similar calculation 
in the nonrelativistic context of superfluid helium has been performed in Ref.\ \cite{2009PhRvB..79j4508A} where, in contrast to our calculation, the sound velocities 
are computed in the superfluid rest frame and without temperature corrections. We do find temperature corrections to the velocity of second sound, which, as we shall see, 
arise from the cubic terms in momentum of the dispersion relation of the Goldstone mode.

The sound wave equations are derived from the hydrodynamic
equations. We start from Eqs.~(\ref{jstheta}), i.e., from the current and entropy conservation and the vorticity equation. [Of course, equivalently, 
one can start from the current conservation plus energy-momentum conservation (\ref{conserve}).] In addition, we need the expression for $d\Psi$ from Eq.~(\ref{dLambda}),
which will allow us to rewrite derivatives of thermodynamic quantities in terms of derivatives of the independent variables. These are the chemical potential 
$\mu=\partial^0\psi$, the superfluid three-velocity ${\bf v}_s$ (more precisely, 
$\nabla\psi=-\mu{\bf v}_s$), the temperature $T=\Theta^0$, and the normal fluid three-velocity ${\bf v}_n$. 
All these variables are now allowed to exhibit small oscillations in space and time about their equilibrium values, $T\to T+\delta T({\bf x},t)$, 
$\mu\to \mu+\delta \mu({\bf x},t)$. We perform the calculation in 
the normal-fluid rest frame from the previous sections, i.e., the superfluid velocity has a (large) static and homogeneous equilibrium value on top of 
which the sound wave oscillations occur, ${\bf v}_s\to {\bf v}_s + \delta{\bf v}_s({\bf x},t)$, while the static and homogeneous part of the normal velocity can be set to zero,
${\bf v}_n\to \delta{\bf v}_n({\bf x},t)$. Of course, we need to keep the
oscillations of the normal fluid velocity $\delta{\bf v}_n({\bf x},t)$
because there is no global rest frame in which they vanish.

We employ the linear approximation in the oscillations. In this case, the temporal component of the vorticity equation is trivially fulfilled.
From the remaining equations one can eliminate the normal velocity, such that one is left with two equations where the sound wave oscillations
are solely expressed in terms of oscillations in $T$ and $\mu$ (oscillations of the superfluid velocity $\nabla\psi$ can be expressed in terms of oscillations
of $\mu$ by applying a time derivative to the whole equation and using $\partial_0\nabla\psi=\nabla\mu$). The derivation of the wave equations is quite lengthy, and we 
explain the details in appendix \ref{AppC1}. One obtains the following system of two equations,
\begin{subequations} \label{hydonetwo}
\bea \label{hydone}
0&\simeq& \frac{w}{s}\left(\frac{\partial n}{\partial T}\partial_0^2\mu+\frac{\partial s}{\partial T}\partial_0^2 T\right) -n_n\Delta\mu-s\Delta T \non[2ex]
&&+\left[\frac{n_s}{\sigma}-\frac{w}{s}\frac{\partial (n_s/\sigma)}{\partial T} +\frac{n_n}{s}\frac{\partial n}{\partial T}-\frac{\partial n}{\partial \mu}
-2\mu\frac{\partial n}{\partial (\nabla\psi)^2}\right]\nabla\psi\cdot\nabla\partial_0\mu
+\left[\frac{n_n}{s}\frac{\partial s}{\partial T}-\frac{\partial s}{\partial \mu}-2\mu\frac{\partial s}{\partial (\nabla\psi)^2}\right]\nabla\psi\cdot\nabla\partial_0 T
\non[2ex]
&&-\left[\frac{n_n}{s}\frac{\partial (n_s/\sigma)}{\partial T}-\frac{\partial (n_s/\sigma)}{\partial \mu}-2\mu\frac{\partial (n_s/\sigma)}
{\partial (\nabla\psi)^2}\right](\nabla\psi\cdot\nabla)^2\mu \, , \allowdisplaybreaks \\[2ex]
0&\simeq& \left(\mu\frac{\partial n}{\partial\mu}+T\frac{\partial n}{\partial T}\right)\partial_0^2\mu+\left(\mu\frac{\partial s}{\partial\mu}
+T\frac{\partial s}{\partial T}\right)\partial_0^2T-n\Delta\mu-s\Delta T \non[2ex]
&& +\left[\frac{n_s}{\sigma}-\mu\frac{\partial(n_s/\sigma)}{\partial\mu}-T\frac{\partial(n_s/\sigma)}{\partial T}+\frac{n_n}{s}\frac{\partial n}{\partial T}-
\frac{\partial n}{\partial \mu}\right]\nabla\psi\cdot\nabla\partial_0\mu +\left(\frac{n_n}{s}\frac{\partial s}{\partial T}
-\frac{\partial s}{\partial \mu}\right)\nabla\psi\cdot\nabla\partial_0 T \non[2ex]
&& -\left[\frac{n_n}{s}\frac{\partial (n_s/\sigma)}{\partial T}-\frac{\partial (n_s/\sigma)}{\partial \mu}\right](\nabla\psi\cdot\nabla)^2\mu
\, ,\label{hydtwo}
\eea
\end{subequations}
where $w\equiv \mu n_n+sT$ is the enthalpy density of the normal fluid.
Each term is a product of a space-time derivative -- in which we can replace $T$ by $\delta T({\bf x},t)$ and  $\mu$ by $\delta \mu({\bf x},t)$ -- 
and a prefactor that only contains the equilibrium values $T$, $\mu$, and $\nabla\psi=-\mu{\bf v}_s$. 

Before coming to the general result, let us write down the wave equations in two limit cases. Firstly, let us set $T=0$. 
In this case, the normal number density vanishes, $n_n=0$, and thus $n=n_s\mu/\sigma$. With this relation and the zero-temperature expression $n_s=\sigma^3/\lambda$ 
(we set $m=0$ for simplicity in this subsection) one finds that all terms on the right-hand side of Eq.~(\ref{hydone}) vanish, and Eq.~(\ref{hydtwo}) can be compactly
written as 
\be
0\simeq (g^{\mu\nu}+2v^\mu v^\nu)\partial_\mu\partial_\nu \mu \, .
%(\sigma^2+2\mu^2)\partial_0^2\mu-\sigma^2\Delta\mu-4\mu\nabla\psi\cdot\nabla\partial_0\mu+2\partial_i\psi\partial_j\psi\partial_i\partial_j\mu \, .
\ee
Again, we recover the sonic metric ${\cal G}^{\mu\nu} = g^{\mu\nu}+2v^\mu v^\nu$, see remark below the generalized pressure (\ref{Psonic}). 
With $\delta \mu= \delta \mu_0 e^{ik\cdot x}$ we obtain ${\cal G}^{\mu\nu}k_\mu k_\nu=0$, which is Eq.\ (4.12) of Ref.\ \cite{Carter:1995if}, 
see also Eq.\ (29) of Ref.\ \cite{Mannarelli:2008jq}. This wave equation has one physical solution $\omega = u_1 |{\bf k}|$, with the velocity of first sound $u_1$. 
The solution is given in Eq.\ (\ref{u1full}) (as we shall see below, this solution is unaltered by temperature effects up to the order we are working).

Secondly, we discuss the limit case without superflow, $\nabla\psi=0$. In this case, 
only the first lines of Eqs.~(\ref{hydone}) and (\ref{hydtwo}) are nonvanishing. Now, with $\delta \mu= \delta \mu_0 e^{i(\omega t-{\bf k}\cdot{\bf x})}$ and
$\delta T = \delta T_0 e^{i(\omega t-{\bf k}\cdot{\bf x})}$
we obtain two equations for the two amplitudes $\delta \mu_0$, $\delta T_0$. Since we are interested in nontrivial solutions, we need to require the determinant of the 
coefficient matrix to vanish. After a bit of algebra, using $n=n_s+n_n$ (which is true for $\nabla\psi=0$) and $\frac{\partial n}{\partial T} 
= \frac{\partial s}{\partial\mu}$, the resulting equation can be written as
\bea \label{soundv0}
0&=&\mu wT\left(\frac{\partial s}{\partial\mu}\frac{\partial n}{\partial T}-\frac{\partial n}{\partial\mu}\frac{\partial s}{\partial T}\right) 
\omega^4 -n_s s^2 T {\bf k}^4
+\left[s^2\mu  \frac{\partial n}{\partial\mu} + (\mu n_n^2 +wn_s)\frac{\partial s}{\partial T} -2\mu s n_n\frac{\partial s}{\partial\mu}\right] 
T \omega^2 {\bf k}^2 \, .
\eea
This result is in exact agreement with the one given in Eqs.~(19) -- (22) of Ref.~\cite{Herzog:2008he}. Now there are two physical solutions,
$\omega = u_{1,2} |{\bf k}|$, with the two sound velocities $u_1$, $u_2$. The reason for the appearance of the second mode is that the presence of the second fluid 
component allows for {\it relative} oscillations between the two fluids. 
The solution of Eq.\ (\ref{soundv0}) is the ${\bf v}_s=0$ limit of the full result (\ref{u12full}).

In general, the full wave equations (\ref{hydonetwo}) yield very complicated results for the sound velocities. However, in our approximation for small temperatures 
up to order $T^6$ in the pressure, one can show that the resulting quartic equation for $\omega$ factorizes into two quadratic equations. This is explained in detail 
in appendix \ref{AppC2}. In this appendix we also explain that our truncation of the low-temperature series does not allow us to 
compute temperature corrections  to the sound velocities of order $T^4$ and higher. The $T^2$ corrections, however, {\it can} be reliably determined. As one can 
see from Eqs.\ (\ref{factorize}), this is
possible because of the $T^6$ terms in the pressure which originate from the $|{\bf k}|^3$ term in the dispersion of the Goldstone mode. 
%In other words, had we 
%restricted ourselves to a linear dispersion $\propto |{\bf k}|$, we would not have been able to calculate any temperature corrections to the sound velocities. 
It turns out that there is a $T^2$ correction only to the second sound $u_2$.
%, i.e., the temperature corrections to the velocity of first sound $u_1$ are at least of order $T^4$. 
The explicit results are 
\begin{subequations}\label{u12full}
\bea \label{u1full}
u_1 &=& \frac{\sqrt{3-{\bf v}_s^2(1+2\cos^2\theta)}\sqrt{1-{\bf v}_s^2}+2|{\bf v}_s|\cos\theta}{3-{\bf v}_s^2} +{\cal O}(T^4) \\[2ex]
u_2 &=& \frac{\sqrt{9(1-{\bf v}_s^2)(1-3{\bf v}_s^2)+{\bf v}_s^2\cos^2\theta}+|{\bf v}_s|\cos\theta}{9(1-{\bf v}_s^2)} \non[2ex]
&&+\frac{4}{63}\left(\frac{\pi T}{\mu}\right)^2
\left[\frac{9(5-4{\bf v}_s^2-46{\bf v}_s^4+36{\bf v}_s^6+9{\bf v}_s^8)-4(5-2{\bf v}_s^2-15{\bf v}_s^4){\bf v}_s^2\cos^2\theta}
{(1-{\bf v}_s^2)(1-3{\bf v}_s^2)^3\sqrt{9(1-{\bf v}_s^2)(1-3{\bf v}_s^2)+{\bf v}_s^2\cos^2\theta}}
-\frac{4(5-2{\bf v}_s^2-15{\bf v}_s^4)|{\bf v}_s|\cos\theta}{(1-{\bf v}_s^2)(1-3{\bf v}_s^2)^3}\right]\non[2ex]
&& +\,{\cal O}(T^4) \, , \label{u2full}
\eea
\end{subequations}
where $\theta$ is the angle between ${\bf v}_s$ and the direction of the sound wave given by the wave vector ${\bf k}$. As a consistency check, we confirm that 
$u_1$ is the ($m=0$ limit of the) linear part of the dispersion of the Goldstone mode from Eq.~(\ref{eps12}), which was computed as one of the poles of the propagator. 
In our approximation, this dispersion does not depend on temperature. We know that in general
the dispersion of the Goldstone mode does become temperature dependent, see for instance Ref.~\cite{Alford:2007qa}, where the melting of the condensate has been taken 
into account, however without any superflow. 

%As for the quantities in Table \ref{table0}, we observe that the results only make sense for superfluid velocities 
%$|{\bf v}_s|<1/\sqrt{3}$, above which 
%we expect superfluidity to break down. Most notably, 
The velocity of second sound $u_2$ becomes complex for certain angles $\theta$ as soon as $|{\bf v}_s|>1/\sqrt{3}$. 
Moreover, the $T^2$ term of $u_2$ is divergent as $|{\bf v}_s|$ approaches $1/\sqrt{3}$. We have seen in Table \ref{table0} that all 
components of the stress-energy tensor and the current exhibit this divergence too. The expressions in that table show that due to this divergence the $T^6$ term 
(say, in the energy density $T^{00}$) becomes comparable to or even larger than the $T^4$ term 
for superfluid velocities close to (and below) the critical velocity $1/\sqrt{3}$, even if $T$ is very small. This suggests that a calculation to all orders 
in $T$ must be performed to predict reliably the behavior in this close-to-critical regime. In analogy, for the speed of second sound close to the critical velocity and 
at nonzero temperatures we would also need a resummed result, and we cannot trust the truncated expression (\ref{u2full}). 

\begin{figure} [t]
\begin{center}
\hbox{\includegraphics[width=0.35\textwidth]{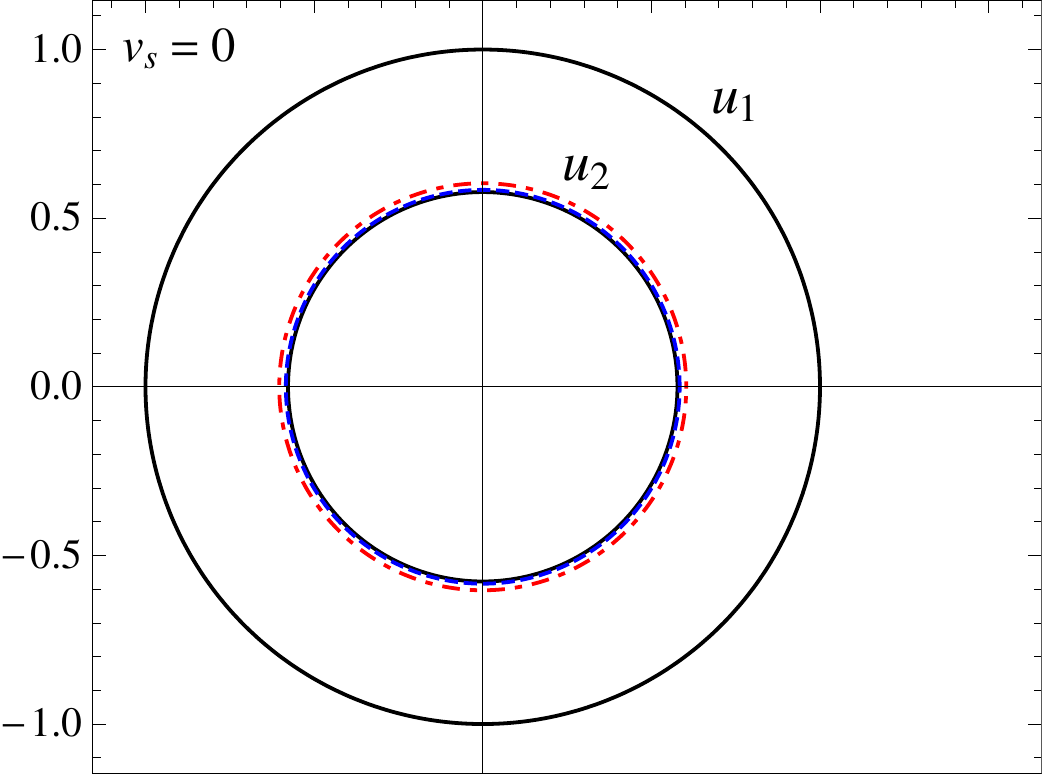}\includegraphics[width=0.319\textwidth]{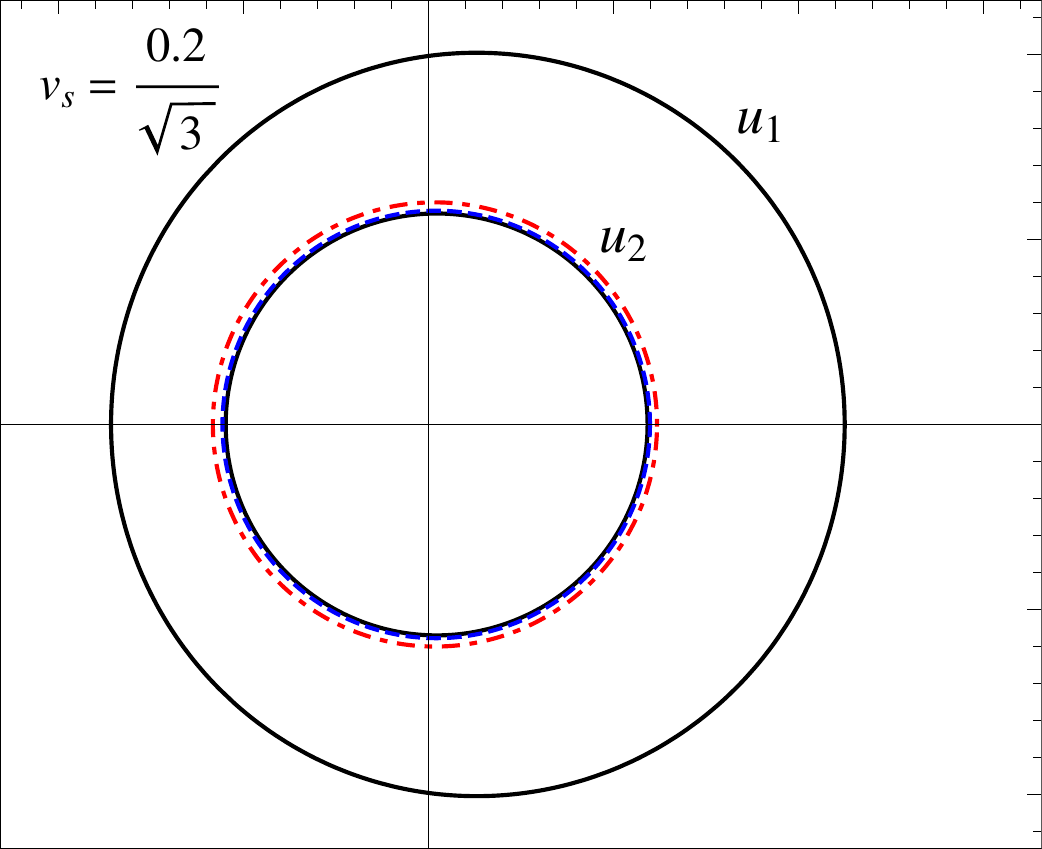}\includegraphics[width=0.319\textwidth]{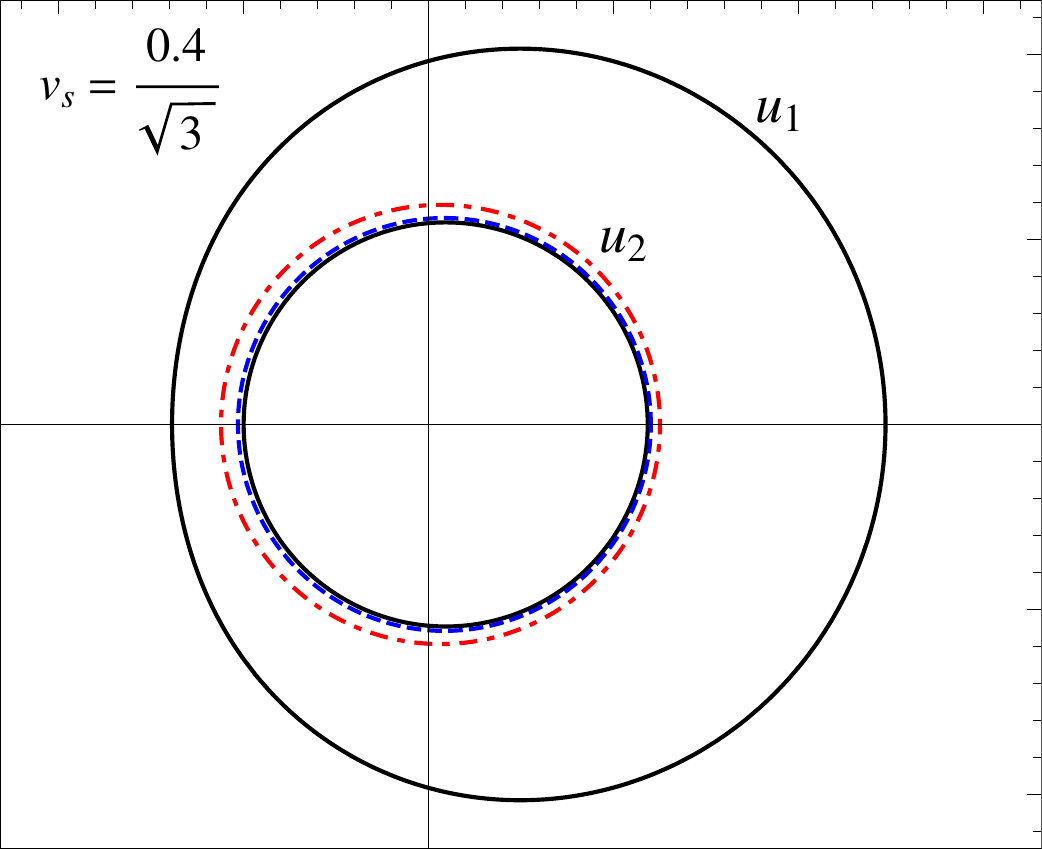}}
\vspace{-0.05cm}
\hbox{\includegraphics[width=0.35\textwidth]{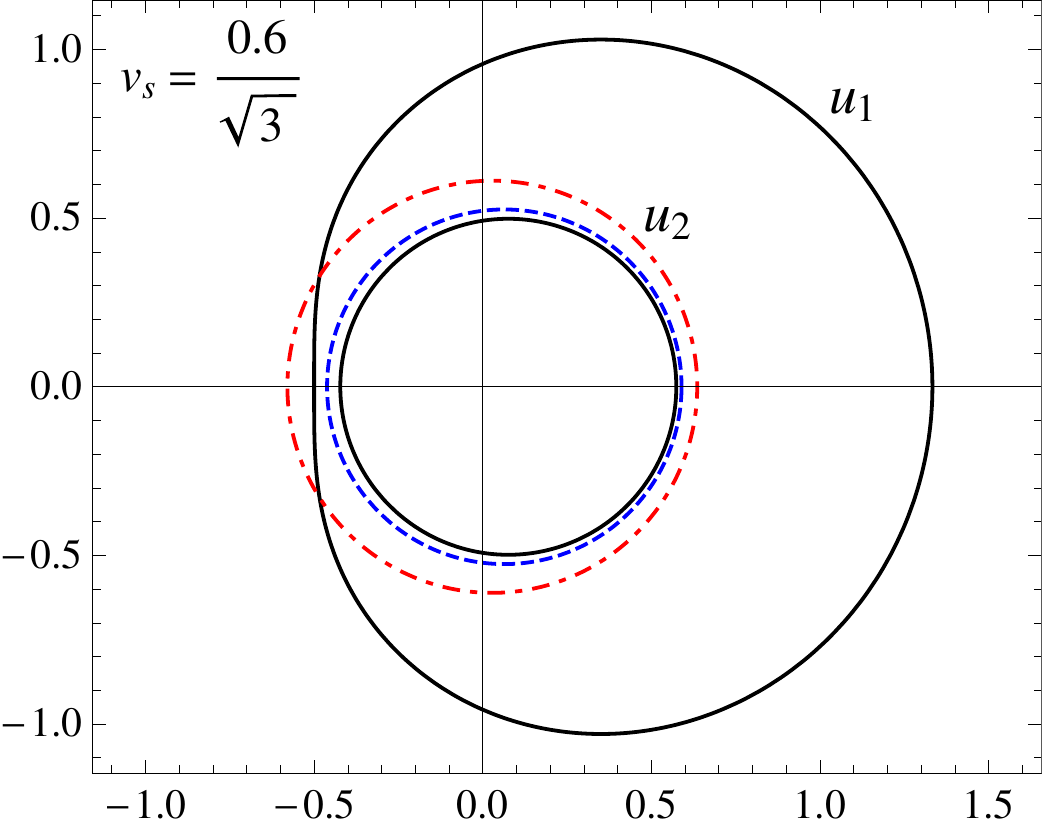}\includegraphics[width=0.319\textwidth]{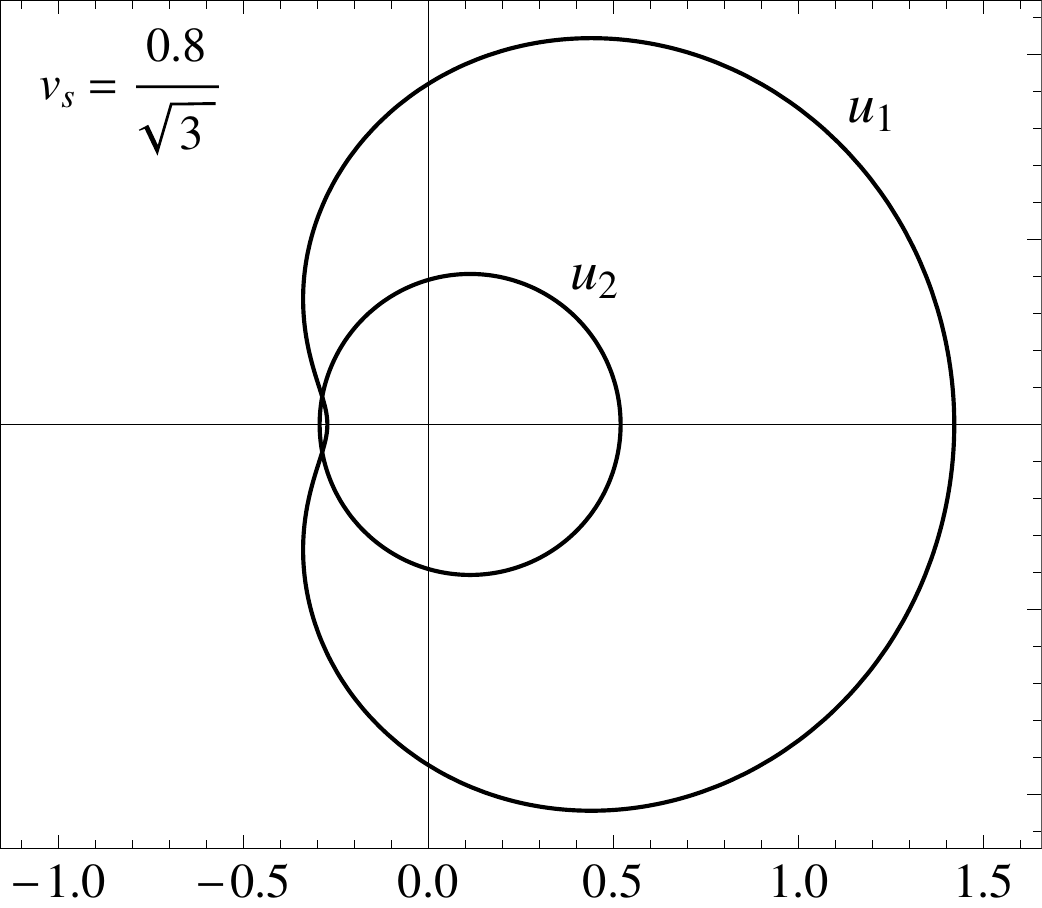}\includegraphics[width=0.319\textwidth]{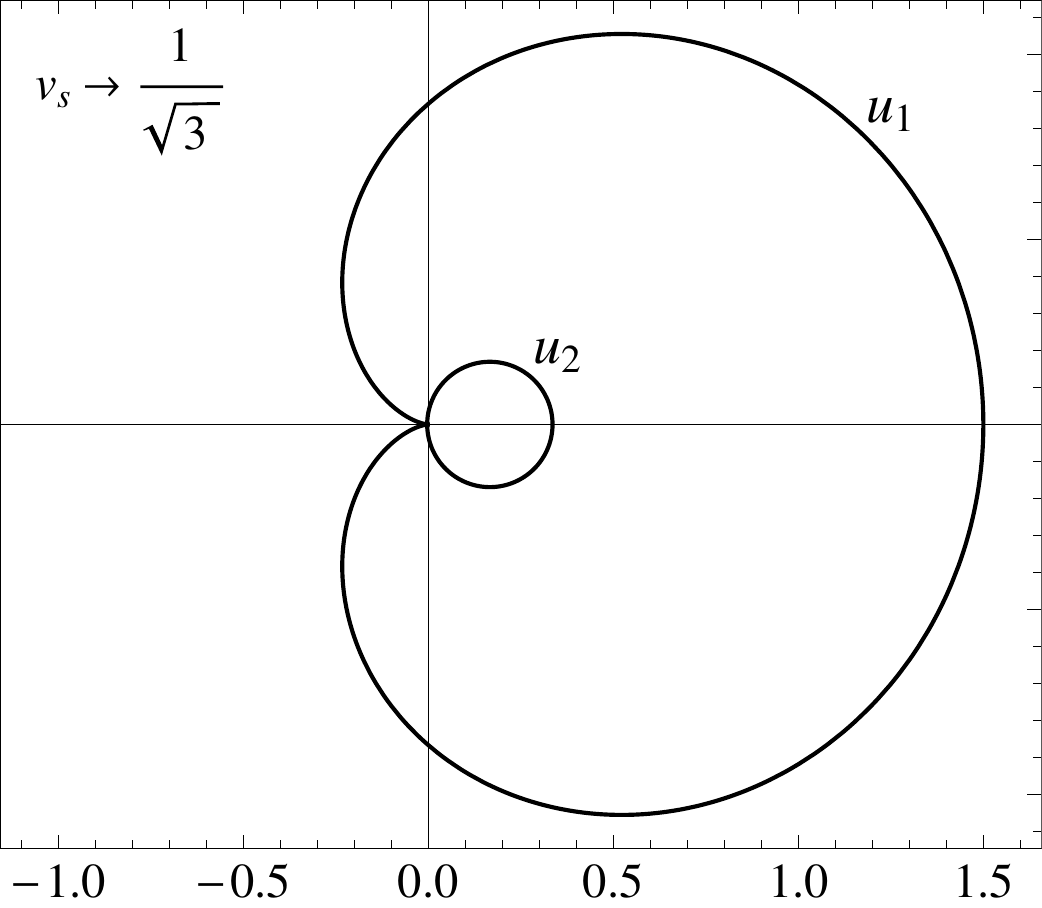}}
\caption{Velocities of first and second sound $u_1$, $u_2$ from Eqs.~(\ref{u12full}) for six different values of the superfluid velocity $|{\bf v}_s|$ between 
0 and $1/\sqrt{3}$. All velocities are measured in the normal-fluid rest frame. In these polar plots,
 the sound velocities for a given angle between the direction of the wave vector and the superflow are given by the radial distance of the curve to the origin; 
the direction of the 
superflow is parallel to the horizontal axis and points to the right; the scale is normalized to the velocity of first sound in the absence of a superflow, as one can 
see in the upper left panel. The speed of first sound does not depend on temperature within our approximation.  The speed of second sound is shown for three different 
temperatures: $T=0$ [(black) solid], $T/\mu=0.02$ [(blue) dashed], $T/\mu=0.04$ 
[(red) dashed-dotted]. For large superfluid velocities the temperature expansion breaks down, and we have only shown the results for $T=0$. } 
\label{figsound}
\end{center}
\end{figure}

We plot the two sound velocities for all angles and for various superfluid velocities in Fig.~\ref{figsound}. Because of the breakdown of the temperature expansion 
we have just explained, the results for nonzero temperature are only shown up to a superfluid velocity where the $T^2$ correction is still smaller than the $T=0$ term. 
We see that both sound velocities are increased when they propagate parallel to the superflow and decreased when 
they propagate in the opposite direction. At $T=0$, where the result can be taken seriously for all $|{\bf v}_s|<1/\sqrt{3}$, the speed of second sound 
decreases significantly when the critical velocity is approached, and goes to zero for all ``backward'' angles $\pi/2<\theta<3\pi/2$
(while the velocity of first sound only goes to zero for propagation exactly antiparallel to the superflow, $\theta=\pi$). 
Interestingly, for a given superfluid velocity, the temperature effect always {\it increases} the speed of 
second sound for all angles. We know that for larger temperatures it must decrease again, because it has to vanish at the critical temperature where there is only 
one fluid in the system. Within our low-temperature approximation we cannot see this decrease.

\section{Summary and outlook}

We have discussed the dissipationless hydrodynamics of a relativistic superfluid, starting from a complex scalar field. Our main goal
has been to relate the field theory with the covariant two-fluid framework of superfluidity, motivated by quark 
superfluidity in the CFL phase and its astrophysical relevance. Our results can be summarized as follows.
\begin{itemize}

\item {\it Microscopic calculation.} We have started from Bose-Einstein condensation in a $\varphi^4$ theory. The condensate has been 
assumed to be static and homogeneous (which corresponds to the simple hydrodynamic scenario of a static, homogeneous superflow). 
A crucial role is played by the phase of the condensate. While small oscillations
of the phase correspond to the excitations of the Goldstone mode, rotations of the phase around the full $U(1)$ circle give rise to 
a chemical potential (speed of the rotation) and a superfluid three-velocity (number of rotations per unit length).
In the presence of a non-zero superflow, the excitations of the Goldstone mode become anisotropic, and we have computed the 
resulting components of the conserved current and stress-energy tensor for nonzero temperatures. We have restricted ourselves to a 
weak-coupling, small-temperature approximation, including terms up to sixth order in the temperature. This has allowed us to present the microscopic results in 
an analytical form. 
For a study of arbitrary temperatures up to the critical temperature, a more elaborate, self-consistent calculation must be performed
numerically.

\item {\it Two-fluid formalism.} In the relativistic two-fluid formalism, the basic variables are the charge current
and the entropy current. In the dissipationless case, both are conserved. For both currents, we can define conjugate momenta. Now, if the 
two fluids are interacting with each other, neither of the currents is (four-)parallel to its own momentum, but also receives a contribution
from the other momentum. This contribution is called entrainment, and it must be computed from an underlying microscopic theory. We have 
explained this formalism in detail, and in particular have shown the connection to other, equivalent, formulations in the literature.
For instance, the entrainment coefficient (and related coefficients) can be expressed in terms of the superfluid and normal-fluid charge densities
$n_n$ and $n_s$
(whose corresponding superfluid and normal-fluid four-currents are, even in the dissipationless case, not separately conserved). 

\item {\it Relationship between them.} There are several concepts and quantities in the two-fluid formalism that are usually not used in field theory,
such as the generalized pressure that depends on Lorentz scalars. Therefore, one important aspect of this work has been a translation of these
quantities into field-theoretic language. For instance, we have proven that, once we assume that the microscopic calculation is performed in the 
rest frame of the normal fluid, it follows that the generalized pressure is given by the effective action. We have also expressed
the coefficients that relate the currents with the conjugate momenta in terms of quantities that are well defined in field theory, 
namely the space-time derivative of the phase of the condensate, the components of the Noether current, and the entropy. As a result, we have 
been able to compute these coefficients explicitly as a function of temperature, chemical potential, and superfluid velocity. Certain combinations of these 
coefficients yield $n_n$ and $n_s$. Our calculation shows for instance that $n_s$ is not simply given by the condensate density: even though we 
have neglected the temperature dependence of the condensate, $n_s$ depends on temperature. It also confirms that $n_n$ is not identical to the phonon number, 
as one might naively expect; while the phonon number goes like $T^3$ for small temperatures (see for instance Ref.\ \cite{Mannarelli:2008jq}), $n_n \propto T^4/\mu$.  

\end{itemize}

As an application, we have computed the angular and temperature dependent velocities of first and second sound in the presence of a superflow. 
To obtain non-zero-temperature corrections it has turned out to be crucial to go beyond the linear approximation of the dispersion of the Goldstone mode.
Cubic corrections in the dispersion give rise to $T^6$ corrections in the pressure and $T^2$ corrections in the velocity of second sound, while the velocity of first sound
remains temperature-independent. We have found that the velocity of second sound increases with small temperatures 
although eventually, beyond our small-temperature approximation, it must go to zero when the temperature approaches the critical point.

Our study opens up several directions for future studies. The most direct extension of our work is to go to higher temperatures, for instance
in the CJT formalism, as we have briefly laid out at the beginning of Sec.~\ref{sec:effact}. Such a calculation would capture the temperature dependence of the 
condensate; it would also go beyond the weak-coupling regime because it necessarily resums higher-order diagrams in order to compute the condensate
self-consistently. Although such an extension requires a numerical approach, we expect that the identifications with the hydrodynamic quantities such as normal-fluid and 
superfluid densities should be straightforward with the help of the present work. 

It would also be interesting to relax our assumptions of a uniform superflow and of vanishing dissipation, although this appears less straightforward. Non-uniform
situations would require us to build a hydrodynamic limit into our calculation, i.e., we would have to approximate the system as locally uniform. Dissipation 
clearly goes beyond our equilibrium treatment of the field theory, and it would be very interesting to see if a similar translation to the two-fluid 
picture also works in that more general case.  

In a more phenomenological context, our
work can be considered as a first step towards a more complete picture of the hydrodynamics of color-flavor locked quark matter. 
There are several steps in this direction for the future. It would be interesting to include a (small) term in the Lagrangian that 
breaks the global $U(1)$ symmetry explicitly and see in which sense superfluidity and the two-fluid picture survive. In a sense, 
this is similar to including dissipative effects because dissipation leads to a non-conservation of the entropy current, while 
explicit symmetry breaking leads to non-conservation of the charge current. It is also not unlike anomalous hydrodynamics, where the axial
anomaly leads to a non-conserved current and which has been discussed in the recent literature \cite{Son:2009tf,Landsteiner:2011cp,Jensen:2012jy}, 
also in the context of superfluidity \cite{Lin:2011mr,Bhattacharya:2011tra,Neiman:2011mj}. 
The case of a broken symmetry, while also of theoretical interest, is relevant for the kaon-condensed CFL phase, a viable candidate phase
of dense quark matter inside a compact star. In this phase the kaon condensate does break a symmetry that is, due to the weak interactions, 
not exact. It would also be interesting to include both superfluid components of the CFL-$K^0$ phase, the one coming from kaon 
condensation and the one coming from the spontaneous breaking of the $U(1)$ associated with baryon number symmetry. One could start from a fermionic formalism here, 
since baryon number is spontaneously broken by the quark Cooper pair condensate.

\begin{acknowledgments}
We thank N.~Andersson, P.~Bedaque, T.~Cohen, M.~Mannarelli, C.~Manuel, and R.~Sharma for valuable comments and discussions. This work has been supported by the Austrian 
science foundation FWF under project no.~P23536-N16, and by U.S.~Department of Energy under contract
% \#DE-FG02-91ER40628,  % Wash U high energy theory
\#DE-FG02-05ER41375, % Mark's nuclear theory
and by the DoE Topical Collaboration 
``Neutrinos and Nucleosynthesis in Hot and Dense Matter'', 
contract \#DE-SC0004955.

\end{acknowledgments}

\appendix

\section{Matsubara sum with anisotropic excitation energies}
\label{App0}

Here we derive the result (\ref{Gammaeps}) for the effective action. The calculation shown here is formulated in a general way, such that 
it is also applicable to the 
stress-energy tensor and the current in Sec.~\ref{sec:Tmunu}. In order to perform the Matsubara sum in Eq.~(\ref{partial}), 
we write the determinant of the inverse propagator in terms of its zeros,
\be
{\rm det}\,S^{-1}(k) = (k_0-\epsilon_{1,{\bf k}}^+)(k_0-\epsilon_{1,{\bf k}}^-)(k_0-\epsilon_{2,{\bf k}}^+)(k_0-\epsilon_{2,{\bf k}}^-) \, .
\ee
In the presence of a superflow $\nabla\psi$, the zeros are very complicated. The reason is the linear term in $k_0$ in the 
off-diagonal elements. With the help of Mathematica we obtain analytical, but very lengthy expressions for $\epsilon_{i,{\bf k}}^\pm$.
For small momenta, we can write 
\begin{subequations} \label{smallkA}
\bea
\epsilon_{1,{\bf k}}^\pm &=& \pm\sqrt{\frac{\sigma^2-m^2}{3\sigma^2-m^2}}\,\zeta_\pm(\uk)|{\bf k}| + {\cal O}(|{\bf k}|^3) \, , 
\label{eps12A}\\[2ex]
\epsilon_{2,{\bf k}}^\pm &=& \pm\sqrt{2}\sqrt{3\sigma^2-m^2+2(\nabla\psi)^2} + {\cal O}(|{\bf k}|) \, , 
\eea
\end{subequations}
where 
\be \label{zetaA}
\zeta^\pm(\uk)\equiv \left[\sqrt{1+2\frac{(\nabla\psi)^2-(\nabla\psi\cdot\uk)^2}{3\sigma^2-m^2}}\mp\frac{2\partial_0\psi\nabla\psi\cdot\uk}{\sqrt{\sigma^2
-m^2}\sqrt{3\sigma^2-m^2}}\right]\left[1+\frac{2(\nabla\psi)^2}{3\sigma^2-m^2}\right]^{-1} \, .
\ee
We now use the Matsubara sum 
\bea \label{matsu}
T\sum_{k_0}\frac{F(k_0,{\bf k})}{{\rm det}\,S^{-1}(k)} &=& -\frac{1}{2}\sum_{e=\pm}\sum_{i=1,2} 
\frac{F(\epsilon_{i,{\bf k}}^e,{\bf k})}
{(\epsilon_{i,{\bf k}}^e-\epsilon_{i,{\bf k}}^{-e})(\epsilon_{i,{\bf k}}^e-\epsilon_{j,{\bf k}}^e)(\epsilon_{i,{\bf k}}^e-\epsilon_{j,{\bf k}}^{-e})}
\coth\frac{\epsilon_{i,{\bf k}}^e}{2T}
\eea
with $j=2$ if $i=1$ and vice versa, and an arbitrary function $F(k_0,{\bf k})$ (without poles in the complex $k_0$ plane). For the effective action,
$F(k_0,{\bf k})$ is given by Eq.~(\ref{Fk0}), for the stress-energy tensor and the current see Eqs.\ (\ref{FF}). Since we are interested in small temperatures, we may neglect the
contribution from the massive mode, i.e., the two of the four terms in the sum where $i=2$. More precisely: later, after writing 
$\coth[\epsilon_{i,{\bf k}}^e/(2T)]=1+2f(\epsilon_{i,{\bf k}}^e)$ with the Bose distribution $f$, we shall only keep the thermal contribution,
which, in the case of the massive mode, is suppressed for small temperatures. For the non-thermal (divergent) contribution, all terms 
have to be kept in principle. However, after renormalization, the contribution is subleading since it contains an additional
factor of the coupling constant $\lambda$ and we shall neglect it. Therefore, after taking the thermodynamic limit, we can write
\bea \label{kintegralA}
\frac{T}{V}\sum_k \frac{F(k_0,{\bf k})}{{\rm det}\,S^{-1}(k)} &\simeq& -\frac{1}{2}\int\frac{d^3{\bf k}}{(2\pi)^3} 
\frac{F(\epsilon_{1,{\bf k}}^+,{\bf k})}{(\epsilon_{1,{\bf k}}^+-
\epsilon_{1,{\bf k}}^-)(\epsilon_{1,{\bf k}}^+-\epsilon_{2,{\bf k}}^+)(\epsilon_{1,{\bf k}}^+-\epsilon_{2,{\bf k}}^-)}
\coth\frac{\epsilon_{1,{\bf k}}^+}{2T}
\non[2ex] 
&&-\,\frac{1}{2}\int\frac{d^3{\bf k}}{(2\pi)^3} \frac{F(\epsilon_{1,{\bf k}}^-,{\bf k})}{(\epsilon_{1,{\bf k}}^--
\epsilon_{1,{\bf k}}^+)(\epsilon_{1,{\bf k}}^--\epsilon_{2,{\bf k}}^+)(\epsilon_{1,{\bf k}}^--\epsilon_{2,{\bf k}}^-)}
\coth\frac{\epsilon_{1,{\bf k}}^-}{2T}
\non[2ex]
&=& - \int\frac{d^3{\bf k}}{(2\pi)^3} \frac{F(\epsilon_{1,{\bf k}}^+,{\bf k})}
{(\epsilon_{1,{\bf k}}^++\epsilon_{1,-{\bf k}}^+)(\epsilon_{1,{\bf k}}^++\epsilon_{2,-{\bf k}}^+)(\epsilon_{1,{\bf k}}^+-\epsilon_{2,{\bf k}}^+)}\,
\coth\frac{\epsilon_{1,{\bf k}}^+}{2T} \, , 
\eea
where, in the last step, we have changed the integration variable of the second integral ${\bf k}\to -{\bf k}$, and have used 
$F(k_0,{\bf k})=F(-k_0,-{\bf k})$ as well as 
\be \label{twin}
\epsilon_{i,-{\bf k}}^+=-\epsilon_{i,{\bf k}}^- \, .
\ee 
This relation is easily checked for the small-momentum expressions (\ref{smallkA}), and also holds for the full results. 
Due to the symmetries with respect to reflection of ${\bf k}$, the poles $\epsilon_{i,{\bf k}}^-$ have thus dropped out of the result, 
and the only physical excitations are $\epsilon_{i,{\bf k}}^+$. Therefore, in the main text, we have simply denoted 
$\epsilon_{i,{\bf k}}\equiv\epsilon_{i,{\bf k}}^+$ and $\zeta(\uk)\equiv \zeta^+(\uk)$.

\section{Renormalization and useful identities for stress-energy tensor}
\label{AppA}

With the function $\Psi_k$ from Eq.~(\ref{PsiK}) we can write the effective action (\ref{effact}) as
\be \label{Geff1}
\frac{T}{V}\Gamma = -U  + \frac{T}{V} \sum_k \Psi_k \, .
\ee 
On the other hand, using Eq.~(\ref{partial}), we have 
\bea \label{Geff2}
\frac{T}{V}\Gamma &=& -U  + \frac{1}{3}\frac{T}{V} \sum_k \left(C_k {\bf k}^2-A_k{\bf k}\cdot\nabla\psi\right) \non[2ex]
&=& -U - \frac{1}{3}(g^{\mu\nu}-u^\mu u^\nu)\frac{T}{V}\sum_k\left(C_k k_\mu k_\nu+A_k k_\mu \partial_\nu\psi\right) \, ,
\eea
with the four-vector $u^\mu=(1,0,0,0)$ and $A_k$, $B_k$, $C_k$ given in Eq.~(\ref{dPsiK}). A useful relation between $A_k$, $B_k$, $C_k$
can be derived with the help of the explicit form of the determinant of the inverse tree-level propagator,
\bea \label{one}
1 &=& \frac{k^4-2k^2(\sigma^2-m^2)-4(k\cdot\partial\psi)^2}{{\rm det}\,S^{-1}} \non[2ex]
&=& -\frac{1}{2}[C_k k^2+B_k \sigma^2 + 2A_k (k\cdot\partial\psi)] +\frac{k^2m^2}{{\rm det}\,S^{-1}} \, .
\eea
Next, we rewrite the stress-energy tensor. With
\be
-\frac{T}{V}\sum_k\Tr\left[S\frac{\partial S^{-1}}{\partial g_{\mu\nu}}\right] = 2\frac{T}{V}\sum_k\frac{\partial\Psi_k}{\partial g_{\mu\nu}}
\ee
we can write the stress-energy tensor from Eq.~(\ref{dlnZ}) as 
\be \label{TmunuA}
T^{\mu\nu} = -\left(2\frac{\partial U}{\partial g_{\mu\nu}}-g^{\mu\nu}U\right)+ 
\frac{T}{V}\sum_k\left[C_kk^\mu k^\nu+B_k\partial^\mu\psi\partial^\nu\psi+A_k(k^\mu\partial^\nu\psi+k^\nu\partial^\nu\psi)
+g^{\mu\nu}+Y^{\mu\nu}\right] \, , 
\ee
where we have used the definition of $\Psi_k$ (\ref{PsiK}) and have added a constant, diagonal tensor $Y^{\mu\nu}$ which has to be determined 
such that the conditions (\ref{condition}) are fulfilled. 
In order to implement these conditions we now set $\nabla\psi=0$ and $\partial_0\psi=\mu$.
In this case, because of the first line of Eq.~(\ref{Geff2}), the pressure $P=\frac{T}{V}\Gamma$ becomes
\be \label{PCk}
P = -U  + \frac{1}{3}\frac{T}{V} \sum_k C_k{\bf k}^2 \, .
\ee 
In order to compute the energy density $\epsilon=-P+\mu n +T s$ we need
\begin{subequations}
\bea
n &=& \frac{\partial P}{\partial\mu} = -\frac{\partial U}{\partial\mu} +\frac{T}{V}\sum_k(B_k\mu+A_kk_0) \, , \\[2ex]
s &=&\frac{\partial P}{\partial T} =-\frac{\partial U}{\partial T} + \frac{P}{T}+\frac{T}{V}\sum_k\left(2+A_k\frac{\mu k_0}{T}
+C_k\frac{k_0^2}{T}\right) \, ,
\eea
\end{subequations}
where we have used the form of the pressure (\ref{Geff1}) and $\partial k_0/\partial T = k_0/T$ (due to the linear temperature-dependence of the 
Matsubara frequencies). Consequently,
\be \label{epsABC}
\epsilon = \frac{T}{V}\sum_k\left(B_k\mu^2+2A_kk_0\mu+C_kk_0^2+2\right) \, .
\ee
On the other hand, the nonzero components of the stress-energy tensor without superflow are, from Eq.~(\ref{TmunuA}),
\be \label{Tijiso}
T^{ij} = \frac{(\mu^2-m^2)^2}{4\lambda}\delta^{ij}
+\frac{T}{V}\sum_k\left(C_k\frac{{\bf k}^2}{3}\delta^{ij} - \delta^{ij} +Y^{ij}\right) \, , 
\ee
and 
\be \label{T00iso}
T^{00} = \frac{(3\mu^2+m^2)(\mu^2-m^2)}{4\lambda}+ \frac{T}{V}\sum_k\left(B_k\mu^2+2A_kk_0\mu+C_kk_0^2+1+Y^{00}\right)  \, ,
\ee
By comparing Eq.~(\ref{Tijiso}) with (\ref{PCk}) and (\ref{T00iso}) with (\ref{epsABC}) we conclude that $Y^{\mu\nu}={\rm diag}(1,1,1,1)$.
Inserting this into Eq.~(\ref{TmunuA}), we can write the renormalized stress-energy tensor as
\be 
T^{\mu\nu} = -\left(2\frac{\partial U}{\partial g_{\mu\nu}}-g^{\mu\nu}U\right)+ 
\frac{T}{V}\sum_k\left[C_kk^\mu k^\nu+B_k\partial^\mu\psi\partial^\nu\psi+A_k(k^\mu\partial^\nu\psi+k^\nu\partial^\mu\psi)
+2u^\mu u^\nu\right] \, , 
\ee
with $u^\mu=(1,0,0,0)$.

\section{Small-temperature expansion}
\label{AppB}

Here we explain the small-temperature expansion needed in Sec.~\ref{sec:smallT} for the effective action, the stress-energy 
tensor, and the current density. We focus on the effective action in this appendix, but the other results are obtained analogously.

Expanding in powers of the temperature corresponds to expanding the integrand in powers of $|{\bf k}|$. In order to obtain the result 
up to $T^6$, we expand the integrand of the momentum integral in Eq.~(\ref{Gammaeps}) as 
\be
\frac{F(\epsilon_{1,{\bf k}},{\bf k})}
{(\epsilon_{1,{\bf k}}+\epsilon_{1,-{\bf k}})(\epsilon_{1,{\bf k}}+\epsilon_{2,-{\bf k}})(\epsilon_{1,{\bf k}}-\epsilon_{2,{\bf k}})}
\simeq a_1|{\bf k}|+\frac{a_2}{\sigma^2}|{\bf k}|^3 \, ,
\ee
and, for the dispersion in the argument of the Bose distribution,
\be
\epsilon_{1,{\bf k}}\simeq c_1|{\bf k}|+\frac{c_2}{\sigma^2}|{\bf k}|^3 \, ,
\ee
where $a_1$, $a_2$, $c_1$, $c_2$ are angular-dependent, dimensionless coefficients. Inserting these expansions, introducing a dimensionless 
integration variable $y=c_1|{\bf k}|/T$, expanding in $T/\sigma$, and performing the resulting integration over $y$ yields
\be 
\frac{T}{V}\Gamma\simeq \frac{(\sigma^2-m^2)^2}{4\lambda} +\frac{2\pi^2T^4}{45}\int\frac{d\Omega}{4\pi}\left[\frac{a_1}{c_1^4}
+\frac{40\pi^2}{7c_1^6}\left(\frac{a_2}{3}-\frac{2a_1c_2}{c_1}\right)\frac{T^2}{\sigma^2}\right] \, , 
\ee
where we have used the integrals
\be
\int_0^\infty dy\frac{y^3}{e^y-1} = \frac{\pi^4}{15} \, , \qquad \int_0^\infty dy\frac{y^5}{e^y-1} = \frac{8\pi^6}{63} \,,  \qquad 
\int_0^\infty dy\frac{y^6 e^y}{(e^y-1)^2} = \frac{16\pi^6}{21} \, .
\ee
For the case without superflow, $\nabla\psi=0$, the angular integral becomes trivial. In this case, with $\partial_0\psi=\mu$, the 
full dispersions are given by Eq.~(\ref{eps0}), and we have 
\be
c_1 = \sqrt{\frac{\mu^2-m^2}{3\mu^2-m^2}} \, ,\qquad c_2 = \frac{\mu^6}{\sqrt{\mu^2-m^2}(3\mu^2-m^2)^{5/2}} \, , 
\ee
and
\be
a_1=\frac{c_1}{4} \, , \qquad a_2 = \frac{3c_1}{4} \, .
\ee
We thus find for the pressure
\bea 
P=\frac{T}{V}\Gamma &\simeq & \frac{(\mu^2-m^2)^2}{4\lambda} + \frac{\pi^2T^4}{90c_1^3}-\frac{4c_2\pi^4T^6}{63\mu^2c_1^6} \non[2ex] 
&=& \frac{(\mu^2-m^2)^2}{4\lambda} + \frac{(3\mu^2-m^2)^{3/2}}{(\mu^2-m^2)^{3/2}}\frac{\pi^2 T^4}{90}-
\frac{\mu^6(3\mu^2-m^2)^{1/2}}{(\mu^2-m^2)^{7/2}}\frac{4\pi^4T^6}{63\mu^2} \, .
\eea
The expressions for the case with superflow are quite lengthy in general, and we give the final results in the limit $m=0$ in the 
main text, see Table \ref{table0}.

\section{Calculation of sound velocities}

\subsection{Derivation of sound wave equations}
\label{AppC1}

We start from the hydrodynamic equations
\begin{subequations}
\bea
0&=& \partial_\mu j^\mu  \, , \label{hyd1} \\[2ex]
0&=& \partial_\mu s^\mu  \, , \label{hyd2}\\[2ex]
0&=& s_\mu(\partial^\mu\Theta^\nu-\partial^\nu\Theta^\mu)  \, .\label{hyd3}
\eea
\end{subequations}
Before we evaluate them, we collect some useful relations. We denote $P\equiv P_n+P_s=\Psi$ and thus can write with Eq.~(\ref{dLambda})
\bea
dP &=& j^\mu d(\partial_\mu\psi) + s^\mu d\Theta_\mu \non[2ex]
&=& nd\mu +sdT -\frac{n_s}{\sigma}\nabla\psi\cdot d\nabla\psi
+\frac{n_n}{s} {\bf s}\cdot d\nabla\psi - {\bf s}\cdot d\left(\frac{n_n}{s}\nabla\psi\right)-{\bf s}\cdot d\left(\frac{w}{s^2}{\bf s}\right) \, ,
\eea
where $j^0=n$, $s^0=s$, $\partial^0\psi=\mu$, $\Theta^0=T$, and we have eliminated ${\bf j}$ and ${\bf \Theta}$ by using 
\be \label{jtheta3}
j^\mu = n_n u^\mu+n_s\frac{\partial^\mu\psi}{\sigma} \, , \qquad \Theta^\mu = -\frac{n_n}{s}\partial^\mu\psi + \frac{w}{s}u^\mu \, , 
\ee
where $w\equiv \epsilon_n+P_n=\mu n_n+sT$ is the enthalpy density of the normal fluid. In the linear approximation, ${\bf s}$ times a space-time derivative is negligible,
because ${\bf s}=s^0{\bf v}_n$ and we neglect products of ${\bf v}_n$ with space-time derivatives. Therefore, we may approximate  
\bea \label{diffP}
dP &\simeq &  nd\mu +sdT -\frac{n_s}{\sigma}\nabla\psi\cdot d\nabla\psi \, .
\eea
This relation is needed to express derivatives of any thermodynamic quantity in terms of derivatives of $T$, $\mu$, and $\nabla\psi$. For instance, we can write
$\partial_0 n = \partial_0\frac{\partial P}{\partial\mu} = \frac{\partial}{\partial\mu}\partial_0 P$ etc.\ and obtain the following useful identities,
\begin{subequations}
\bea
\partial_0 n &=& \frac{\partial n}{\partial\mu}\partial_0\mu + \frac{\partial s}{\partial\mu}\partial_0T - 
\frac{\partial (n_s/\sigma)}{\partial\mu}\nabla\psi\cdot\nabla\mu \, , \label{d0n} \\[2ex]
\partial_0 s &=& \frac{\partial n}{\partial T}\partial_0\mu + \frac{\partial s}{\partial T}\partial_0T 
- \frac{\partial (n_s/\sigma)}{\partial T}\nabla\psi\cdot\nabla\mu \, , \label{d0s}\\[2ex]
\partial_0\left(\frac{n_s}{\sigma}\partial_i\psi\right) &=& 
-\frac{\partial n}{\partial(\partial_i\psi)}\partial_0\mu-\frac{\partial s}{\partial(\partial_i\psi)}\partial_0T 
+\frac{\partial (n_s/\sigma)}{\partial(\partial_i\psi)}\nabla\psi\cdot\nabla\mu +\frac{n_s}{\sigma}\partial_i\mu \, , \label{d0psi} \\[2ex]
\nabla\cdot\left(\frac{n_s}{\sigma}\nabla\psi\right) &=& -\frac{\partial n}{\partial(\partial_i\psi)}\partial_i\mu-\frac{\partial s}{\partial(\partial_i\psi)}\partial_iT 
+\frac{\partial (n_s/\sigma)}{\partial(\partial_i\psi)}\partial_j\psi\partial_i\partial_j\psi +\frac{n_s}{\sigma}\Delta\psi \, , \label{divpsi} 
\eea
\end{subequations}
where $\partial_0\psi = \mu$ has been used.
With these preparations we can discuss the hydrodynamic equations. The current conservation (\ref{hyd1}) obviously becomes
\be \label{hyd11}
0\simeq \partial_0 n + n_n\nabla\cdot {\bf v}_n -\nabla\cdot\left(\frac{n_s}{\sigma}\nabla\psi\right) \, ,
\ee
where we have used ${\bf u}\simeq {\bf v}_n$ and 
$\nabla\cdot(n_n{\bf v}_n)\simeq n_n\nabla\cdot {\bf v}_n$. 
Inserting Eqs.~(\ref{d0n}) and (\ref{divpsi}) into this equation, taking the time derivative of the 
result, and multiplying the whole equation by $\mu$ yields
\bea \label{hyd1d0}
0&\simeq& \mu\frac{\partial n}{\partial\mu}\partial_0^2\mu+\mu\frac{\partial s}{\partial\mu}\partial_0^2T -n_s\frac{\mu}{\sigma}\Delta\mu
+\mu n_n\nabla\cdot\partial_0{\bf v}_n 
-\mu\frac{\partial (n_s/\sigma)}{\partial\mu}\nabla\psi\cdot\nabla\partial_0\mu \non[2ex]
&&+\mu\frac{\partial n}{\partial(\partial_i\psi)}\partial_0\partial_i \mu
+\mu\frac{\partial s}{\partial(\partial_i\psi)}\partial_0\partial_i T -\mu\frac{\partial (n_s/\sigma)}{\partial(\partial_i\psi)}\partial_j\psi\partial_i\partial_j\mu \, .
\eea
Due to the linear approximation, all expressions have the form (equilibrium quantity) $\times$ (second space-time derivative), since 
products of two first space-time derivatives are of higher order. 

The entropy conservation (\ref{hyd2}) reads
\be
0 \simeq \partial_0 s + s\nabla\cdot {\bf v}_n \, .
\ee
Inserting Eq.~(\ref{d0s}), taking the time derivative of the result and multiplying the whole equation by $T$ yields
\be\label{hyd2d0}
0\simeq T\frac{\partial n}{\partial T}\partial_0^2\mu+T\frac{\partial s}{\partial T}\partial_0^2T - T\frac{\partial (n_s/\sigma)}{\partial T}\nabla\psi\cdot
\nabla\partial_0\mu +sT\nabla\cdot\partial_0{\bf v}_n \, .
\ee
It is convenient for the following to add Eqs.~(\ref{hyd1d0}) and (\ref{hyd2d0}),
\bea \label{sum12}
0&\simeq& \left(\mu\frac{\partial n}{\partial\mu}+T\frac{\partial n}{\partial T}\right)\partial_0^2\mu+\left(\mu\frac{\partial s}{\partial\mu}
+T\frac{\partial s}{\partial T}\right)\partial_0^2T-\left[\mu\frac{\partial (n_s/\sigma)}{\partial\mu}+T\frac{\partial (n_s/\sigma)}{\partial T}\right]\nabla\psi\cdot
\nabla\partial_0\mu+w\nabla\cdot\partial_0{\bf v}_n \non[2ex]
&&+2\mu\frac{\partial n}{\partial(\nabla\psi)^2}\nabla\psi\cdot\nabla\partial_0 \mu
+2\mu\frac{\partial s}{\partial(\nabla\psi)^2}\nabla\psi\cdot\nabla\partial_0 T -2\mu\frac{\partial (n_s/\sigma)}{\partial(\nabla\psi)^2}(\nabla\psi\cdot\nabla)^2\mu 
-n_s\frac{\mu}{\sigma}\Delta\mu\, ,
\eea
where we have rewritten the derivative with respect to $\partial_i\psi$ in terms of the derivative with respect to $(\nabla\psi)^2$.

Finally, we need the vorticity equation (\ref{hyd3}). The temporal component becomes
\be
0 = {\bf s}\cdot(\nabla T+\partial_0{\bf \Theta}) \, .
\ee
We can neglect this equation completely, since in both terms the normal fluid velocity is multiplied with a space-time derivative. The spatial components are
\be \label{vortspatial}
0\simeq s(\partial_0{\bf \Theta}+\nabla T) \simeq s\nabla T+w\partial_0{\bf v}_n+s\partial_0\left(\frac{n_n}{s}\nabla\psi\right) \, ,
\ee
where ${\bf \Theta}$ from Eq.~(\ref{jtheta3}) has been used. The last term needs some rearrangements,
\bea
s\partial_0\left(\frac{n_n}{s}\nabla\psi\right) &=& -\frac{n_n}{s}\nabla\psi\,\partial_0s+\partial_0(n_n\nabla\psi) = -\frac{n_n}{s}\nabla\psi\,\partial_0s
+\partial_0(n\nabla\psi)-\partial_0\left(n_s\frac{\mu}{\sigma}\nabla\psi\right) \non[2ex]
&=& -\frac{n_n}{s}\nabla\psi\,\partial_0s +\nabla\psi\partial_0 n +n\nabla\mu -\frac{n_s}{\sigma}\nabla\psi\partial_0\mu -\mu\partial_0\left(\frac{n_s}{\sigma}
\nabla\psi\right) \, .
\eea
Inserting Eqs.~(\ref{d0n}), (\ref{d0s}), and (\ref{d0psi}) into this relation, the result into Eq.~(\ref{vortspatial}), and taking the divergence of the 
resulting equation, we arrive at
\bea \label{vortdi}
0&\simeq& n_n\Delta\mu+s\Delta T+w\nabla\cdot\partial_0{\bf v}_n -\frac{n_s}{\sigma}\nabla\psi\cdot\nabla\partial_0\mu 
-\left[\frac{n_n}{s}\frac{\partial n}{\partial T}-\frac{\partial n}{\partial \mu}-2\mu\frac{\partial n}{\partial (\nabla\psi)^2}\right]\nabla\psi\cdot\nabla\partial_0\mu
\non[2ex]
&&-\left[\frac{n_n}{s}\frac{\partial s}{\partial T}-\frac{\partial s}{\partial \mu}-2\mu\frac{\partial s}{\partial (\nabla\psi)^2}\right]\nabla\psi\cdot\nabla\partial_0 T
+\left[\frac{n_n}{s}\frac{\partial (n_s/\sigma)}{\partial T}-\frac{\partial (n_s/\sigma)}{\partial \mu}-2\mu\frac{\partial (n_s/\sigma)}
{\partial (\nabla\psi)^2}\right](\nabla\psi\cdot\nabla)^2\mu \, .
\eea
The normal fluid velocity can now be eliminated by solving this relation for $\nabla\cdot\partial_0{\bf v}_n$ and inserting the result into the other two equations: 
inserting it into Eq.~(\ref{hyd2d0}) yields, after multiplying the whole equation with $w/(Ts)$, Eq.~(\ref{hydone}) in the main text, while inserting it
into Eq.~(\ref{sum12}) yields Eq.~(\ref{hydtwo}). These are the sound wave equations from which the sound velocities are computed as follows.

\subsection{Solution of sound wave equations}
\label{AppC2}

Replacing the chemical potential and the temperature in all space-time derivatives of the sound wave equations 
(\ref{hydonetwo}) by $\delta \mu = \delta\mu_0e^{i(\omega t-{\bf k}\cdot{\bf x})}$ and $\delta T = \delta T_0e^{i(\omega t-{\bf k}\cdot{\bf x})}$, 
the sound wave equations become
\begin{subequations} \label{muTeqs}
\bea
0 &\simeq& \left[a_1\tilde{\omega}^2+(a_2+a_4\mu^2{\bf v}_s^2\cos^2\theta)+a_3\mu|{\bf v}_s|\tilde{\omega}\cos\theta  \right]\delta\mu_0 
+(b_1\tilde{\omega}^2+b_2+b_3\mu|{\bf v}_s|\tilde{\omega}\cos\theta )\,\delta T_0 \, , \\[2ex]
0 &\simeq& \left[A_1\tilde{\omega}^2+(A_2+A_4\mu^2{\bf v}_s^2\cos^2\theta)+A_3\mu|{\bf v}_s|\tilde{\omega}\cos\theta\right]\delta\mu_0 
+(B_1\tilde{\omega}^2+B_2+B_3\mu|{\bf v}_s|\tilde{\omega}\cos\theta  )\,\delta T_0 \, , 
\eea
\end{subequations}
where $\cos\theta\equiv \uk\cdot\hat{\bf v}_s$, $\tilde{\omega}\equiv \omega/|{\bf k}|$, and we have abbreviated
\begin{subequations}
\bea
a_1 &\equiv& \frac{w}{s}\frac{\partial n}{\partial T} \, , \qquad a_2 \equiv -n_n \,  , \qquad 
a_3\equiv \frac{n_s}{\sigma}-\frac{w}{s}\frac{\partial (n_s/\sigma)}{\partial T} +\frac{n_n}{s}\frac{\partial n}{\partial T}-\frac{\partial n}{\partial \mu}
-2\mu\frac{\partial n}{\partial (\nabla\psi)^2} \, , \non[2ex]
a_4&\equiv&-\left[\frac{n_n}{s}\frac{\partial (n_s/\sigma)}{\partial T}-\frac{\partial (n_s/\sigma)}{\partial \mu}-2\mu\frac{\partial (n_s/\sigma)}
{\partial (\nabla\psi)^2}\right] \, , \\[2ex]
b_1 &\equiv& \frac{w}{s}\frac{\partial s}{\partial T} \, , \qquad b_2\equiv-s \, , \qquad b_3 \equiv \frac{n_n}{s}\frac{\partial s}{\partial T}-\frac{\partial s}{\partial \mu}
-2\mu\frac{\partial s}{\partial (\nabla\psi)^2} \, , \\[2ex]
A_1 &\equiv& \mu\frac{\partial n}{\partial\mu}+T\frac{\partial n}{\partial T} \, , \qquad A_2 \equiv -n \, ,\qquad 
A_3\equiv \frac{n_s}{\sigma}-\mu\frac{\partial(n_s/\sigma)}{\partial\mu}-T\frac{\partial(n_s/\sigma)}{\partial T}+\frac{n_n}{s}\frac{\partial n}{\partial T}-
\frac{\partial n}{\partial \mu} \, , \non[2ex]
A_4&\equiv&-\left[\frac{n_n}{s}\frac{\partial (n_s/\sigma)}{\partial T}-\frac{\partial (n_s/\sigma)}{\partial \mu}\right] \\[2ex]
B_1 &\equiv& \mu\frac{\partial s}{\partial\mu}+T\frac{\partial s}{\partial T} \, , \qquad B_2\equiv-s \, , \qquad B_3 \equiv \frac{n_n}{s}\frac{\partial s}{\partial T}
-\frac{\partial s}{\partial \mu} \, .
\eea
\end{subequations}
In general, the determinant of the coefficient matrix of the system of two equations (\ref{muTeqs}) yields a complicated 
quartic equation for $\tilde{\omega}$. However, we can simplify the result as follows. First one can check, for instance by explicit calculation, that the temperature 
dependence of the various coefficients is
\be
a_i = a_i^{(4)} T^4+ a_i^{(6)}   T^6 \, ,\quad A_i = A_i^{(0)}  + A_i^{(4)} T^4+ A_i^{(6)}  T^6 \, , \quad b_j= b_j^{(3)} T^3+ b_j^{(5)}  T^5 
\, , \quad B_j= B_j^{(3)} T^3+ B_j^{(5)}  T^5\, ,
\ee
where $i=1,2,3,4$, $j=1,2,3$, and where the prefactors in front of the various powers of $T$ depend on the superfluid velocity ${\bf v}_s$. 
Since we have computed the pressure up to 
order $T^6$, all terms of order $T^7$ and higher must be neglected in these expressions. We see in particular that only the $A_i$'s contribute in the limit 
$T=0$. Now, for the determinant we encounter two kinds of products, namely  $A_i b_j$ and $a_i B_j$,
\begin{subequations}
\bea
A_ib_j &=& T^3[A_i^{(0)}+A_i^{(4)}T^4+A_i^{(6)}T^6][b_j^{(3)}+b_j^{(5)}T^2] = T^3 A_i^{(0)}[b_j^{(3)}+b_j^{(5)}T^2] + {\cal O}(T^7) \, , \\[2ex] 
a_iB_j &=& T^7[a_i^{(4)}+a_i^{(6)}T^2][B_j^{(3)}+B_j^{(5)}T^2] = {\cal O}(T^7) \, .
\eea
\end{subequations}
The first line shows that the ${\cal O}(T^7)$ terms are unknown in our expansion because a $b_j^{(7)}T^7$ term in $b_j$ would give rise to a $T^7$ term in $A_ib_j$,
but we have not computed $b_j^{(7)}$. Therefore, we must neglect all products of the form given in the second line since they are all of order $T^7$ and higher. 
In other words, it is consistent 
with our approximation to set $a_i\simeq B_i\simeq 0$ and use the $T=0$ results for $A_i$. 
In this case the quartic equation for $\omega$ factorizes into two quadratic equations,
\begin{subequations} \label{factorize}
\bea
0&\simeq& A_1^{(0)}\tilde{\omega}^2 +(A_2^{(0)}+A_4^{(0)}\mu^2{\bf v}_s^2\cos^2\theta)+A_3^{(0)}\mu|{\bf v}_s|\tilde{\omega}\cos\theta \, , \\[2ex]
0&\simeq& [b_1^{(3)}+b_1^{(5)}T^2]\tilde{\omega}^2 +[b_2^{(3)}+b_2^{(5)}T^2]+[b_3^{(3)}+b_3^{(5)}T^2]\mu|{\bf v}_s|\tilde{\omega}\cos\theta \, .
\eea
\end{subequations}
After dividing out the overall factor $T^3$ of the quartic equation, the highest remaining power of temperature is 2, i.e., our approximation allows us
to reliably compute the sound velocities up to $T^2$. From the first equation we see that one of the solutions has no $T^2$ correction. This is the 
velocity of first sound. The second equation yields the velocity of second sound which does have a $T^2$ correction. We see that the coefficients $b_i^{(5)}$ 
are needed to compute this 
correction. These coefficients arise from the $T^5$ terms in the entropy, i.e., from the $T^6$ terms in the pressure. Had we truncated our expansion of the pressure 
at order $T^4$, the velocity of second 
sound would have turned out to be independent of temperature. There are two physical solutions of Eqs.\ (\ref{factorize}), $\omega= u_{1,2} |{\bf k}|$. 
For the explicit calculation of the two sound velocities $u_1$, $u_2$ we need $n_s$ and $n_n$ from Eqs.\ (\ref{nsnn}) and the entropy $s$
(obtained by taking the derivative with respect to temperature of $\Psi=T_\perp$ from Table \ref{table0}). The result is given 
in Eqs.\ (\ref{u12full}) in the main part of the paper.

\bibliography{refs}

\begin{thebibliography}{60}
\expandafter\ifx\csname natexlab\endcsname\relax\def\natexlab#1{#1}\fi
\expandafter\ifx\csname bibnamefont\endcsname\relax
  \def\bibnamefont#1{#1}\fi
\expandafter\ifx\csname bibfnamefont\endcsname\relax
  \def\bibfnamefont#1{#1}\fi
\expandafter\ifx\csname citenamefont\endcsname\relax
  \def\citenamefont#1{#1}\fi
\expandafter\ifx\csname url\endcsname\relax
  \def\url#1{\texttt{#1}}\fi
\expandafter\ifx\csname urlprefix\endcsname\relax\def\urlprefix{URL }\fi
\providecommand{\bibinfo}[2]{#2}
\providecommand{\eprint}[2][]{\url{#2}}

\bibitem[{\citenamefont{Tisza}(1938)}]{tisza38}
\bibinfo{author}{\bibfnamefont{L.}~\bibnamefont{Tisza}},
  \bibinfo{journal}{Nature} \textbf{\bibinfo{volume}{141}},
  \bibinfo{pages}{913} (\bibinfo{year}{1938}).

\bibitem[{\citenamefont{{Kapitza}}(1938)}]{1938Natur.141...74K}
\bibinfo{author}{\bibfnamefont{P.}~\bibnamefont{{Kapitza}}},
  \bibinfo{journal}{Nature} \textbf{\bibinfo{volume}{141}}, \bibinfo{pages}{74}
  (\bibinfo{year}{1938}).

\bibitem[{\citenamefont{{Allen} and {Misener}}(1938)}]{1938Natur.141...75A}
\bibinfo{author}{\bibfnamefont{J.~F.} \bibnamefont{{Allen}}} \bibnamefont{and}
  \bibinfo{author}{\bibfnamefont{A.~D.} \bibnamefont{{Misener}}},
  \bibinfo{journal}{Nature} \textbf{\bibinfo{volume}{141}}, \bibinfo{pages}{75}
  (\bibinfo{year}{1938}).

\bibitem[{\citenamefont{Landau}(1941)}]{landau41}
\bibinfo{author}{\bibfnamefont{L.}~\bibnamefont{Landau}},
  \bibinfo{journal}{Phys. Rev.} \textbf{\bibinfo{volume}{60}},
  \bibinfo{pages}{356} (\bibinfo{year}{1941}).

\bibitem[{\citenamefont{{Khalatnikov} and
  {Lebedev}}(1982)}]{1982PhLA...91...70K}
\bibinfo{author}{\bibfnamefont{I.~M.} \bibnamefont{{Khalatnikov}}}
  \bibnamefont{and} \bibinfo{author}{\bibfnamefont{V.~V.}
  \bibnamefont{{Lebedev}}}, \bibinfo{journal}{Physics Letters A}
  \textbf{\bibinfo{volume}{91}}, \bibinfo{pages}{70} (\bibinfo{year}{1982}).

\bibitem[{\citenamefont{{Lebedev} and
  {Khalatnikov}}(1982)}]{1982ZhETF..83.1601L}
\bibinfo{author}{\bibfnamefont{V.~V.} \bibnamefont{{Lebedev}}}
  \bibnamefont{and} \bibinfo{author}{\bibfnamefont{I.~M.}
  \bibnamefont{{Khalatnikov}}}, \bibinfo{journal}{Zh.\ Eksp.\ Teor.\ Fiz.}
  \textbf{\bibinfo{volume}{83}}, \bibinfo{pages}{1601} (\bibinfo{year}{1982}),
  \bibinfo{note}{[Sov.\ Phys.\ JETP, {\bf 56}, 923 (1982)]}.

\bibitem[{\citenamefont{Carter}(1989)}]{carter89}
\bibinfo{author}{\bibfnamefont{B.}~\bibnamefont{Carter}}, in
  \emph{\bibinfo{booktitle}{Relativistic Fluid Dynamics (Noto 1987)}}, edited
  by \bibinfo{editor}{\bibfnamefont{A.}~\bibnamefont{Anile}} \bibnamefont{and}
  \bibinfo{editor}{\bibfnamefont{M.}~\bibnamefont{Choquet-Bruhat}}
  (\bibinfo{publisher}{Springer-Verlag}, \bibinfo{year}{1989}), pp.
  \bibinfo{pages}{1--64}.

\bibitem[{\citenamefont{{Carter} and
  {Khalatnikov}}(1992)}]{1992PhRvD..45.4536C}
\bibinfo{author}{\bibfnamefont{B.}~\bibnamefont{{Carter}}} \bibnamefont{and}
  \bibinfo{author}{\bibfnamefont{I.~M.} \bibnamefont{{Khalatnikov}}},
  \bibinfo{journal}{\prd} \textbf{\bibinfo{volume}{45}}, \bibinfo{pages}{4536}
  (\bibinfo{year}{1992}).

\bibitem[{\citenamefont{Andersson and Comer}(2005)}]{Andersson:2006nr}
\bibinfo{author}{\bibfnamefont{N.}~\bibnamefont{Andersson}} \bibnamefont{and}
  \bibinfo{author}{\bibfnamefont{G.}~\bibnamefont{Comer}},
  \bibinfo{journal}{Living Rev.Rel.} \textbf{\bibinfo{volume}{10}},
  \bibinfo{pages}{1} (\bibinfo{year}{2005}), \eprint{gr-qc/0605010}.

\bibitem[{\citenamefont{{London}}(1938)}]{1938Natur.141..643L}
\bibinfo{author}{\bibfnamefont{F.}~\bibnamefont{{London}}},
  \bibinfo{journal}{Nature} \textbf{\bibinfo{volume}{141}},
  \bibinfo{pages}{643} (\bibinfo{year}{1938}).

\bibitem[{\citenamefont{Carter and Langlois}(1995)}]{Carter:1995if}
\bibinfo{author}{\bibfnamefont{B.}~\bibnamefont{Carter}} \bibnamefont{and}
  \bibinfo{author}{\bibfnamefont{D.}~\bibnamefont{Langlois}},
  \bibinfo{journal}{Phys.Rev.} \textbf{\bibinfo{volume}{D51}},
  \bibinfo{pages}{5855} (\bibinfo{year}{1995}), \eprint{hep-th/9507058}.

\bibitem[{\citenamefont{Comer and Joynt}(2003)}]{Comer:2002dm}
\bibinfo{author}{\bibfnamefont{G.}~\bibnamefont{Comer}} \bibnamefont{and}
  \bibinfo{author}{\bibfnamefont{R.}~\bibnamefont{Joynt}},
  \bibinfo{journal}{Phys.Rev.} \textbf{\bibinfo{volume}{D68}},
  \bibinfo{pages}{023002} (\bibinfo{year}{2003}), \eprint{gr-qc/0212083}.

\bibitem[{\citenamefont{Nicolis}(2011)}]{Nicolis:2011cs}
\bibinfo{author}{\bibfnamefont{A.}~\bibnamefont{Nicolis}}
  (\bibinfo{year}{2011}), \eprint{1108.2513}.

\bibitem[{\citenamefont{Son}(2002)}]{Son:2002zn}
\bibinfo{author}{\bibfnamefont{D.~T.} \bibnamefont{Son}}
  (\bibinfo{year}{2002}), \eprint{hep-ph/0204199}.

\bibitem[{\citenamefont{Manuel et~al.}(2005)\citenamefont{Manuel, Dobado, and
  Llanes-Estrada}}]{Manuel:2004iv}
\bibinfo{author}{\bibfnamefont{C.}~\bibnamefont{Manuel}},
  \bibinfo{author}{\bibfnamefont{A.}~\bibnamefont{Dobado}}, \bibnamefont{and}
  \bibinfo{author}{\bibfnamefont{F.~J.} \bibnamefont{Llanes-Estrada}},
  \bibinfo{journal}{JHEP} \textbf{\bibinfo{volume}{09}}, \bibinfo{pages}{076}
  (\bibinfo{year}{2005}), \eprint{hep-ph/0406058}.

\bibitem[{\citenamefont{Alford et~al.}(2007)\citenamefont{Alford, Braby, Reddy,
  and Sch{\"a}fer}}]{Alford:2007rw}
\bibinfo{author}{\bibfnamefont{M.~G.} \bibnamefont{Alford}},
  \bibinfo{author}{\bibfnamefont{M.}~\bibnamefont{Braby}},
  \bibinfo{author}{\bibfnamefont{S.}~\bibnamefont{Reddy}}, \bibnamefont{and}
  \bibinfo{author}{\bibfnamefont{T.}~\bibnamefont{Sch{\"a}fer}},
  \bibinfo{journal}{Phys. Rev.} \textbf{\bibinfo{volume}{C75}},
  \bibinfo{pages}{055209} (\bibinfo{year}{2007}), \eprint{nucl-th/0701067}.

\bibitem[{\citenamefont{Manuel and Llanes-Estrada}(2007)}]{Manuel:2007pz}
\bibinfo{author}{\bibfnamefont{C.}~\bibnamefont{Manuel}} \bibnamefont{and}
  \bibinfo{author}{\bibfnamefont{F.}~\bibnamefont{Llanes-Estrada}},
  \bibinfo{journal}{JCAP} \textbf{\bibinfo{volume}{0708}}, \bibinfo{pages}{001}
  (\bibinfo{year}{2007}), \eprint{arXiv:0705.3909 [hep-ph]}.

\bibitem[{\citenamefont{Mannarelli and Manuel}(2008)}]{Mannarelli:2008jq}
\bibinfo{author}{\bibfnamefont{M.}~\bibnamefont{Mannarelli}} \bibnamefont{and}
  \bibinfo{author}{\bibfnamefont{C.}~\bibnamefont{Manuel}},
  \bibinfo{journal}{Phys. Rev.} \textbf{\bibinfo{volume}{D77}},
  \bibinfo{pages}{103014} (\bibinfo{year}{2008}), \eprint{0802.0321}.

\bibitem[{\citenamefont{Alford et~al.}(2008{\natexlab{a}})\citenamefont{Alford,
  Braby, and Schmitt}}]{Alford:2008pb}
\bibinfo{author}{\bibfnamefont{M.~G.} \bibnamefont{Alford}},
  \bibinfo{author}{\bibfnamefont{M.}~\bibnamefont{Braby}}, \bibnamefont{and}
  \bibinfo{author}{\bibfnamefont{A.}~\bibnamefont{Schmitt}},
  \bibinfo{journal}{J. Phys.} \textbf{\bibinfo{volume}{G35}},
  \bibinfo{pages}{115007} (\bibinfo{year}{2008}{\natexlab{a}}),
  \eprint{0806.0285}.

\bibitem[{\citenamefont{Mannarelli and Manuel}(2010)}]{Mannarelli:2009ia}
\bibinfo{author}{\bibfnamefont{M.}~\bibnamefont{Mannarelli}} \bibnamefont{and}
  \bibinfo{author}{\bibfnamefont{C.}~\bibnamefont{Manuel}},
  \bibinfo{journal}{Phys.Rev.} \textbf{\bibinfo{volume}{D81}},
  \bibinfo{pages}{043002} (\bibinfo{year}{2010}), \eprint{0909.4486}.

\bibitem[{\citenamefont{Alford et~al.}(1999)\citenamefont{Alford, Rajagopal,
  and Wilczek}}]{Alford:1998mk}
\bibinfo{author}{\bibfnamefont{M.~G.} \bibnamefont{Alford}},
  \bibinfo{author}{\bibfnamefont{K.}~\bibnamefont{Rajagopal}},
  \bibnamefont{and} \bibinfo{author}{\bibfnamefont{F.}~\bibnamefont{Wilczek}},
  \bibinfo{journal}{Nucl. Phys.} \textbf{\bibinfo{volume}{B537}},
  \bibinfo{pages}{443} (\bibinfo{year}{1999}), \eprint{hep-ph/9804403}.

\bibitem[{\citenamefont{Andersson}(1998)}]{Andersson:1997xt}
\bibinfo{author}{\bibfnamefont{N.}~\bibnamefont{Andersson}},
  \bibinfo{journal}{Astrophys. J.} \textbf{\bibinfo{volume}{502}},
  \bibinfo{pages}{708} (\bibinfo{year}{1998}), \eprint{gr-qc/9706075}.

\bibitem[{\citenamefont{Jaikumar et~al.}(2008)\citenamefont{Jaikumar, Rupak,
  and Steiner}}]{Jaikumar:2008kh}
\bibinfo{author}{\bibfnamefont{P.}~\bibnamefont{Jaikumar}},
  \bibinfo{author}{\bibfnamefont{G.}~\bibnamefont{Rupak}}, \bibnamefont{and}
  \bibinfo{author}{\bibfnamefont{A.~W.} \bibnamefont{Steiner}},
  \bibinfo{journal}{Phys.Rev.} \textbf{\bibinfo{volume}{D78}},
  \bibinfo{pages}{123007} (\bibinfo{year}{2008}), \eprint{0806.1005}.

\bibitem[{\citenamefont{Andersson et~al.}(2010)\citenamefont{Andersson,
  Haskell, and Comer}}]{Andersson:2010sh}
\bibinfo{author}{\bibfnamefont{N.}~\bibnamefont{Andersson}},
  \bibinfo{author}{\bibfnamefont{B.}~\bibnamefont{Haskell}}, \bibnamefont{and}
  \bibinfo{author}{\bibfnamefont{G.}~\bibnamefont{Comer}},
  \bibinfo{journal}{Phys.Rev.} \textbf{\bibinfo{volume}{D82}},
  \bibinfo{pages}{023007} (\bibinfo{year}{2010}), \eprint{1005.1163}.

\bibitem[{\citenamefont{Alford et~al.}(2012{\natexlab{a}})\citenamefont{Alford,
  Mahmoodifar, and Schwenzer}}]{Alford:2010fd}
\bibinfo{author}{\bibfnamefont{M.}~\bibnamefont{Alford}},
  \bibinfo{author}{\bibfnamefont{S.}~\bibnamefont{Mahmoodifar}},
  \bibnamefont{and}
  \bibinfo{author}{\bibfnamefont{K.}~\bibnamefont{Schwenzer}},
  \bibinfo{journal}{Phys.Rev.} \textbf{\bibinfo{volume}{D85}},
  \bibinfo{pages}{024007} (\bibinfo{year}{2012}{\natexlab{a}}),
  \eprint{1012.4883}.

\bibitem[{\citenamefont{Alford et~al.}(2012{\natexlab{b}})\citenamefont{Alford,
  Mahmoodifar, and Schwenzer}}]{Alford:2011pi}
\bibinfo{author}{\bibfnamefont{M.~G.} \bibnamefont{Alford}},
  \bibinfo{author}{\bibfnamefont{S.}~\bibnamefont{Mahmoodifar}},
  \bibnamefont{and}
  \bibinfo{author}{\bibfnamefont{K.}~\bibnamefont{Schwenzer}},
  \bibinfo{journal}{Phys.Rev.} \textbf{\bibinfo{volume}{D85}},
  \bibinfo{pages}{044051} (\bibinfo{year}{2012}{\natexlab{b}}),
  \eprint{1103.3521}.

\bibitem[{\citenamefont{{Epstein}}(1988)}]{1988ApJ...333..880E}
\bibinfo{author}{\bibfnamefont{R.~I.} \bibnamefont{{Epstein}}},
  \bibinfo{journal}{\apj} \textbf{\bibinfo{volume}{333}}, \bibinfo{pages}{880}
  (\bibinfo{year}{1988}).

\bibitem[{\citenamefont{{Lee}}(1995)}]{1995AA...303..515L}
\bibinfo{author}{\bibfnamefont{U.}~\bibnamefont{{Lee}}},
  \bibinfo{journal}{Astron. Astrophys.} \textbf{\bibinfo{volume}{303}},
  \bibinfo{pages}{515} (\bibinfo{year}{1995}).

\bibitem[{\citenamefont{{Langlois} et~al.}(1998)\citenamefont{{Langlois},
  {Sedrakian}, and {Carter}}}]{1998MNRAS.297.1189L}
\bibinfo{author}{\bibfnamefont{D.}~\bibnamefont{{Langlois}}},
  \bibinfo{author}{\bibfnamefont{D.~M.} \bibnamefont{{Sedrakian}}},
  \bibnamefont{and} \bibinfo{author}{\bibfnamefont{B.}~\bibnamefont{{Carter}}},
  \bibinfo{journal}{Mon.Not.Roy.Astron.Soc.} \textbf{\bibinfo{volume}{297}},
  \bibinfo{pages}{1189} (\bibinfo{year}{1998}),
  \eprint{arXiv:astro-ph/9711042}.

\bibitem[{\citenamefont{{Haskell} et~al.}(2012)\citenamefont{{Haskell},
  {Andersson}, and {Comer}}}]{Haskell:2012vp}
\bibinfo{author}{\bibfnamefont{B.}~\bibnamefont{{Haskell}}},
  \bibinfo{author}{\bibfnamefont{N.}~\bibnamefont{{Andersson}}},
  \bibnamefont{and} \bibinfo{author}{\bibfnamefont{G.~L.}
  \bibnamefont{{Comer}}}, \bibinfo{journal}{\prd}
  \textbf{\bibinfo{volume}{86}}, \bibinfo{eid}{063002} (\bibinfo{year}{2012}),
  \eprint{1204.2894}.

\bibitem[{\citenamefont{Glampedakis et~al.}(2012)\citenamefont{Glampedakis,
  Andersson, and Lander}}]{Glampedakis:2011yw}
\bibinfo{author}{\bibfnamefont{K.}~\bibnamefont{Glampedakis}},
  \bibinfo{author}{\bibfnamefont{N.}~\bibnamefont{Andersson}},
  \bibnamefont{and} \bibinfo{author}{\bibfnamefont{S.~K.}
  \bibnamefont{Lander}}, \bibinfo{journal}{Mon.Not.Roy.Astron.Soc.}
  \textbf{\bibinfo{volume}{420}}, \bibinfo{pages}{1263} (\bibinfo{year}{2012}),
  \eprint{1106.6330}.

\bibitem[{\citenamefont{Alford et~al.}(2008{\natexlab{b}})\citenamefont{Alford,
  Schmitt, Rajagopal, and Schafer}}]{Alford:2007xm}
\bibinfo{author}{\bibfnamefont{M.~G.} \bibnamefont{Alford}},
  \bibinfo{author}{\bibfnamefont{A.}~\bibnamefont{Schmitt}},
  \bibinfo{author}{\bibfnamefont{K.}~\bibnamefont{Rajagopal}},
  \bibnamefont{and} \bibinfo{author}{\bibfnamefont{T.}~\bibnamefont{Schafer}},
  \bibinfo{journal}{Rev.Mod.Phys.} \textbf{\bibinfo{volume}{80}},
  \bibinfo{pages}{1455} (\bibinfo{year}{2008}{\natexlab{b}}),
  \eprint{0709.4635}.

\bibitem[{\citenamefont{Bedaque and Sch{\"a}fer}(2002)}]{BedaqueSchaefer}
\bibinfo{author}{\bibfnamefont{P.~F.} \bibnamefont{Bedaque}} \bibnamefont{and}
  \bibinfo{author}{\bibfnamefont{T.}~\bibnamefont{Sch{\"a}fer}},
  \bibinfo{journal}{Nucl. Phys.} \textbf{\bibinfo{volume}{A697}},
  \bibinfo{pages}{802} (\bibinfo{year}{2002}), \eprint{hep-ph/0105150}.

\bibitem[{\citenamefont{Son}(2001{\natexlab{a}})}]{Son:2001xd}
\bibinfo{author}{\bibfnamefont{D.~T.} \bibnamefont{Son}}
  (\bibinfo{year}{2001}{\natexlab{a}}), \eprint{hep-ph/0108260}.

\bibitem[{\citenamefont{Alford et~al.}(2008{\natexlab{c}})\citenamefont{Alford,
  Braby, and Schmitt}}]{Alford:2007qa}
\bibinfo{author}{\bibfnamefont{M.~G.} \bibnamefont{Alford}},
  \bibinfo{author}{\bibfnamefont{M.}~\bibnamefont{Braby}}, \bibnamefont{and}
  \bibinfo{author}{\bibfnamefont{A.}~\bibnamefont{Schmitt}},
  \bibinfo{journal}{J. Phys.} \textbf{\bibinfo{volume}{G35}},
  \bibinfo{pages}{025002} (\bibinfo{year}{2008}{\natexlab{c}}),
  \eprint{arXiv:0707.2389 [nucl-th]}.

\bibitem[{\citenamefont{Luttinger and Ward}(1960)}]{Luttinger:1960ua}
\bibinfo{author}{\bibfnamefont{J.~M.} \bibnamefont{Luttinger}}
  \bibnamefont{and} \bibinfo{author}{\bibfnamefont{J.~C.} \bibnamefont{Ward}},
  \bibinfo{journal}{Phys. Rev.} \textbf{\bibinfo{volume}{118}},
  \bibinfo{pages}{1417} (\bibinfo{year}{1960}).

\bibitem[{\citenamefont{Baym}(1962)}]{Baym:1962sx}
\bibinfo{author}{\bibfnamefont{G.}~\bibnamefont{Baym}}, \bibinfo{journal}{Phys.
  Rev.} \textbf{\bibinfo{volume}{127}}, \bibinfo{pages}{1391}
  (\bibinfo{year}{1962}).

\bibitem[{\citenamefont{Cornwall et~al.}(1974)\citenamefont{Cornwall, Jackiw,
  and Tomboulis}}]{Cornwall:1974vz}
\bibinfo{author}{\bibfnamefont{J.~M.} \bibnamefont{Cornwall}},
  \bibinfo{author}{\bibfnamefont{R.}~\bibnamefont{Jackiw}}, \bibnamefont{and}
  \bibinfo{author}{\bibfnamefont{E.}~\bibnamefont{Tomboulis}},
  \bibinfo{journal}{Phys. Rev.} \textbf{\bibinfo{volume}{D10}},
  \bibinfo{pages}{2428} (\bibinfo{year}{1974}).

\bibitem[{\citenamefont{Alford et~al.}(in preparation)\citenamefont{Alford,
  Mallavarapu, Schmitt, and Stetina}}]{future}
\bibinfo{author}{\bibfnamefont{M.~G.} \bibnamefont{Alford}},
  \bibinfo{author}{\bibfnamefont{S.~K.} \bibnamefont{Mallavarapu}},
  \bibinfo{author}{\bibfnamefont{A.}~\bibnamefont{Schmitt}}, \bibnamefont{and}
  \bibinfo{author}{\bibfnamefont{S.}~\bibnamefont{Stetina}} (\bibinfo{year}{in
  preparation}).

\bibitem[{\citenamefont{Son}(2001{\natexlab{b}})}]{Son:2000ht}
\bibinfo{author}{\bibfnamefont{D.}~\bibnamefont{Son}},
  \bibinfo{journal}{Int.J.Mod.Phys.} \textbf{\bibinfo{volume}{A16S1C}},
  \bibinfo{pages}{1284} (\bibinfo{year}{2001}{\natexlab{b}}),
  \eprint{hep-ph/0011246}.

\bibitem[{\citenamefont{Gusakov and Andersson}(2006)}]{Gusakov:2006ga}
\bibinfo{author}{\bibfnamefont{M.~E.} \bibnamefont{Gusakov}} \bibnamefont{and}
  \bibinfo{author}{\bibfnamefont{N.}~\bibnamefont{Andersson}},
  \bibinfo{journal}{Mon.Not.Roy.Astron.Soc.} \textbf{\bibinfo{volume}{372}},
  \bibinfo{pages}{1776} (\bibinfo{year}{2006}), \eprint{astro-ph/0602282}.

\bibitem[{\citenamefont{Gusakov}(2007)}]{Gusakov:2007px}
\bibinfo{author}{\bibfnamefont{M.~E.} \bibnamefont{Gusakov}},
  \bibinfo{journal}{Phys. Rev.} \textbf{\bibinfo{volume}{D76}},
  \bibinfo{pages}{083001} (\bibinfo{year}{2007}), \eprint{0704.1071}.

\bibitem[{\citenamefont{Herzog et~al.}(2009)\citenamefont{Herzog, Kovtun, and
  Son}}]{Herzog:2008he}
\bibinfo{author}{\bibfnamefont{C.}~\bibnamefont{Herzog}},
  \bibinfo{author}{\bibfnamefont{P.}~\bibnamefont{Kovtun}}, \bibnamefont{and}
  \bibinfo{author}{\bibfnamefont{D.}~\bibnamefont{Son}},
  \bibinfo{journal}{Phys.Rev.} \textbf{\bibinfo{volume}{D79}},
  \bibinfo{pages}{066002} (\bibinfo{year}{2009}), \eprint{0809.4870}.

\bibitem[{\citenamefont{Herzog and Yarom}(2009)}]{Herzog:2009md}
\bibinfo{author}{\bibfnamefont{C.~P.} \bibnamefont{Herzog}} \bibnamefont{and}
  \bibinfo{author}{\bibfnamefont{A.}~\bibnamefont{Yarom}},
  \bibinfo{journal}{Phys.Rev.} \textbf{\bibinfo{volume}{D80}},
  \bibinfo{pages}{106002} (\bibinfo{year}{2009}), \eprint{0906.4810}.

\bibitem[{\citenamefont{Andersson and Comer}(2011)}]{Andersson:2011zza}
\bibinfo{author}{\bibfnamefont{N.}~\bibnamefont{Andersson}} \bibnamefont{and}
  \bibinfo{author}{\bibfnamefont{G.}~\bibnamefont{Comer}},
  \bibinfo{journal}{Int.J.Mod.Phys.} \textbf{\bibinfo{volume}{D20}},
  \bibinfo{pages}{1215} (\bibinfo{year}{2011}).

\bibitem[{\citenamefont{{Israel}}(1981)}]{1981PhyA..106..204I}
\bibinfo{author}{\bibfnamefont{W.}~\bibnamefont{{Israel}}},
  \bibinfo{journal}{Physica A Statistical Mechanics and its Applications}
  \textbf{\bibinfo{volume}{106}}, \bibinfo{pages}{204} (\bibinfo{year}{1981}).

\bibitem[{\citenamefont{Weldon}(1982)}]{Weldon:1982aq}
\bibinfo{author}{\bibfnamefont{H.~A.} \bibnamefont{Weldon}},
  \bibinfo{journal}{Phys.Rev.} \textbf{\bibinfo{volume}{D26}},
  \bibinfo{pages}{1394} (\bibinfo{year}{1982}).

\bibitem[{\citenamefont{{Taylor} et~al.}(2009)\citenamefont{{Taylor}, {Hu},
  {Liu}, {Pitaevskii}, {Griffin}, and {Stringari}}}]{2009PhRvA..80e3601T}
\bibinfo{author}{\bibfnamefont{E.}~\bibnamefont{{Taylor}}},
  \bibinfo{author}{\bibfnamefont{H.}~\bibnamefont{{Hu}}},
  \bibinfo{author}{\bibfnamefont{X.-J.} \bibnamefont{{Liu}}},
  \bibinfo{author}{\bibfnamefont{L.~P.} \bibnamefont{{Pitaevskii}}},
  \bibinfo{author}{\bibfnamefont{A.}~\bibnamefont{{Griffin}}},
  \bibnamefont{and}
  \bibinfo{author}{\bibfnamefont{S.}~\bibnamefont{{Stringari}}},
  \bibinfo{journal}{\pra} \textbf{\bibinfo{volume}{80}}, \bibinfo{eid}{053601}
  (\bibinfo{year}{2009}), \eprint{0905.0257}.

\bibitem[{\citenamefont{{Arahata} and {Nikuni}}(2009)}]{2009PhRvA..80d3613A}
\bibinfo{author}{\bibfnamefont{E.}~\bibnamefont{{Arahata}}} \bibnamefont{and}
  \bibinfo{author}{\bibfnamefont{T.}~\bibnamefont{{Nikuni}}},
  \bibinfo{journal}{\pra} \textbf{\bibinfo{volume}{80}}, \bibinfo{eid}{043613}
  (\bibinfo{year}{2009}), \eprint{0907.2743}.

\bibitem[{\citenamefont{{Hu} et~al.}(2010)\citenamefont{{Hu}, {Taylor}, {Liu},
  {Stringari}, and {Griffin}}}]{2010NJPh...12d3040H}
\bibinfo{author}{\bibfnamefont{H.}~\bibnamefont{{Hu}}},
  \bibinfo{author}{\bibfnamefont{E.}~\bibnamefont{{Taylor}}},
  \bibinfo{author}{\bibfnamefont{X.-J.} \bibnamefont{{Liu}}},
  \bibinfo{author}{\bibfnamefont{S.}~\bibnamefont{{Stringari}}},
  \bibnamefont{and}
  \bibinfo{author}{\bibfnamefont{A.}~\bibnamefont{{Griffin}}},
  \bibinfo{journal}{New Journal of Physics} \textbf{\bibinfo{volume}{12}},
  \bibinfo{pages}{043040} (\bibinfo{year}{2010}), \eprint{1001.0772}.

\bibitem[{\citenamefont{{Bertaina} et~al.}(2010)\citenamefont{{Bertaina},
  {Pitaevskii}, and {Stringari}}}]{2010PhRvL.105o0402B}
\bibinfo{author}{\bibfnamefont{G.}~\bibnamefont{{Bertaina}}},
  \bibinfo{author}{\bibfnamefont{L.}~\bibnamefont{{Pitaevskii}}},
  \bibnamefont{and}
  \bibinfo{author}{\bibfnamefont{S.}~\bibnamefont{{Stringari}}},
  \bibinfo{journal}{Physical Review Letters} \textbf{\bibinfo{volume}{105}},
  \bibinfo{eid}{150402} (\bibinfo{year}{2010}), \eprint{1002.0195}.

\bibitem[{\citenamefont{{Meppelink} et~al.}(2009)\citenamefont{{Meppelink},
  {Koller}, and {van der Straten}}}]{2009PhRvA..80d3605M}
\bibinfo{author}{\bibfnamefont{R.}~\bibnamefont{{Meppelink}}},
  \bibinfo{author}{\bibfnamefont{S.~B.} \bibnamefont{{Koller}}},
  \bibnamefont{and} \bibinfo{author}{\bibfnamefont{P.}~\bibnamefont{{van der
  Straten}}}, \bibinfo{journal}{\pra} \textbf{\bibinfo{volume}{80}},
  \bibinfo{eid}{043605} (\bibinfo{year}{2009}), \eprint{0909.3455}.

\bibitem[{\citenamefont{{Arahata} et~al.}(2011)\citenamefont{{Arahata},
  {Nikuni}, and {Griffin}}}]{2011PhRvA..84e3612A}
\bibinfo{author}{\bibfnamefont{E.}~\bibnamefont{{Arahata}}},
  \bibinfo{author}{\bibfnamefont{T.}~\bibnamefont{{Nikuni}}}, \bibnamefont{and}
  \bibinfo{author}{\bibfnamefont{A.}~\bibnamefont{{Griffin}}},
  \bibinfo{journal}{\pra} \textbf{\bibinfo{volume}{84}}, \bibinfo{eid}{053612}
  (\bibinfo{year}{2011}), \eprint{1108.2382}.

\bibitem[{\citenamefont{{Adamenko} et~al.}(2009)\citenamefont{{Adamenko},
  {Nemchenko}, {Slipko}, and {Wyatt}}}]{2009PhRvB..79j4508A}
\bibinfo{author}{\bibfnamefont{I.~N.} \bibnamefont{{Adamenko}}},
  \bibinfo{author}{\bibfnamefont{K.~E.} \bibnamefont{{Nemchenko}}},
  \bibinfo{author}{\bibfnamefont{V.~A.} \bibnamefont{{Slipko}}},
  \bibnamefont{and} \bibinfo{author}{\bibfnamefont{A.~F.~G.}
  \bibnamefont{{Wyatt}}}, \bibinfo{journal}{\prb}
  \textbf{\bibinfo{volume}{79}}, \bibinfo{eid}{104508} (\bibinfo{year}{2009}).

\bibitem[{\citenamefont{Son and Surowka}(2009)}]{Son:2009tf}
\bibinfo{author}{\bibfnamefont{D.~T.} \bibnamefont{Son}} \bibnamefont{and}
  \bibinfo{author}{\bibfnamefont{P.}~\bibnamefont{Surowka}},
  \bibinfo{journal}{Phys.Rev.Lett.} \textbf{\bibinfo{volume}{103}},
  \bibinfo{pages}{191601} (\bibinfo{year}{2009}), \eprint{0906.5044}.

\bibitem[{\citenamefont{Landsteiner et~al.}(2011)\citenamefont{Landsteiner,
  Megias, and Pena-Benitez}}]{Landsteiner:2011cp}
\bibinfo{author}{\bibfnamefont{K.}~\bibnamefont{Landsteiner}},
  \bibinfo{author}{\bibfnamefont{E.}~\bibnamefont{Megias}}, \bibnamefont{and}
  \bibinfo{author}{\bibfnamefont{F.}~\bibnamefont{Pena-Benitez}},
  \bibinfo{journal}{Phys.Rev.Lett.} \textbf{\bibinfo{volume}{107}},
  \bibinfo{pages}{021601} (\bibinfo{year}{2011}), \eprint{1103.5006}.

\bibitem[{\citenamefont{Jensen}(2012)}]{Jensen:2012jy}
\bibinfo{author}{\bibfnamefont{K.}~\bibnamefont{Jensen}},
  \bibinfo{journal}{Phys.Rev.} \textbf{\bibinfo{volume}{D85}},
  \bibinfo{pages}{125017} (\bibinfo{year}{2012}), \eprint{1203.3599}.

\bibitem[{\citenamefont{Lin}(2012)}]{Lin:2011mr}
\bibinfo{author}{\bibfnamefont{S.}~\bibnamefont{Lin}},
  \bibinfo{journal}{Nucl.Phys.} \textbf{\bibinfo{volume}{A873}},
  \bibinfo{pages}{28} (\bibinfo{year}{2012}), \eprint{1104.5245}.

\bibitem[{\citenamefont{Bhattacharya et~al.}(2011)\citenamefont{Bhattacharya,
  Bhattacharyya, Minwalla, and Yarom}}]{Bhattacharya:2011tra}
\bibinfo{author}{\bibfnamefont{J.}~\bibnamefont{Bhattacharya}},
  \bibinfo{author}{\bibfnamefont{S.}~\bibnamefont{Bhattacharyya}},
  \bibinfo{author}{\bibfnamefont{S.}~\bibnamefont{Minwalla}}, \bibnamefont{and}
  \bibinfo{author}{\bibfnamefont{A.}~\bibnamefont{Yarom}}
  (\bibinfo{year}{2011}), \eprint{1105.3733}.

\bibitem[{\citenamefont{Neiman and Oz}(2011)}]{Neiman:2011mj}
\bibinfo{author}{\bibfnamefont{Y.}~\bibnamefont{Neiman}} \bibnamefont{and}
  \bibinfo{author}{\bibfnamefont{Y.}~\bibnamefont{Oz}}, \bibinfo{journal}{JHEP}
  \textbf{\bibinfo{volume}{1109}}, \bibinfo{pages}{011} (\bibinfo{year}{2011}),
  \eprint{1106.3576}.

\end{thebibliography}

\end{document}